\documentclass[
  journal=pasa,
  manuscript=research-paper, 
  year=202X,
  volume=YY,
]{cup-journal}

\PassOptionsToPackage{authoryear}{natbib}
\usepackage{natbib}

\usepackage{amsmath}
\usepackage{amssymb,microtype,siunitx,booktabs}
\usepackage{longtable}
\usepackage{booktabs}
\usepackage{multirow}
\usepackage{graphicx}
\usepackage{caption}
\usepackage{subcaption}

\newcommand{\ms}{\ensuremath{\mathrm{m\,s^{-1}}}}

\usepackage{xcolor}
\usepackage{soul}        
\sethlcolor{yellow} 

\title[Hot-cold Jupiters]{Hot Jupiter - Cold Jupiter: A complex sibling relation}

\author{Adriana Errico}
\affiliation{University of Southern Queensland, Centre for Astrophysics, West Street, Toowoomba, QLD 4350, Australia}
\email[Adriana Errico]{aberrico@gmail.com}

\author{Robert A. Wittenmyer}
\affiliation{University of Southern Queensland, Centre for Astrophysics, West Street, Toowoomba, QLD 4350, Australia}

\author{Jonathan Horner}
\affiliation{University of Southern Queensland, Centre for Astrophysics, West Street, Toowoomba, QLD 4350, Australia}

\author{Brad Carter}
\affiliation{University of Southern Queensland, Centre for Astrophysics, West Street, Toowoomba, QLD 4350, Australia}

\author{Valeria López}
\affiliation{Astronomy Department, Williams College, Williamstown, MA, 01267 USA}
\affiliation{University of Southern Queensland, Centre for Astrophysics, West Street, Toowoomba, QLD 4350, Australia}

\received {dd Mmm YYYY}
\revised  {dd Mmm YYYY}
\accepted {dd Mmm YYYY}
\published{22 September 202X}


\begin{document}

\begin{abstract}

A handful of planetary systems hosting a Hot Jupiter have been subsequently found to also host long-period giant planets.  These ``cold Jupiters,'' giant planets residing beyond the snow line ($\sim$3\,au), play an important role in the dynamical evolution of the system as a whole.  In this work, we investigate the detectability of cold Jupiters around a sample of 28 well-studied Hot Jupiter host stars to estimate the occurrence rate of this distinctive system architecture.  We perform extensive simulations using the combination of all publicly available radial velocity (RV) data for those stars with synthetic RV data.  The synthetic data test observing strategies along three axes: cadence, duration, and measurement precision.  For each scenario, we determine detection limits based on the semi-major axis at which a 1 Jupiter mass planet would be recovered 50\% of the time.  We find the following: 1) the existing RV data are remarkably insensitive to these Hot Jupiter/Cold Jupiter pairs;  2) the total baseline over which an observational campaign is carried out is the dominant factor in our ability to detect cold Jupiters; and 3) the results are relatively insensitive to the individual RV measurement precision.  We conclude that metre-class telescopes with lower RV precision are ideally suited to surveying Hot Jupiter-cold Jupiter systems.

\end{abstract}

\section{INTRODUCTION}
\label{sec:int}

Prior to the discovery of the first planets orbiting other stars, it was generally accepted that other planetary systems would resemble the Solar system. With the discovery of the first exoplanets, that assumption was revealed to be wholly incorrect -- with the first detected planets \citep[the pulsar planets, Draugh, Poltergeist and Phobetor/PSR B1257 +12 b, c, and d; and the first "Hot Jupiter" Dimidium/51 Pegasi b;][]{wolszczan1992planetary,wz94,mayor1995jupiter} proving utterly different to anything humanity had expected to find. Over the decades that followed, we discovered that the diversity of exoplanets was far greater than was ever anticipated based solely on our knowledge of the Solar system \citep[e.g.][]{Diverse1,Diverse2,Diverse3}\footnote{For a detailed overview of the links between our knowledge of the Solar system and Exoplanetary Science, we direct the interested reader to \cite{SSRev}, and references therein.}.

The discovery and characterisation of exoplanets has now matured to the point where, in an age where more than 5000 planets are known\footnote{At the time of writing (2026 Jan 2), the number of confirmed exoplanets listed on the NASA Exoplanet Archive \citep{Jessie25} to date stands at 6,087.}, we can make meaningful observational tests of key questions in planet formation. One such question is: How common (or unusual) are systems like the Solar system?  This mystery has tickled the imaginations of humans since they first looked up at the stars and wondered whether there was another Earth out there. For the first time, the current generation of scientists has the data to quantitatively address this question. This endeavour is of vital importance because the coming decades will see the advent of extremely large telescopes and flagship spacecraft missions designed to detect truly Earth-like planets in Earth-like orbits, and the search for life beyond the Solar system will begin in earnest \citep[e.g.][]{HabRev,FujiBS,SchBiosig,HarHab}. 

One issue confounding our attempts to place the Solar system in the context of the `typical' exoplanetary system is that the techniques used to find the vast majority of known exoplanets are strongly biased towards the discovery of planetary systems that are very different to our own \citep[see e.g.][and references therein]{perryman2018exoplanet}. It is far easier to find massive planets orbiting close to their host stars than to find giant planets moving on decades-long orbits, like those seen in the Solar system. Similarly, detecting planets similar to the Solar system's terrestrial worlds remains challenging -- although as technology has improved over the years, such planets are now within our reach. Despite these challenges, however, a number of studies have been carried out attempting to bridge this gap, and to determine how common are planets analogous to Jupiter \citep[e.g.][]{wittenmyer2016anglo,ToastyWit}. The California Legacy Survey \citep{rosenthal2021rvsearch} and the Kepler Giant Planet Survey \citep{WeissKep} are currently carrying out ongoing programs of RV observations to search for such planets.

Earth-size planets appear to be extremely common \citep[e.g.][]{wittenmyer2011frequencyexo,dressing2015occurrence, zhu2018about, kunimoto2020searching}, and several tens of rocky planets are known to orbit in their host stars' habitable zones \citep[e.g.][]{gillon2017seven, vanderburg2020habitable, gilbert2023second}. However, these orbital properties are merely a necessary, but not sufficient, condition for a planet to be genuinely Earth-like \citep[see e.g.][and references therein]{HabRev,FoF4,Walt19,SpaceWeatherHabitability,Milank20,PamMilk}. Occurrence rate studies suggest that there may be an embarrassment of riches in terms of potentially Earth-like planets, with some results indicating that frequency (eta-Earth) to be approaching unity \citep{bryson2021occurrence} -- though large uncertainties remain due to incompleteness. Choosing the best candidates will be a critical task in the years to come.  

The presence of "Jupiter analogues": giant planets moving on low-eccentricity orbits beyond $\sim$5\,au \citep[e.g.][]{wittenmyer2011frequency, zechmeister2013planet, wittenmyer2016anglo, bonomo2023cold}, is thought to be a factor influencing the habitability of inner rocky planets \citep[see e.g.][]{Wetherill94,Wetherill95,FoF1,FoF2,FoF3,Graz16,Ray17,PamMilk}. The low eccentricities of such giant planets could indicate a relatively benign dynamical history for the system, enhancing the probability of an interior rocky planet remaining on a long-term stable, near-circular orbit like our Earth. Distant giant planets with high-eccentricity orbits are also of interest; cold giants that remain beyond the snow line are expected to have retained low-eccentricity orbits, unless perturbed by interactions with other planets. Such planet-planet scattering can result in ejections, or one planet being hurled into a highly elliptical orbit, and would also significantly destabilise the small bodies in the system -- leading to periods of extreme impact rates on any terrestrial planet analogues \citep[e.g.][]{lhb1,lhb2,nesv18}, but potentially also delivering a wealth of volatile material to those planets \citep[e.g.][]{LV1,LV2,LV3,LV4,Ray17}. These dynamical fingerprints are important evidence for determining the frequency of various outcomes of post-formation processes.     

Systems containing highly eccentric cold giant planets are of particular interest as they tell an intriguing story of their misspent youth.  \citet{errico2022HD83443} reported the discovery of a giant planet with a 22-year orbital period and an eccentricity of $e=0.76$ in the HD\,83443 system, which was already known to host a Hot Jupiter \citep{butler2002on}. This system configuration, with a Hot Jupiter and an eccentric distant cold Jupiter, is quite rare, with the HD\,83443 system being only the eighth such system detected \footnote{The other seven being: HD 118203, HD 187123, HD 217107, HD 68988, Pr0211, WASP-53, WASP-8. Star properties are summarised in Table \ref{tab:star_properties}.}. The eccentric outer planet is thought to have originated from a dynamical event which scattered the Hot Jupiter inward while sending the outer planet into its highly elliptical orbit. The discovery naturally raises the question: How common is this outcome? 


\begin{table*}
  \caption{Stellar properties of the eight stars currently known to host a hot Jupiter -- cold Jupiter pair.}
  
  \label{tab:star_properties}
  \begin{tabular}{lccccccc}
    \hline
    Star & Spectral type & Mass  & Radius  & $T_{\rm eff}$ & [Fe/H] & Planet & Parameter \\    
     &  & (M$_\odot$) & (R$_\odot$) & (K) &  & Reference & Reference \\
    \hline
    HD 68988  &  G1V         & &  &  &  & \cite{rosenthal2021rvsearch} & \cite{grieves2018} \\
     &          & 1.18 & 1.20  & 5978 & +0.35 &  & \cite{sweetcat2024} \\
    \hline
    HD 83443  & K0V        &  &   &  &  & \cite{errico2022HD83443} & \cite{cannon1993}\\ 
              &          & 0.97 & 1.02  & 5503 & +0.34 &  & \cite{sweetcat2024} \\
    \hline 
    HD 118203 & G0V       &  &   &  &  & \cite{maciejewski2024tracking} & \cite{grieves2018} \\
              &          & 1.35 & 1.99  & 5872 & +0.27 &  &  \cite{sweetcat2024} \\
    \hline 
    HD 187123 & G2V      &  &  &  &  & \cite{wright2009ten} & \cite{gray2001} \\
              &          & 1.05 & 1.15  & 5837 & +0.12 &  &  \cite{sweetcat2024} \\
              \hline 
    HD 217107 & G5V      &  &   &  &  & \cite{vogt2005five} & \cite{cannon1993}\\
              &          & 1.06 & 1.12  & 5653 & +0.35 &  &  \cite{sweetcat2024} \\
    \hline 
    Pr0211    & G9V        &  &  &  &  & \cite{malavolta2016gaps} & \cite{kraus2007} \\
              &          & 0.935 & 0.827 & 5300 & +0.18 &  &  \cite{malavolta2016gaps} \\
              \hline 
    WASP-53   & K3V        &  &  &  &  & \cite{triaud2017peculiar} & \cite{triaud2017peculiar} \\
              &          & 0.80 & 0.93  & 4993 & +0.23 &  & \cite{sweetcat2024}  \\
              \hline 
    WASP-8    &  G8V      &  &   &  &  & \cite{knutson2014friends} & \cite{salz2015} \\
              &          & 0.99 & 1.01  & 5627 & +0.27 &  & \cite{sweetcat2024} \\
    \hline
  \end{tabular}
\end{table*}

In this work, we examine all known Hot Jupiter systems which have publicly available radial velocity data, and ask how many "HD\,83443-like" systems with an outer cold giant planet (''acquaintances of Hot Jupiters'') could be lurking undetected. We model how future observations will impact the likelihood of finding cold giant planets of different masses at a variety of distances, and examine the roles of observation cadence and instrumental precision on these results. Section 2 describes the extant observational data used to perform the simulations that are described in Section 3. Section 4 gives our results for 28 Hot Jupiter systems: both the current state of detectability for such Hot/Cold Jupiter systems, and the degree to which such planets would be detected under various observing scenarios. Finally in Section 5 we give our conclusions.

\section{Observations} \label{sec:obs}

We selected the Hot Jupiters from the NASA Exoplanet Archive with the following criteria: orbital period $<$ 10 days; semi-major axis $<$ 0.1 au; and mass $>$ 0.3 Jupiter mass (\(M_{\rm Jup}\)). Because we are interested in radial velocity sensitivity to distant cold giant planets, we filtered our sample to include only those 37 Hot Jupiters that were discovered using that method. Whilst hundreds of Hot Jupiters are known from transit surveys, including them here would introduce a severe bias: such planets tend to be confirmed with only a short span of radial velocities, and hence their data would be useless in constraining the presence of long-period planets. We then obtained all publicly-available data from the Data \& Analysis Center for Exoplanets (DACE database), a web platform at the University of Geneva dedicated to extrasolar planets data visualisation, exchange and analysis)\footnote{ \url{https://dace.unige.ch/dashboard/}} and from the HARPS RVBank \citep{trifonov2020harps} for the planets in our sample. 

\begin{figure*}
\centering
\includegraphics[width=\textwidth]{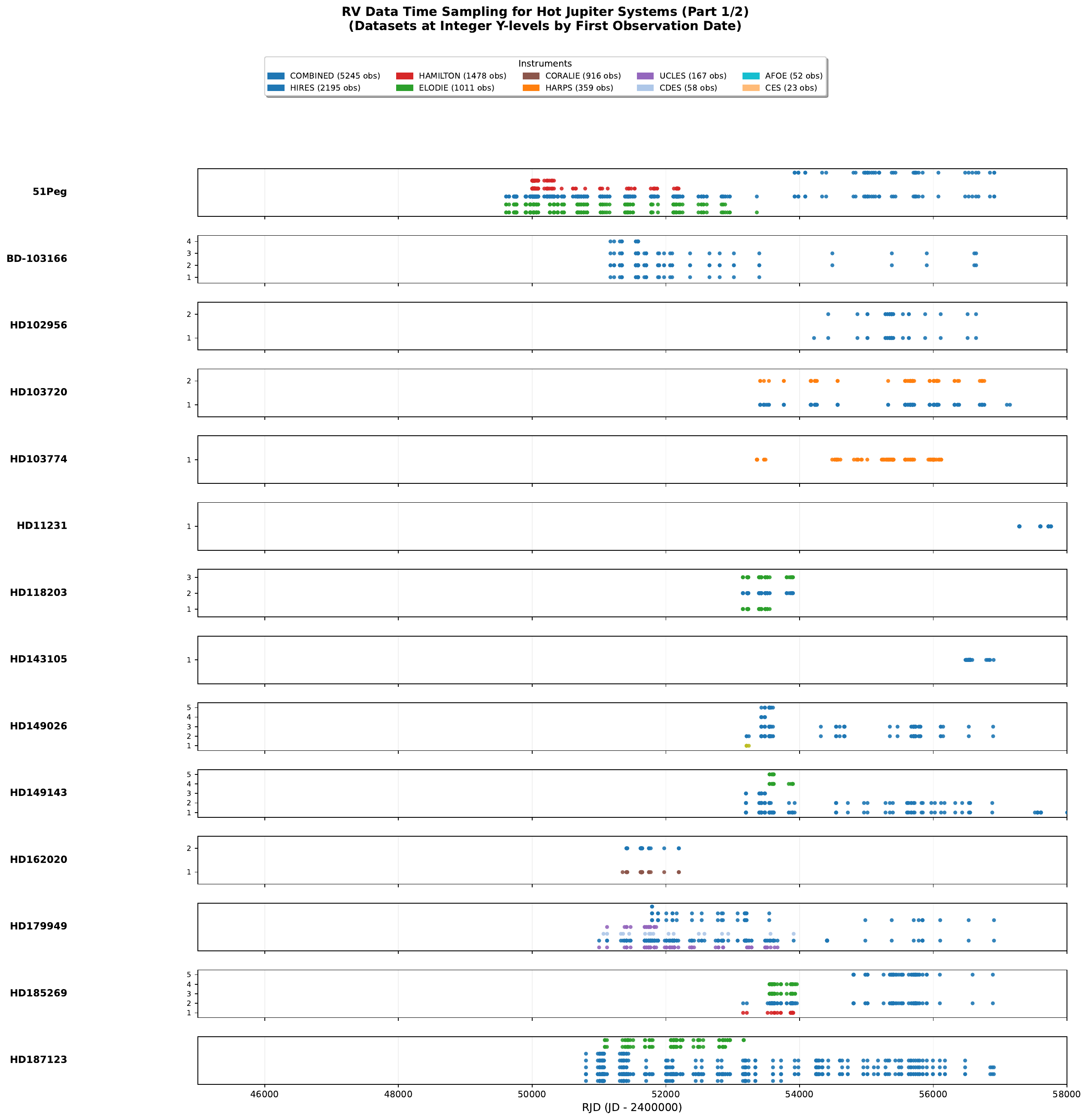}  
\caption{The distribution of RV data used in this work over time for the selected targets (part 1). The instrumental precision of this data, along with additional data, are presented in Table~\ref{tab:observations}.}
\label{fig:data_view_1}
\end{figure*}

\begin{figure*}
\centering
\includegraphics[width=\textwidth]{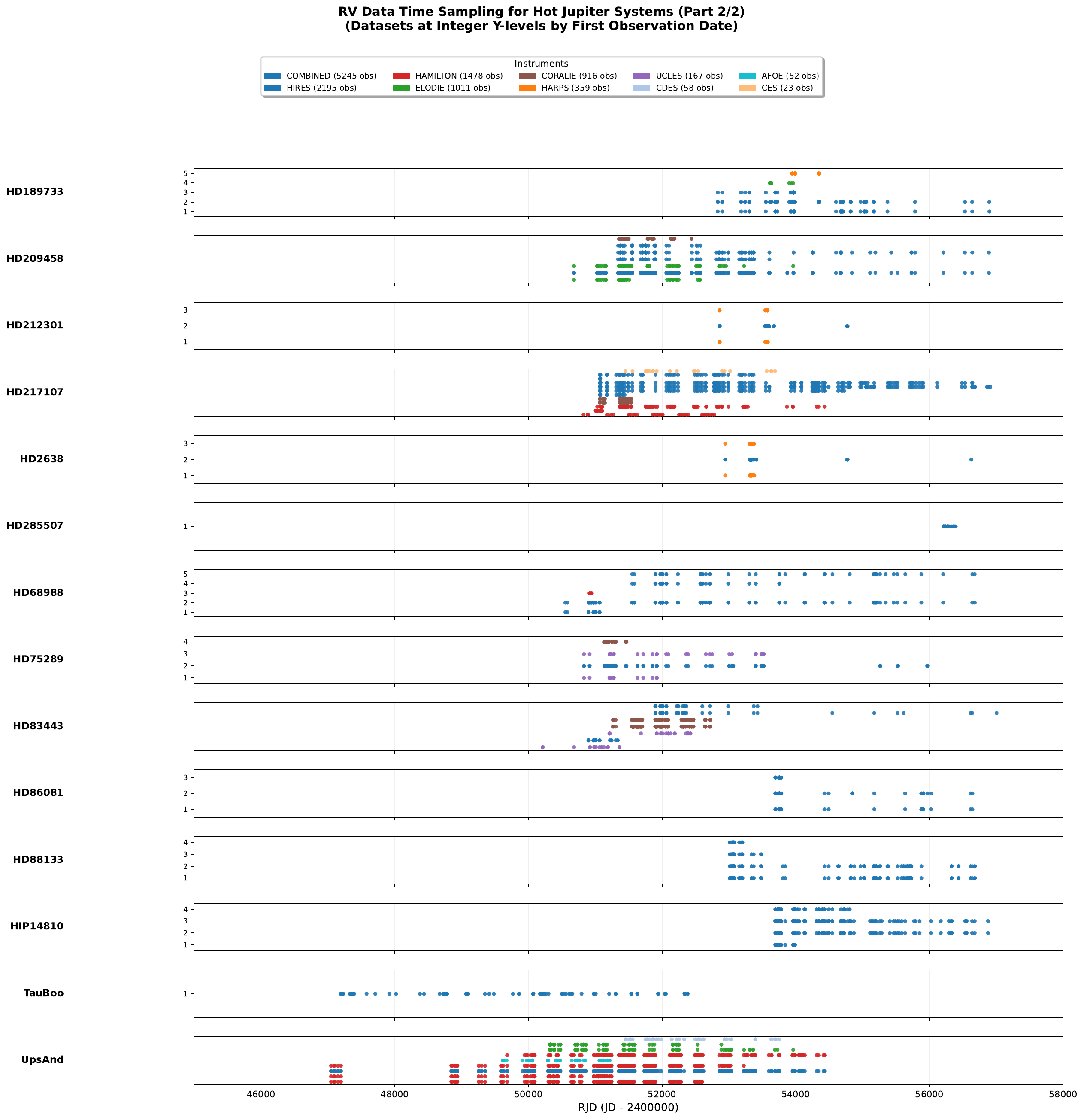}  
\caption{The distribution of RV data used in this work over time for the selected targets (part 2). The instrumental precision of this data, along with additional data, are presented in Table~\ref{tab:observations}.}
\label{fig:data_view_2}
\end{figure*}

Of the 37 Hot Jupiters discovered using the RV method, just 28 fulfil the criteria of having publicly available data -- and thus these 28 planets form the final sample considered in this work. The observational data are summarised in Table~\ref{tab:observations}, and can be graphically visualised in \ref{fig:data_view_1} and \ref{fig:data_view_2}. As expected, the data come from disparate sources and time periods, hence they have offsets and gaps as visualised in Figures \ref{fig:data_view_1} and \ref{fig:data_view_2}. Such features can affect the detectability of further planets; these subtleties have been explored in detail by other work \citep[e.g.][]{witt13, lagrange23, li25}. The main characteristics of the spectrographs selected for this analysis are summarised in Table \ref{tab:spectrographs}

We note that we have carried out additional observations of HD\,83443 using \textsc{Minerva}-Australis \citep{wittenmyer2018understanding,addison2019minerva, addison2021TOI257b,errico2022HD83443} in the time since the 18-year Anglo-Australian Planet Search \citep{tinney01, butler01, twojupiters} ceased operation. Those data amount to 22 observations between 8 February 2019 to 22 February 2021.

\onecolumn

\begin{longtable}{lllcccll}
\caption{Details of the archival data for the planet host systems featured in this work. The start and end dates for each data set are provided, along with the span across which observations were made (the observational baseline). The number of spectra and median precision of that data (in m/s) are given for each star, along with details of the instruments used. A graphical representation of the time coverage, including gaps and overlaps, for each target is given in Figs}~\ref{fig:data_view_1} and ~\ref{fig:data_view_2}.
\label{tab:observations} 
\\
\textbf{Target} & \textbf{Start Date} & \textbf{End Date} & \textbf{Time Span} &  \textbf{\# Spectra} & \textbf{Precision} &\textbf{Instrument} & \textbf{ Citation 
} \\
& & & \textbf{(d)} &  & \textbf{(m/s)} & &  \\

\midrule
\endfirsthead

\multicolumn{8}{c}{{\bfseries \tablename\ \thetable{} -- continued from previous page}} \\
\toprule
\textbf{Target} & \textbf{Start Date} & \textbf{End Date} & \textbf{Time Span} &  \textbf{\# Spectra} & \textbf{Precision} &\textbf{Instrument} & \textbf{ Citation 
} \\
& & & \textbf{(d)} &  & \textbf{(m/s)} & &  \\
\midrule
\endhead

\midrule \multicolumn{7}{r}{{Continued on next page}} \\
\endfoot

\bottomrule
\endlastfoot
BD-10 3166 & 1998 Dec 24 & 2000 Feb 11 & 464 & 84 & 3.69 & HIRES 
& \cite{butler2000planetary}\\ 
{} & 1998 Dec 24 & 2005 Jan 27 & 2226 & & & HIRES 
& \cite{butler2006catalog} \\
{} & 1998 Dec 24 & 2013 Dec 14 & 5469 & & & HIRES 
& \cite{butler2017lces} \\ \midrule
HD\,102956 & 2007 Apr 26 & 2013 Dec 13 & 2423 & 29 & 1.28 & HIRES 
& \cite{Johnson2010hot}\\ \midrule
HD\,103720 & 2005 Feb 9 & 2014 April 16 & 3353 & 153 & 1.92 & HARPS-pre \footnote{We note here that HARPS underwent a fibre upgrade in 2015 \citep[as described in][]{locurto2015}, with the result that data from before and after the change needs to be considered as data from two separate instruments. To make these data explicit, we therefore denote observations made before the change as HARPS-pre, and data after the change as HARPS-post.} & \cite{moutou2015harps}\\ 
{} & 2005 Feb 9 & 2018 March 26 & 4793 & & & HARPS-post & \cite{trifonov2020harps}\\ \midrule
HD\,103774 &  2004 Dec 21 & 2012 July 11 & 2760 & 122 & 3.96 & HARPS-pre & \cite{locurto2013harps}\\
{} & 2004 Dec 21 & 2018 April 2 & 4550 & & & HARPS-post & \cite{trifonov2020harps} \\ \midrule
HD\,11231 & 2015 Sep 21 & 2017 Jan 9 & 476 & 48 & 2.12 & HARPS-post & \cite{haswell2020dispersed}\\ \midrule  
HD\,118203 & 2004 May 25 & 2005 July 1 & 402 & 99 & 14.72  & ELODIE 
& \cite{dasilva2006elodie}\\
{}  & 2004 May 25 & 2006 June 14 & 750 & & & ELODIE 
& \cite{118203DRS}\\ \midrule
HD\,143105 & 2013 Jul 7 & 2014 set 2 & 422 & 100 & 9.18 & SOPHIE+ & \cite{hebrard2016sophie}\\ \midrule
HD\,120136 (Tau Boo) & 1987 June 12 & 2002 Apr 6 & 5412 & 98 & 24.07 & Lick 
& \cite{butler1997three,butler2006catalog} \\ \midrule
HD\,149026 & 2004 July 20 & 2004 Aug 24 & 35 & 98 & 1.86 &  HIRES 
& \cite{sato2005n2k}\\
{} & 2005 Feb 27 & 2005 Aug 20 & 174 & & & HIRES 
& \cite{butler2006catalog} \\
{} & 2005 Feb 25 & 2014 Aug 26 & 3469 & & & HIRES 
& \cite{butler2017lces} \\ \midrule
HD\,149143 & 2005 June 28 & 2006 June 14 & 351 & 121 & 4.37 & ELODIE & \cite{dasilva2006elodie}\\
{} & 2005 June 28 & 2005 Aug 31 & 64 & & & ELODIE 
& \cite{dasilva2006elodie}\\
{} & 2004 July 11 & 2005 April 23 & 286 & & & HIRES 
& \cite{butler2006catalog} \\
{} & 2004 July 11 & 2014 Aug 23 & 3685 & & & HIRES 
& \cite{butler2017lces}\\ \midrule
HD\,162020 & 1999 June 24 & 2001 Oct 13 & 843 & 46 & 10.26 & CORALIE & \cite{udry2002coralie}\\ \midrule
HD\,179949 & 1998 Sep 11 & 2006 June 25 & 2844 & 203 & 4.66 & 2.7m Tull & \cite{tinney2001first}\\
{} & 2000 Sep 5 & 2000 Sep 8 & 3 & & & HIRES & \cite{tinney2001first}\\
{} & 2000 Sep 5 & 2005 June 25 & 1754 & & & HIRES 
& \cite{butler2006catalog}\\
{} & 2000 Sep 5 & 2014 Sep 8 & 5116 & & & HIRES 
& \cite{butler2017lces}\\
{} & 1998 Nov 3 & 2000 Nov 7 & 735 & & & UCLES 
& \cite{tinney2001first}\\
{} & 1998 July 7 & 2005 Oct 27 & 2669 & & & UCLES 
& \cite{butler2006catalog}\\ \midrule
HD\,185269 & 2005 June 30 & 2006 Aug 12 & 409 & 176 & 4.67 & ELODIE &\cite{johnson2006eccentric}\\
{} & 2005 June 30 & 2006 July 17 & 383 & & & ELODIE 
& \cite{dasilva2006elodie}\\
{} & 2004 May 30 & 2006 June 18 & 749 & & & Lick/Hamilton 
& \cite{johnson2006eccentric}\\
{} & 2008 Dec 5 & 2014 Aug 22 & 2086 & & & HIRES 
& \cite{butler2017lces}\\ \midrule
HD\,187123 & 1998 Sep 30 & 2004 June 10 & 2079 & 453 & 4.19 & ELODIE & \cite{butler1998planet}\\
{} & 1998 Sep 30 & 2003 Sep 5 & 1801 & & & ELODIE 
& \cite{naef2004elodie}\\
{} & 1997 Dec 23 & 1999 Sep 18 & 634 & & & HIRES 
& \cite{vogt2000six}\\
{} & 1997 Dec 23 & 2005 Dec 19 & 2918 & & & HIRES 
& \cite{butler2006catalog}\\
{} & 1997 Dec 23 & 2013 July 3 & 5671 & & & HIRES & \cite{Feng15} \\
{} & 1997 Dec 23 & 2014 Sep 6 & 6101 & & & HIRES 
& \cite{butler2017lces}\\ \midrule 
HD\,189733 & 2005 Aug 28 & 2006 Aug 12 & 349 & 309 & 1.74 & ELODIE & \cite{bouchy2005elodie}\\
{} & 2006 July 30 & 2007 Aug 29 & 395 & & & HARPS 
& \cite{triaud2009rossiter}\\
{} & 2003 July 12 & 2006 Aug 21 & 1136 & & & HIRES 
& \cite{butler2006catalog}\\
{} & 2003 July 12 & 2014 Aug 24 & 4061 & & & HIRES 
& \cite{butler2017lces}\\ \midrule
HD\,209458 & 1999 June 26 & 2002 June 14 & 1084 & 551 & 7.33 & CORALIE & \cite{henry2000transiting}\\
{} & 1997 Aug 21 & 2006 Aug 13 & 3280 & & & ELODIE 
& \cite{naef2004elodie}\\
{} & 1997 Aug 21 & 2002 Oct 17 & 1884 & & & ELODIE 
& \cite{naef2004elodie}\\
{} & 1999 June 11 & 2004 Dec 31 & 2030 & & & HIRES 
& \cite{wittenmyer2005system}\\
{} & 1999 June 11 & 2002 Oct 28 & 1235 & & & HIRES 
& \cite{wittenmyer2005system}\\
{} & 1999 June 11 & 2005 Aug 22 & 2264 & & & HIRES 
& \cite{butler2006catalog}\\
{} & 1999 June 11 & 2014 Aug 19 & 5548 & & & HIRES 
& \cite{butler2017lces}\\ \midrule
HD\,212301 & 2003 Aug 5 & 2005 July 28 & 723 & 82 & 3.59 & HARPS-Pre & \cite{locurto2006harps}\\ \midrule
HD\,217107 & 1999 Sep 28 & 2005 Nov 15 & 2240 & 789 & 4.60 & Lick/Hamilton & \cite{fischer1999stars}\\ 
{} & 1998 Sep 11 & 1999 Dec 26 & 471 & & & CORALIE 
&\cite{naef2001coralie}\\
{} & 1998 Sep 11 & 1999 Dec 26 & 471 & & & CORALIE & \cite{Feng15}\\
{} & 1998 July 11 & 1998 Dec 15 & 96 & & & Lick/Hamilton 
& \cite{fischer1999planetary}\\
{} & 1998 Aug 2 & 2007 Nov 23 & 3400 & & & Lick/Hamilton 
& \cite{wright2009ten}\\ 
{} & 1998 Jan 10 & 2003 May 17 & 51 & & & Lick/Hamilton 
& \cite{Feng15}\\
{} & 1998 Sep 12 & 1998 Sep 19 & 7 & & & HIRES 
& \cite{fischer1999planetary}\\
{} & 1998 Sep 12 & 1999 Sep 17 & 21 & & & HIRES 
& \cite{vogt2000six}\\
{} & 1998 Sep 12 & 2004 Dec 28 & 2299 & & & HIRES 
& \cite{vogt2005stars}\\
{} & 1998 Sep 12 & 2008 Sep 12 & 3653 & & & HIRES 
& \cite{wright2009ten}\\
{} & 1998 Sep 12 & 2013 Dec 14 & 5572 & & & HIRES & \cite{Feng15}\\
{} & 1998 Sep 12 & 2014 Sep 8 & 5840 & & & HIRES 
& \cite{butler2017lces}\\ \midrule
HD\,217014 (51Peg) & 1994 Sep 15 & 2004 Dec 21 & 3751 & 730 & 6.45 & ELODIE & \cite{mayor1995jupiter}\\
{} & 1994 Sep 15 & 2003 Sep 5 & 3277 & & & ELODIE 
& \cite{naef2004elodie}\\
{} & 1995 Oct 12 & 1996 Aug 31 & 324 & & & Lick/Hamilton 
&\cite{marcy1997planet}\\
{} & 1995 Oct 12 & 2001 Oct 7 & 2187 & & & Lick/Hamilton 
& \cite{johnson2006eccentric}\\
{} & 2006 July 10 & 2014 Sep 13 & 2987 & & & HIRES 
& \cite{butler2017lces}\\ \midrule
HD\,2638 & 2003 Oct 31 & 2005 Jan 7 & 434 & 94 & 1.59 & HARPS-Pre & \cite{moutou2005harps}\\ \midrule
HD\,285507 & 2012 Sep 26 & 2013 Apr 4 & 32 & 190 & 12.15 & TRES & \cite{quinn2014HD285507}\\ \midrule
HD\,68988 & 1998 April 10 & 1998 May 13 & 33 & 98 & 2.63 & HAMILTON 
& \cite{vogt2002ten}\\
{} & 1997 April 13 & 1998 Sep 8 & 513 & & & HIRES 
& \cite{vogt2002ten}\\
{} & 1997 April 13 & 2006 Jan 18 & 2202 & & & HIRES 
& \cite{butler2006catalog}\\
{} & 1997 April 13 & 2014 Jan 18 & 5124 & & & HIRES 
& \cite{butler2017lces}\\ \midrule
HD\,75289 & 1998 Nov 13 & 1999 Oct 14 & 335 & 151 & 6.59 &  CORALIE 
& \cite{Udry2000fcoralie}\\
{} & 1998 Jan 16 & 2001 Jan 10 & 1090 & & & UCLES 
& \cite{butler2001two}\\
{} & 1998 Jan 16 & 2005 May 28 & 2689 & & & UCLES 
& \cite{butler2006catalog}\\ \midrule
HD\,83443 & 1999 March 25 & 2003 March 19 & 1455 & 147 & 2.75 & CORALIE & \cite{butler2002on}\\
{} & 1998 March 25 & 1999 June 5 & 437 & & & HIRES 
& \cite{vogt2002ten}\\
{} & 1998 March 25 & 2005 Feb 25 & 1529 & & & HIRES 
& \cite{butler2006catalog}\\
{} & 1998 March 25 & 2014 Dec 11 & 5105 & & & HIRES 
& \cite{butler2017lces}\\
{} & 2019 Feb 8 & 2019 Feb 23 & 15 & & & Minerva-ThAr & \cite{errico2022HD83443}\\
{} & 2021 Feb 17 & 2021 Feb 22 & 5 & & & Minerva-Iodine & \cite{errico2022HD83443}\\
{} & 2003 Dec 28 & 2015 May 1 & 4142 & & & HARPS-pre & \cite{moutou2015harps}\\
{} & 2015 Dec 10 & 2016 May 28 & 170 & & & HARPS-post & \cite{trifonov2020harps}\\ 
{} & 1999 Feb 2 & 2015 March 13 & 5883 & & & UCLES & \cite{tinney2001first}\\ \midrule
HD\,86081 & 2005 Nov 19 & 2006 Feb 14 & 87 & 73 & 2.06 & HIRES 
& \cite{butler2006catalog}\\
{} & 2005 Nov 19 & 2013 Dec 14 & 2947 & & & HIRES 
& \cite{butler2017lces}\\ \midrule
HD\,88133 & 2004 Jan 10 & 2004 July 13 & 185 & 114 & 2.19 & HIRES & \cite{fischer2005n2k}\\
{} & 2004 Jan 10 & 2005 April 23 & 469 & & & HIRES 
& \cite{butler2006catalog}\\
{} & 2004 Jan 10 & 2014 Jan 19 & 3662 & & & HIRES 
& \cite{butler2017lces}\\ \midrule
HD\,9826 (Ups And) & 1994 Sep 22 & 1999 Feb 4 & 1596 & 1143 & 8.57 &  AFOE & \cite{butler1997three}\\
{} & 1999 Oct 1 & 2006 Jan 11 & 2294 & & & Hobby-Eberly 
& \cite{wittenmyer2007long}\\
{} & 1996 Aug 23 & 2006 Aug 13 & 3641 & & & ELODIE & \cite{mayor1995jupiter}\\
{} & 1996 Aug 23 & 2003 Sep 3 & 2566 & & & ELODIE 
& \cite{naef2004elodie}\\
{} & 1987 Sep 8 & 1999 March 5 & 4196 & & & Lick/Hamilton 
& \cite{fischer1999planetary}\\
{} & 1987 Sep 8 & 2002 Nov 19 & 5551 & & & Lick/Hamilton 
& \cite{fischer2003planetary}\\
{} & 1987 Sep 8 & 2004 Aug 3 & 6174 & & & Lick/Hamilton 
&\cite{butler2006catalog}\\
{} & 1994 Nov 24 & 2007 Nov 24 & 4748 & & & Lick/Hamilton 
& \cite{wright2009ten}\\ \midrule
HIP\,14810 & 2005 Nov 19 & 2006 Sep 6 & 291 & 176 & 1.12 & HIRES & \cite{butler2006catalog}\\
{} & 2005 Nov 19 & 2008 Dec 5 & 1112 & & & HIRES 
& \cite{wright2009ten}\\
{} & 2005 Nov 19 & 2014 Aug 5 & 3178 & & & HIRES 
& \cite{butler2017lces}\\ \midrule
\end{longtable}

\twocolumn

\begin{table*}
\caption{Instrumental parameters of selected Telescopes.}
\label{tab:spectrographs}
\begin{tabular}{lcccc}
\hline
Instrument                  & Wavelength range (\,nm)   & Resolving power (R)                             & Precision (\ms)     & Telescope \& Size \\
\hline
VLT/ESPRESSO                & 380-788                   & 70{,}000 (4-UT), 140{,}000--190{,}000 (1-UT)    & 0.1                 & VLT UT, 8.2\,m (single UT) / equiv.\ 16\,m (4-UT) \\
Keck/HIRES                  & 370-800                   & $\sim$60{,}000                                  & 3                   & Keck, 10\,m \\
\textsc{Minerva}-Australis  & 480-630                   & $\gtrsim$75{,}000                               & 10                  & 4$\times$0.7\,m CDK700 \\
\hline
\end{tabular}
\end{table*}

Table \ref{tab:stellar_params} summarises the stellar parameters for our 28 target stars. Table \ref{tab:planet_params} shows the planetary parameters of the Hot Jupiter for each of our targets. The host stars considered here are generally mature Solar-type dwarfs, reflecting the selection biases of radial velocity surveys.  The specific selection functions of the multifarious RV surveys included herein are beyond the scope of this work - but in general they tend to select solar type, quiet, single stars as represented in Table~\ref{tab:stellar_params}.  Such stars are therefore well-represented in the exoplanet demographics literature, with cold giant planets found to orbit $\sim$10\% of them \citep[e.g.][]{cumming08, wittenmyer2011frequency, zechmeister2013planet, bryan16, wittenmyer2016anglo, nielsen19, poleski21, gan24}. The effect of the presence of a Hot Jupiter on the occurrence rate of cold Jupiters is unexplored and is the main aim of this work.

\begin{table*}
\caption{Stellar parameters of selected host stars. For consistency, all parameters listed here are sourced from \citet{sweetcat2024}, though we note that they have been rounded for ease of reading.}
\label{tab:stellar_params}
\begin{tabular}{lccccc}
\hline
Star & \(V_{\rm mag}\) & Mass (\(M_\odot\)) & \(T_{\rm eff}\) (K) & \(\log g\) (cgs) & [Fe/H]  \\
\hline 
51 Peg (HD 217014) & 5.46   & 1.07  & 5810 & 4.33  & +0.21  \\
BD-10 3166         & 10.01  & 0.94  & 5417 & 4.32  & +0.34  \\
HD 102956          & 7.85   & 1.48  & 5010 & 3.21  & +0.10  \\
HD 103720          & 9.49   & 0.75  & 4914 & 4.21  & –0.02  \\
HD 103774          & 7.12   & 1.43  & 6586 & 4.48  & +0.31   \\
HD 11231           & 8.57   & 1.49  & 6643 & 4.32  & +0.19  \\
HD 118203          & 8.06   & 1.35  & 5872 & 4.05  & +0.27   \\
HD 143105          & 6.75   & 1.35  & 6381 & 4.41  & +0.17   \\
HD 149026          & 8.14   & 1.32  & 6166 & 4.35  & +0.37  \\
HD 149143          & 7.89   & 1.31  & 5958 & 4.20  & +0.33 \\
HD 162020          & 9.12   & 0.72  & 4751 & 4.26  & -0.06  \\
HD 179949          & 6.25   & 1.24  & 6282 & 4.49  & +0.23   \\
HD 185269          & 6.68   & 1.37  & 6063 & 4.09  & +0.16   \\
HD 187123          & 7.83   & 1.05  & 5837 & 4.37  & +0.12  \\
HD 189733          & 7.68   & 0.75  & 4969 & 4.30  & -0.08  \\
HD 209458          & 7.63   & 1.12  & 6126 & 4.50  & +0.04   \\
HD 212301          & 7.76   & 1.22  & 6261 & 4.49  & +0.20   \\
HD 217107          & 6.18   & 1.06  & 5653 & 4.28  & +0.35   \\
HD 2638            & 9.38   & 0.84  & 5230 & 4.35  & +0.15   \\
HD 285507          & 10.50  & 0.69  & 4523 & 4.14  & +0.10   \\
HD 68988           & 8.19   & 1.18  & 5978 & 4.54  & +0.35   \\
HD 75289           & 6.36   & 1.25  & 6181 & 4.40  & +0.30   \\
HD 83443           & 8.24   & 0.97  & 5503 & 4.30  & +0.34   \\
HD 86081           & 8.70   & 1.29  & 6057 & 4.22  & +0.27   \\
HD 88133           & 8.06   & 1.23  & 5469 & 3.91  & +0.36   \\
HIP 14810          & 8.50   & 1.00  & 5638 & 4.40  & +0.28  \\
$\tau$ Boo         & 4.49   & 1.46  & 6735 & 4.79  & +0.39  \\
$\upsilon$ And     & 4.10   & 1.34  & 6275 & 4.29  & +0.20   \\
\hline
\end{tabular}
\end{table*}

\begin{table*}
\caption{Orbital parameters of hot Jupiter exoplanets used in this work. For each system, we present the parameters as derived in the discovery work, for consistency. Where the discovery work does not provide a given parameter, we give the value presented by the next available work, for completeness. Where no uncertainties were presented, we give the value as published. We note that these numbers are presented solely for the benefit of the reader, as our analysis fully refit each system rather than assuming a prior published orbit for known planets.}
\label{tab:planet_params}
\begin{tabular}{lcccccc}
\hline
Planet & $M$ sin $i$ (\(M_{\rm Jup}\)) & \(a\) (AU) & \(e\) & Period (days) & \(K\) (m/s) & Citation \\
\hline
51 Peg b         & 0.472$\pm$0.039 & 0.0527$\pm$0.0030 & 0.013$\pm$0.012 & 4.230785$\pm$0.000036 & 55.94$\pm$0.69     & \cite{mayor1995jupiter} \\ [4pt] \hline
HD 189733 b      & 1.15$\pm$0.04  & 0.0313$\pm$0.0004  & 0 & 2.219$\pm$0.0005 & 205$\pm$6    & \cite{bouchy2005elodie}   \\ [4pt] \hline
HD 209458 b      & 0.63 & 0.0467  &               & 3.5250$\pm$0.003 &      & \cite{Charbonneau2000}\\
                 &      &         & 0.00$\pm$0.04 &                  & 81.5$\pm$5.5       & \cite{henry2000transiting}\\ [4pt] \hline
$\tau$ Boo b     & 3.87 & 0.0462  & 0.018$\pm$0.016 & 3.3128$\pm$0.0002 & 469$\pm$5    & \cite{butler1997three}   \\ [4pt] \hline
$\upsilon$ And b & 0.68 & 0.057  & 0.109$\pm$0.040 & 4.611$\pm$0.005 & 74.1$\pm$4.0     & \cite{butler1997three}  \\ [4pt] \hline
HD 75289 b       & 0.42 & 0.046  & 0.024$\pm$0.021  & 3.5098$\pm$0.0007 & 54$\pm$1     & \cite{Udry2000fcoralie}     \\ [4pt] \hline
HD 83443 b       & 0.34 & 0.0375  & 0.0520$\pm$0.0500 & 2.98553$\pm$0.00040 & 57.5$\pm$0.2     & \cite{butler2002on}   \\ [4pt] \hline
HD 179949 b      & 0.84$\pm$0.05 & 0.045$\pm$0.004  & 0.05$\pm$0.03  & 3.093$\pm$0.001 & 102.2$\pm$3.0    & \cite{tinney01}   \\ [4pt] \hline
HD 187123 b      & 0.52 & 0.042  & 0.03$\pm$0.03  & 3.097$\pm$0.03 & 72.0$\pm$2.0     & \cite{butler1998planet}   \\ [4pt] \hline
HD 149026 b      & 0.36$\pm$0.03 & 0.042  & 0  & 2.8766$\pm$0.001 & 43.3$\pm$1.2     & \cite{sato2005n2k}   \\ [4pt] \hline
BD-10 3166 b     & 0.48$\pm$0.03 & 0.046  & 0.05$\pm$0.05 & 3.487$\pm$0.001 & 60.6$\pm$1.01   & \cite{butler2000planetary} \\  [4pt] \hline
HD 102956 b      & 0.96$\pm$0.05 & 0.081$\pm$0.002  & 0.048$\pm$0.027 & 6.4950$\pm$0.0004 & 73.7$\pm$1.9   & \cite{Johnson2010hot} \\ [4pt] \hline
HD 103720 b      & 0.620$\pm$0.025 & 0.0498$\pm$0.0008 & 0.086$\pm$0.024 & 4.5557$\pm$0.0001 & 89$\pm$2     & \cite{moutou2015harps}\\ [4pt] \hline
HD 103774 b      & 0.367$\pm$0.022& 0.070$\pm$0.001 & 0.09$\pm$0.04 & 5.8881$\pm$0.0005 & 34.3$\pm$1.8   & \cite{locurto2013harps} \\  [4pt] \hline
DMPP 2-b (HD 11231 b)       & 0.437$^{+0.030}_{-0.059}$ & 0.0664$\pm$0.0005 & 0.078 & 5.2072$^{+0.0002}_{-0.0055}$ & 40.26$^{+2.69}_{-5.34}$   & \cite{haswell2020dispersed} \\ [4pt] \hline
HD 118203 b      & 2.13 & 0.07 & 0.309$\pm$0.014 & 6.1335$\pm$0.0006 &    & \cite{dasilva2006elodie} \\
                 &      &      &                 &                   & 217.0$\pm$3.0 & \cite{stassun2017} \\ [4pt] \hline
HD 143105 b      & 1.21$\pm$0.06 & 0.0379$\pm$0.0009 & <0.07 & 2.1974$\pm$0.0003  & 144.0$\pm$2.6  &\cite{hebrard2016sophie} \\  [4pt] \hline
HD 149143 b      & 1.33 & 0.053 & 0.016$\pm$0.01 & 4.072$\pm$0.70 & 149.6$\pm$3.0  & \cite{fischer2006} \\ [4pt] \hline
HD 162020 b\footnote{We note here that the mass determined for HD 162020 b in the discovery work is greater than 13 M$_{Jup}$, and, as such, that object might well be considered to be a brown dwarf rather than a giant planet. Regardless of its true nature, we have included the system in this work for completeness. It should also be noted that more recent work \citep{stassun2017} obtains a planetary mass for this object, at $M$ sin $i$ of 9.840$\pm$2.750 M$_{\rm Jup}$.}      & 14.4 & 0.074 & 0.277$\pm$0.002 & 8.428198$\pm$0.000056 & 1813$\pm$4 & \cite{udry2002coralie} \\  [4pt] \hline
HD 185269 b      & 0.94 & 0.077 & 0.30$\pm$0.04  & 6.838$\pm$0.001  & 91$\pm$4.5 & \cite{johnson2006n2k}\\  [4pt] \hline
HD 212301 b      & 0.450 & 0.03600 & 0.000 & 2.245715$\pm$0.000028 &    & \cite{locurto2006harps} \\ 
                 &       &         &       &                       & 59.5$\pm$0.7 & \cite{stassun2017} \\[4pt] \hline
HD 217107 b      & 1.37$\pm$0.14 & 0.074$\pm$0.002 & 0.13$\pm$0.02 & 7.1269$\pm$0.00022 & 140.7$\pm$2.6  & \cite{vogt2005five} \\ [4pt] \hline
HD 2638 b        & 0.48 & 0.044 & 0.0 & 3.4442$\pm$0.0002  & 67.4$\pm$0.4   & \cite{moutou2005harps} \\ [4pt] \hline
HD 285507 b      & 0.917$\pm$0.033 &     & 0.086$\pm$0.019 & 6.0881$\pm$0.0018  & 125.8$\pm$2.3  &\cite{quinn2014HD285507} \\ 
                 &                 & 0.060$\pm$0.000 & & & & \cite{stassun2017} \\ [4pt] \hline
HD 68988 b       & 1.9 & 0.071 & 0.14$\pm$0.03 & 6.276$\pm$0.002  & 187$\pm$6  & \cite{vogt2002ten} \\ [4pt] \hline
HD 86081 b       & 1.50 & 0.035 & 0.0080$\pm$0.0040 & 2.1375$\pm$0.0002   & 207.7$\pm$0.8  & \cite{johnson2006n2k} \\ [4pt] \hline
HD 88133 b       & 0.29 & 0.046 & 0.11$\pm$0.05  & 3.415$\pm$0.001  & 35.7$\pm$2.2   & \cite{fischer2005n2k}\\ [4pt] \hline
HIP 14810 b      & 3.84$\pm$0.54 & 0.0692$\pm$0.0040 & 0.1480$\pm$0.0060 & 6.6740$\pm$0.0020  & 420.7$\pm$3.0  & \cite{butler2006catalog} \\
\hline 
\end{tabular}
\end{table*}

\section{Methodology} \label{sec:methodology}

In this section, we describe the two main analysis procedures employed to assess the current and future detectability of long-period "acquaintances" to the known Hot Jupiters studied in this work. First, we examine the sensitivity afforded by the existing data. Then, we perform a suite of simulations under various scenarios of instrumental precision, observing cadence, and temporal baseline. Taken together, these analyses reveal the extent of our knowledge about such cold giant companions and inform the observational strategies required to probe the outer reaches of these planetary systems.

\subsection{Existing Data}

\begin{figure}[htb]
  \centering
  \includegraphics[width=\textwidth]{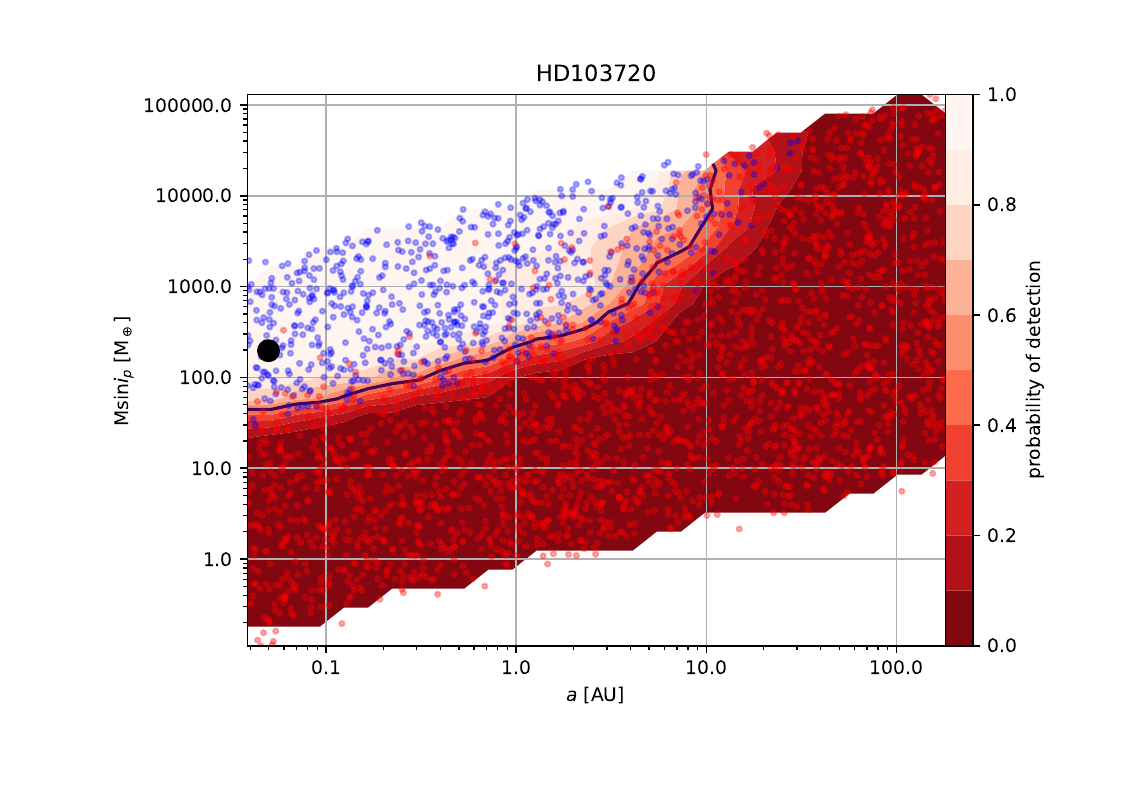} 
  \caption{RVSearch completeness contour plot for HD\,103720. The large black dot represents the periodic signal identified by \texttt{RVSearch} (i.e., the known planet). The small points denote injected synthetic planets. Blue points correspond to recovered signals, while red points were not recovered. Red contours show the detection probability, averaged over small bins in semi-major axis and $m \sin i$ space. The black line marks the 50\% detection probability contour.}
  \label{fig:HD103720_example}
\end{figure}

To assess the sensitivity of the existing data to cold Jupiters, we conducted injection-recovery tests using \texttt{RVSearch} \citep{rosenthal2021rvsearch}, a Python package built on top of RadVel \citep{fulton2018radvel}. The algorithm injects synthetic planetary signals, with orbital periods and minimum masses ($m \sin i$) drawn from log-uniform distributions, eccentricities that are drawn from a beta distribution \citep{kipping2013arametrizing}, and the argument of periapsis drawn from uniform distribution between 0 and 360$^\circ$. It is assumed that the orbits of the planets are co-planar in this injection analysis. \texttt{RVSearch} then attempts to recover each injected signal, estimating the detection probability for planets across a range of parameters. As a graphic output, the algorithm produces a completeness contour plot featuring two clearly distinct regions, as illustrated for HD\,103720 in Figure~\ref{fig:HD103720_example}.

\texttt{RVSearch} has become a widely adopted tool in the exoplanet community, particularly for evaluating survey completeness and population-level detection efficiencies \citep{howardfulton2016,fulton2018radvel,rosenthal2021rvsearch}. Its performance has been validated across diverse RV datasets \citep[e.g.][]{RVref1,RVref2,RVref3}, and it has been instrumental in quantifying detection biases and uncovering the parameter space accessible to current surveys \citep[e.g.][]{RVref4,RVref5,RVref6}. A detailed technical description of the algorithm, including its treatment of multi-instrument datasets and inter-dataset offsets, is provided in \citep{rosenthal2021rvsearch}, to which we refer the reader for full details. For our analysis, we employed the standard \texttt{RVSearch} configuration, with 3000 injected trial signals per star to ensure convergence of the completeness estimates. Tests with larger numbers of injections yielded consistent results, supporting the robustness of this choice, but proved to be prohibitively computationally intensive for such injection numbers to be used for our full analysis. Poorly sampled or highly eccentric orbital configurations are naturally incorporated into the injection–recovery framework, and their contribution to the overall completeness is therefore properly accounted for.

%
%

\subsection{Simulated Data}

To inform the observational strategy, we tested nine scenarios in a 3x3 grid of (low, medium, high) cadence\footnote{low cadence - 4 observations per year (approximately every two months)\\
medium cadence - 8 observations per year (approximately every month)\\
high cadence - 16 observations per year (approximately twice a month)\\} and (low, medium, high) RV measurement precision. We define these categories of measurement precision by representative instruments: "low" -- \textsc{Minerva}-Australis \citep{addison2019minerva}, "medium" -- Keck/HIRES \citep{vogt1994}, "high" -- VLT/ESPRESSO \citep{pepe2014ESPRESSO}.  We wish to test the relative importance of precision and cadence for this science goal, and these values of cadence and precision for the simulated observations are chosen to bracket the range of typical RV survey programmes.  Clearly, one would like to observe at the highest cadence and the highest precision, but the realities of telescope time allocation preclude such a scenario. In general, precision and cadence (i.e. the amount of time available) are inversely related. Small telescopes, such as those that make up the \textsc{Minerva}-Australis array \citep{addison2019minerva}, can dedicate a great deal of time to such a project, whereas larger telescopes with more in-demand instruments (e.g. Keck/HIRES, VLT/ESPRESSO) would naturally be under far more competitive pressure, limiting the available time.  

\begin{figure*}[ht]
\centering
\setlength{\tabcolsep}{4pt} 
\renewcommand{\arraystretch}{0.1} 

\begin{tabular}{c c c}
    & \textbf{Low cadence -- 3 years} & \textbf{High cadence -- 12 years} \\

    \raisebox{7.0em}{\rotatebox{90}{\textbf{\textsc{Minerva}-Australis}}} &
    \subcaptionbox{}{\includegraphics[width=0.42\textwidth]{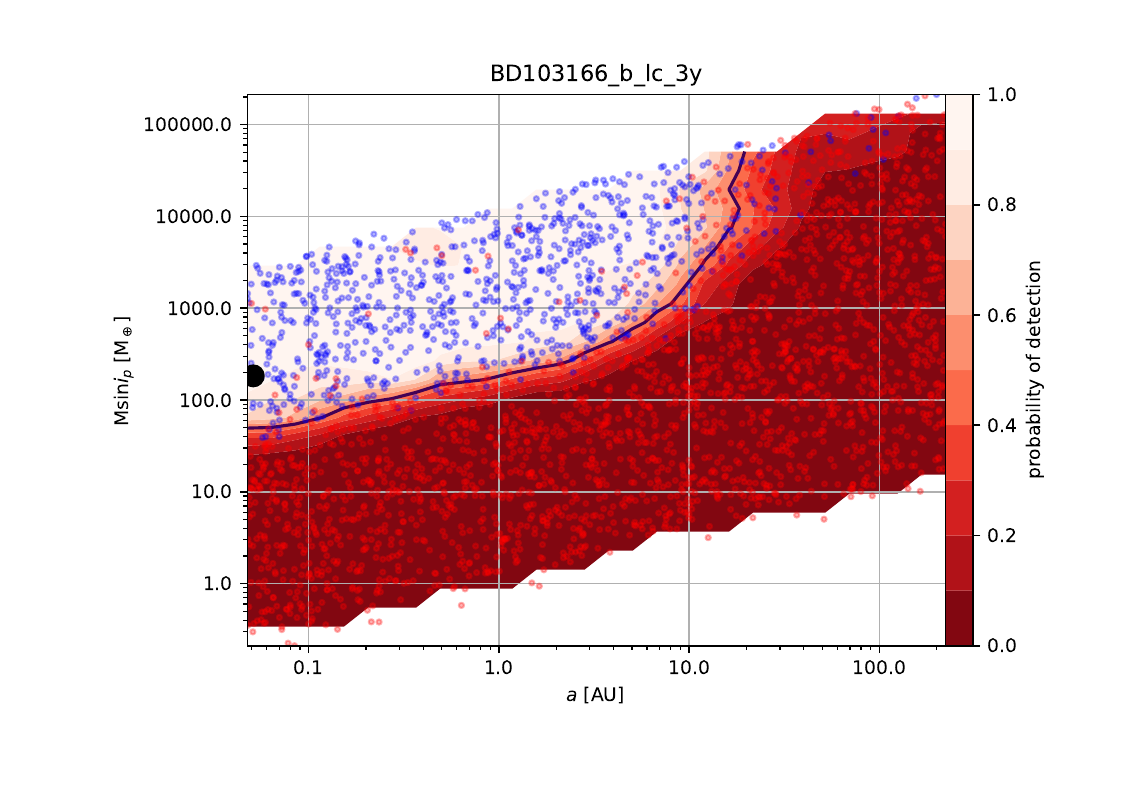}} &
    \subcaptionbox{}{\includegraphics[width=0.42\textwidth]{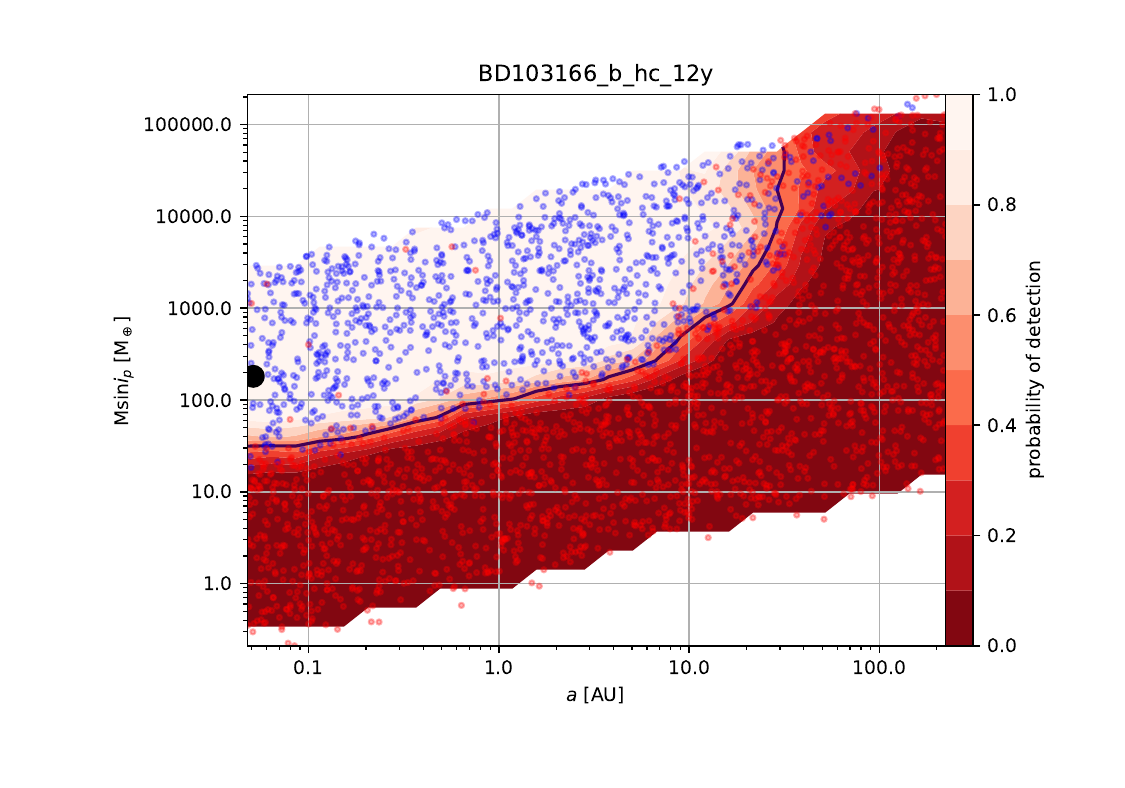}} \\

    \raisebox{7.0em}{\rotatebox{90}{\textbf{Keck/HIRES}}} &
    \subcaptionbox{}{\includegraphics[width=0.42\textwidth]{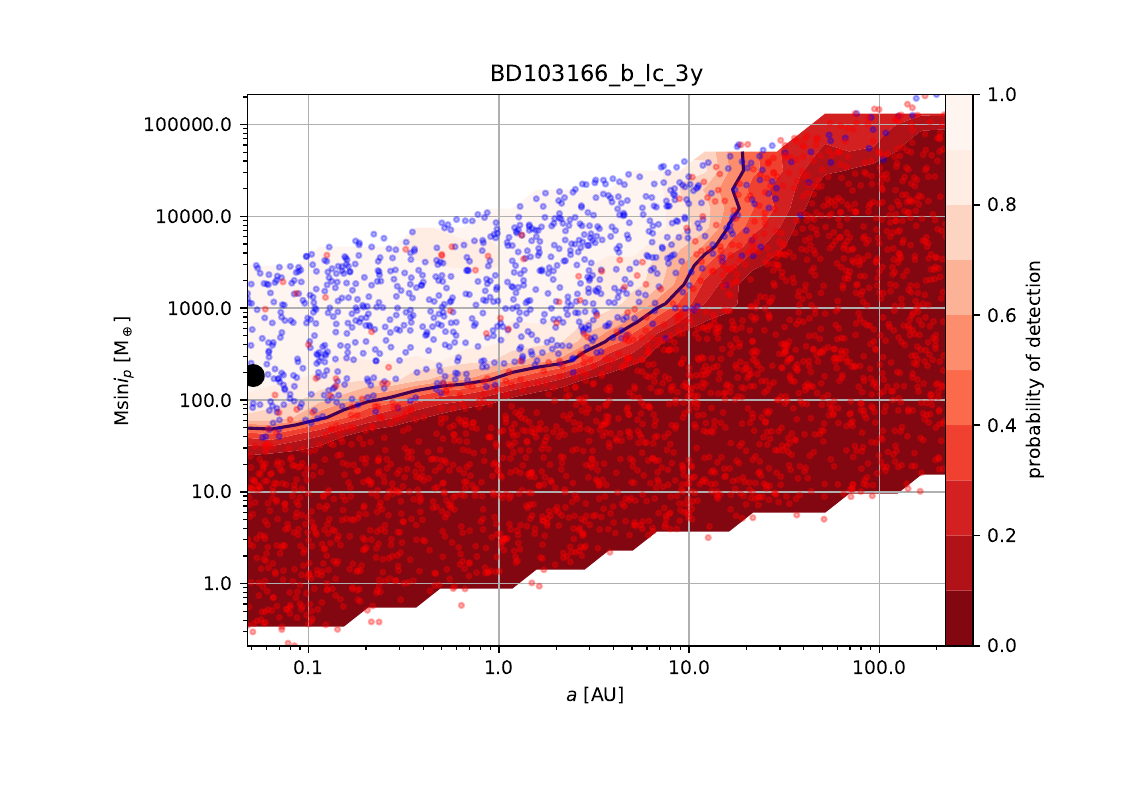}} &
    \subcaptionbox{}{\includegraphics[width=0.42\textwidth]{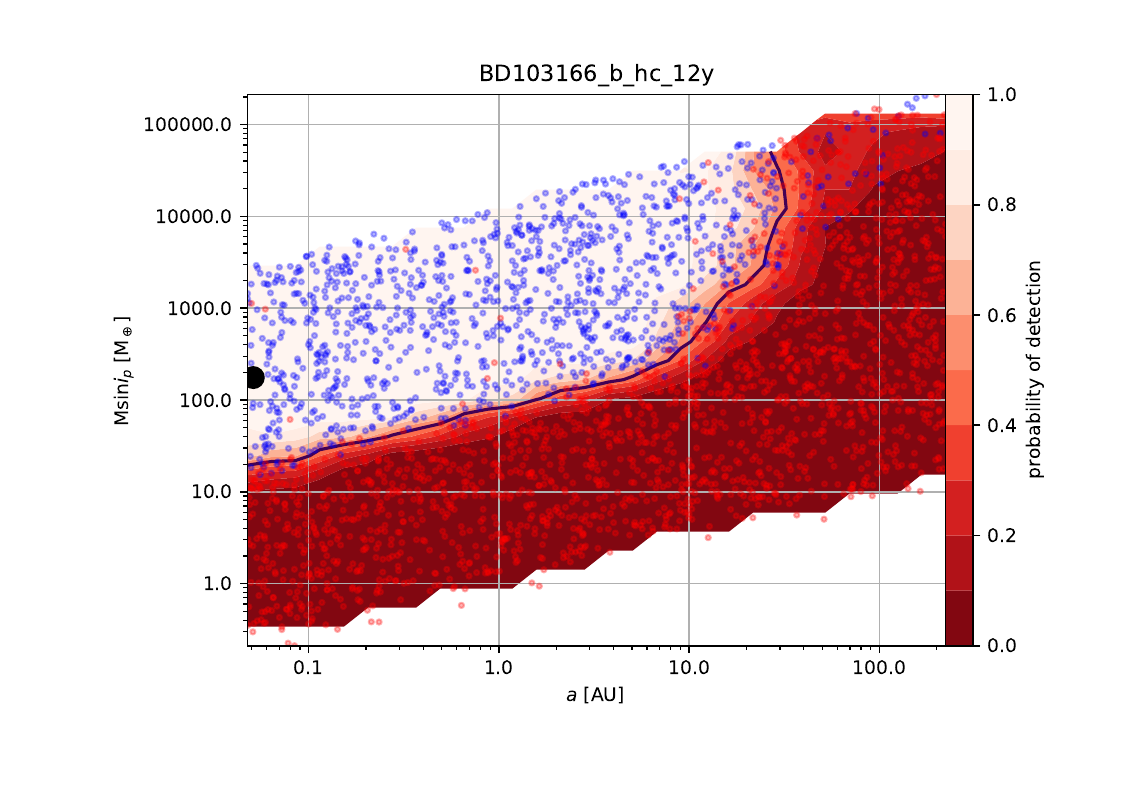}} \\

    \raisebox{7.0em}{\rotatebox{90}{\textbf{VLT/ESPRESSO}}} &
    \subcaptionbox{}{\includegraphics[width=0.42\textwidth]{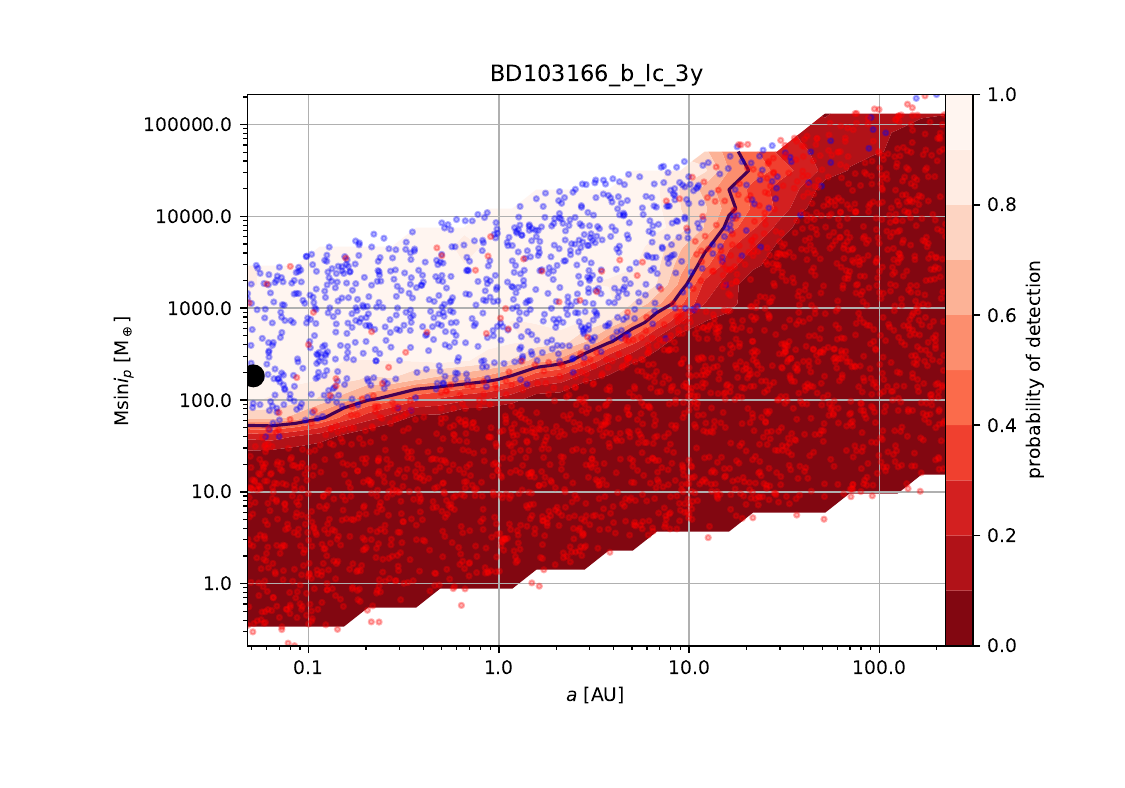}} &
    \subcaptionbox{}{\includegraphics[width=0.42\textwidth]{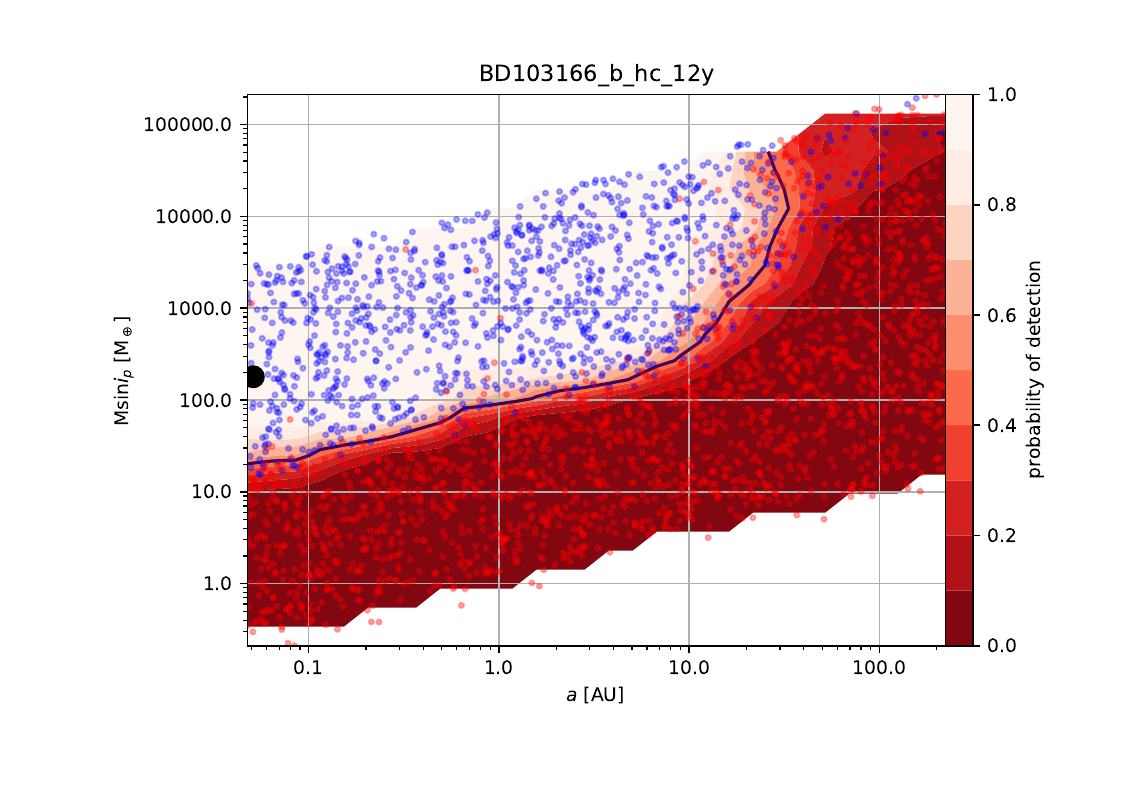}} \\
\end{tabular}

\caption{Comparison of different instruments and observing strategies. 
It can be noted that the line corresponding to 50\% detection shifts downward and to the right as both the cadence and duration increase. 
Although subtle, differences are also seen between telescopes due to varying measurement precision.}
\label{fig:high_vs_low_3_tel}
\end{figure*}

To test the relative importance of these factors, we simulate observing campaigns for the 28 stars examined herein. We also test three different temporal baselines for the new observations: 3, 6, and 12 years. The purpose is to determine the degree to which the observational baseline can overcome lower cadence and/or instrumental precision. The final result is a set of 27 total observing scenarios for each of our 28 selected systems. Since it is not possible to include such a large number of figures, we include Figure \ref{fig:high_vs_low_3_tel} to illustrate the improvement in detection when comparing low- and high-cadence observations for the three selected telescopes. Again, it is obvious that the highest cadence, at the highest precision, for the longest time, will always produce the best results. Our purpose here is to provide a matrix of possibilities to inform and optimise observing strategies to address the science goal of measuring the occurrence rate of Hot Jupiter-Cold Jupiter pairs. 

We generated simulated RVs using the fitted parameters of the known planet(s) to produce Keplerian orbits, to which noise was added.  The value of the noise was drawn with replacement from the residuals of a fit to the known planet(s) performed with the DACE tool \citep{fulton2018radvel}.  
We draw the uncertainty of each simulated measurement from a Gaussian distribution with means and standard deviation as follows in units of \ms: "low precision": $\mu$ = 10 and $\sigma$ = 2; "medium precision": $\mu$ = 3 and $\sigma$ = 0.6; "high precision": $\mu$ = 0.5 and $\sigma$ = 0.1.  These were chosen to approximate, respectively, instruments akin to MINERVA and \textsc{Minerva}-Australis \citep{swift2015minerva, wilson2019FirstRV,addison2019minerva}, Keck/HIRES \citep{vogt1994}, and VLT/ESPRESSO \citep{pepe2014ESPRESSO}.  Hereafter we refer to these instruments by name as exemplars of the three levels of measurement precision tested.

\section{Results}\label{sec:results_3x9}

In this section, we present the results of our detection sensitivity analysis for the currently available data, then show the results of the simulated scenarios to illustrate the effectiveness of various instruments and observing strategies.  

\subsection{Sensitivity of existing data to cold giant companions}

\begin{figure*}[htbp]
    \centering
        \makebox[0.49\textwidth][c]{\textbf{(a) Linear scale}}%
        \hfill
        \makebox[0.43\textwidth][c]{\textbf{(b) Log scale}}
        \par\vspace{2mm}
    \begin{subfigure}[t]{0.39\textwidth}
        \centering 
        \includegraphics[width=\linewidth]{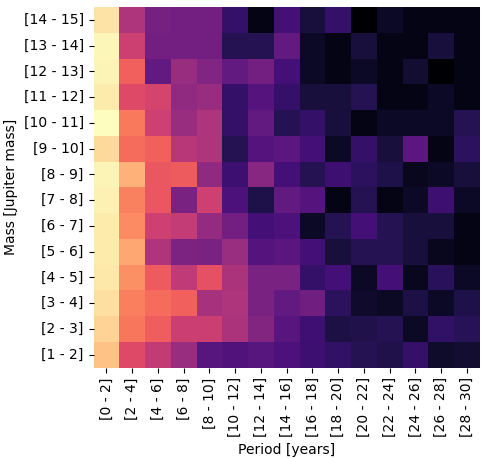}
        \label{fig:left}
    \end{subfigure}%
    \hfill
    \begin{subfigure}[t]{0.61\textwidth}
        \centering
        \raisebox{-14.1mm}{\includegraphics[width=\linewidth]{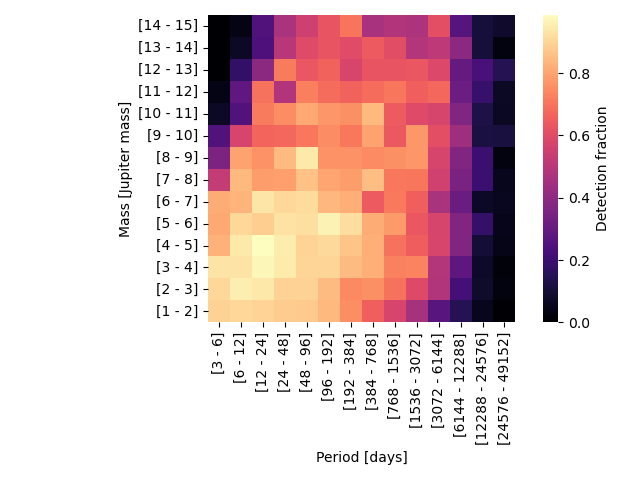}}
        \label{fig:right}
    \end{subfigure}
     \caption{Heat maps showing the detection efficiency for exoplanet candidates injected to the existing observational radial velocity data for 28 known Hot Jupiter host stars. A total of 3,000 unique potential candidate companions were injected, one at a time, for each star considered, with masses and orbital periods each drawn from a log-normal distribution. In both panels, the colour of a square denotes the detection efficiency for a planet in that square -- the mean of the detection efficiencies (number of detections divided by the number of injected planets) across the sample of 28 stars. The left panel shows the data as a function of orbital period, whilst the right shows the same data as a function of log(period). It is clear that planets are most easily detected at short orbital periods, with true Jupiter and Saturn analogues (P$\sim$ 12 and 29 years, respectively) challenging to detect based on current data. The ``void'' at the top left of the right hand panel is an artefact of the injection-recovery process; the signals injected by \texttt{RVSearch} all had radial velocities of less than 1000 ms$^{-1}$. As a result, no planets were generated in the region of the `void’, and so recovery of planets in that region was impossible as there were none to recover.}
    \label{fig:lin-log_1}
\end{figure*}

Despite these being well-studied and bright Hot Jupiter hosts, the existing data are remarkably ineffective at detecting cold Jupiters (Table~\ref{tab:1Jmass_a_real:data}).  Figure \ref{fig:lin-log_1} shows the detection sensitivity for cold Jupiters averaged over our 28 Hot Jupiter systems. From these results, we see that in general, 49.91 percent of cold Jupiters beyond 3 au ($P\sim$5 years) could be detected with the currently available data.  To aid in interpreting the heat maps in this Section, we note that the typical uncertainty of the location of the \texttt{RVSearch} 50\% recovery boundary (as denoted by the thick black line in Figure~\ref{fig:HD103720_example}) is about 10\% in semimajor axis space. Furthermore, it is worth noting that \texttt{RVSearch} correctly identified the known Hot Jupiter in 98.8\% of trials.

%
%

\citet{zink23} presented a completeness-corrected occurrence rate for Hot Jupiter-Cold Jupiter pairs of nearly 100\%, with a steeply rising power-law toward longer periods. The extremely low completeness shown in our sample is consistent with this interpretation: that such "acquaintances of Hot Jupiters" should be common (approaching 100\%) but a great many have so far been missed. To better understand these systems and their architectures, it is therefore profitable to continue to observe these systems. In the next subsection, we show the results of our simulated observing strategies designed to identify the most efficient route to reveal the population of these cold companions lurking in the outer darkness.   

\begin{table*}
\caption{Semi-major axes (in AU) at which 50\% of 1~$M_\mathrm{J}$ planets would be detected, should they be present in the system in question, based upon an injection-recovery analysis of the currently available radial velocity data. The data reveals significant differences in our ability to detect Cold Jupiters across this population of 28 stars for which long-period radial velocity data are available.}
\label{tab:1Jmass_a_real:data}
\centering
\begin{tabular}{lc}
\toprule
\textbf{Target}       & \textbf{semi-major axis} (AU) \\ 
\midrule
51Peg                 & 5.85  \\ 
BD-10 3166              & 2.69  \\ 
HD102956              & 1.28  \\ 
HD103720              & 2.31  \\ 
HD103774              & 0.98  \\ 
HD11231               & 0.10  \\ 
HD118203              & 0.57  \\ 
HD143105              & 0.44  \\ 
HD149026              & 2.01  \\ 
HD149143              & 2.60  \\ 
HD162020              & 1.07  \\ 
HD179949              & 1.17  \\ 
HD185269              & 3.49  \\ 
HD187123              & 9.98  \\ 
HD189733              & 0.52  \\ 
HD209458              & 6.21  \\ 
HD212301              & 0.72  \\ 
HD217107              & 9.09  \\ 
HD2638                & 0.60  \\ 
HD285507              & 0.59  \\ 
HD68988               & 2.81  \\ 
HD75289               & 4.90  \\ 
HD83443               & 7.57  \\ 
HD86081               & 1.11  \\ 
HD88133               & 3.76  \\ 
HIP14810              & 2.27  \\ 
TauBoo                & 0.10  \\ 
UpsAnd                & 2.34  \\ 
\bottomrule
\end{tabular}
\end{table*}

\subsection{Nine Future Observation Scenarios for Three Different Exoplanet Observation Facilities}

In Figure~\ref{fig:AU_for_1Jmass}, we summarise the results of our simulated scenarios for each of the three observatories considered in this work (MINERVA/\textsc{Minerva}-Australis, Keck/HIRES, and VLT/ESPRESSO). In each panel, we present a three-by-three grid showing the semi-major axis at which the facility has a 50\% chance of recovering a $1$ \(M_{\rm Jup}\) planet for each combination of observing cadence (low, medium, high) and baseline of additional observations (3, 6, or 12 years)\footnote{These values correspond to the semi-major axis of the thick black contour in the exemplar results from \texttt{RVSearch}, shown in Figure~\ref{fig:HD103720_example}, for a mass of $1$ \(M_{\rm Jup}\)}. Each box represents the mean semi-major axis across the 28 stars in our sample. The three facilities in the figure are characterised by assumed instrumental RV precisions of $10\,\ms$ for \textsc{Minerva}-Australis -- Figure~\ref{fig:a_heatmap_Minerva}; $3\,\ms$ for Keck/HIRES -- Figure~\ref{fig:a_heatmap_Keck}; and $0.5\,\ms$ for VLT/ESPRESSO -- Figure~\ref{fig:a_heatmap_ESPRESSO}.
/hl{For reference, the complete data for each telescope, for each combination of cadence and duration are given in Tables~\ref{tab:1Jmass_a_Minerva}, \ref{tab:1Jmass_a_Keck} and \ref{tab:1Jmass_a_ESPRESSO}

\begin{figure*}
    \centering
    \begin{subfigure}[t]{0.3\textwidth}
        \centering
        \includegraphics[width=\linewidth]{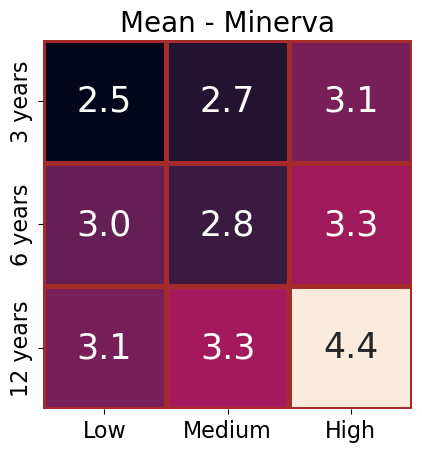}
        \caption{\textsc{Minerva}-Australis (assumed RV precision $10\,\ms$)}
        \label{fig:a_heatmap_Minerva}
    \end{subfigure}
    \hfill
    \begin{subfigure}[t]{0.3\textwidth}
        \centering
        \includegraphics[width=\linewidth]{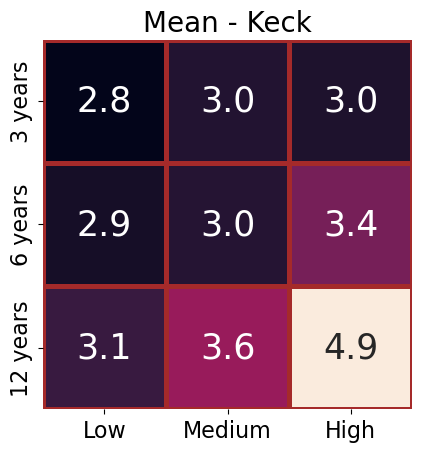}
        \caption{Keck/HIRES (assumed RV precision $3\,\ms$)}
        \label{fig:a_heatmap_Keck}
    \end{subfigure}
    \hfill
    \begin{subfigure}[t]{0.37\textwidth}
        \centering
        \includegraphics[width=\linewidth]{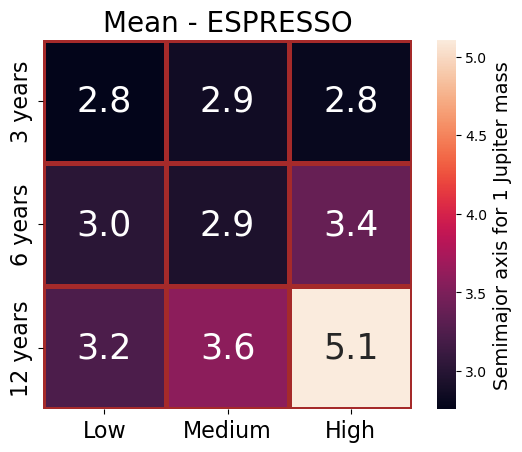}
        \caption{VLT/ESPRESSO (assumed RV precision $0.5\,\ms$)}
        \label{fig:a_heatmap_ESPRESSO}
    \end{subfigure}
    \caption{Cadence vs duration matrices showing the mean semi-major axis at which a $1$ \(M_{\rm Jup}\) planet was detectable 50\% of the time for our simulated scenarios, based on the 28 stars in our sample. The three sub-figures (a, b, and c) present results for three exoplanet facilities - \textsc{Minerva}-Australis, Keck/HIRES, and VLT/ESPRESSO, assuming mean RV precisions for those facilities of 10, 3 and 0.5 $\ms$, respectively. In a given sub-figure, we show how the ability of each facility to detect Cold Jupiters is affected by different observation scenarios -- temporal baselines for new observations of 3, 6, and 12 years (y-axis), and various observation cadences through that period (low, medium, and high cadence; x-axis). The colour and value of each tile gives the distance (in AU) at which the facility have a 50\% chance of being able to detect a $1$ \(M_{\rm Jup}\) planet, averaged across our sample of 28 target stars. It is clear that the ability of each facility to detect such planets improves with higher observation cadence and longer baselines, as expected. Whilst increased RV precision yields some improvements in the detection distance, these improvements are somewhat less pronounced than might be expected.}
    \label{fig:AU_for_1Jmass}
\end{figure*}

The first detail to notice is that for lower cadence and shorter duration, the semi-major axis at which a $1$ \(M_{\rm Jup}\) planet is detectable is smaller than for scenarios with higher cadence and longer duration, as one would intuitively expect. The second detail regarding the comparison of telescopes is that the highest-precision instrument detects a $1$ \(M_{\rm Jup}\) planet at a larger semi-major axis than lower-precision instruments for a given combination of cadence and duration; this is again consistent with expectations. However, it is perhaps surprising to note that the gains made from greatly improved RV precision are only small, compared to those accrued as a result of longer baselines or higher observation cadence.

Indeed, it is particularly interesting that the detection semi-major axis achieved by the facility with the lowest RV precision (\textsc{Minerva}-Australis) for combinations of higher cadence and longer durations equals or exceeds the performance of more precise instruments at lower cadence and/or shorter durations. These results show that the disadvantage of lower measurement precision can be compensated for by higher cadence -- demonstrating the ongoing value of ''regular'' RV observations in an extreme-precision (EPRV) world. In other words, an VLT/ESPRESSO-like EPRV instrument may deliver measurement precision 10-20 times better than a \textsc{Minerva}-Australis-like facility, but, when those facilities are used to search for Cold Jupiters, the ''lesser'' (and more accessible) facility achieves the desired outcome.  

For each star (and each scenario), \texttt{RVSearch} produces a table containing the orbital parameters of the injected signals (planets) and whether they were recovered. We considered the space defined by planet mass and period and divided it into "boxes" (for example, intervals of one Jupiter mass and two years). As an illustration, the first box (located at the bottom left of the heat map in Figure \ref{fig:lin-log_1}) spans from 1 to 2 Jupiter masses and from 0 to 2 years.  The right panel (Figure \ref{fig:lin-log_1}) shows the same result on a log scale.

It is interesting to directly compare the characteristics of the real observational data (presented in Figure~\ref{fig:lin-log_1}) and the simulated data (Figure~\ref{fig:lin-log_2}). The high degree of similarity between the two figures is encouraging -- revealing that the simulated data has very similar characteristics to the real data. In other words, this suggests that the methodology behind the creation of the simulated data is robust, and accurately reflects the reality of the observations made.

For each mass–period box, we computed the ratio of detected to injected signals (this ratio is always less than or equal to 1) and visualised the results as a heat map. This process yielded 756 heat maps, which we then averaged box-by-box across all stars, resulting in 9 heat maps that represent the average detection rates over the 28 stars.

\begin{figure*}[htbp]
    \centering
        \makebox[0.49\textwidth][c]{\textbf{(a) Linear scale}}%
        \hfill
        \makebox[0.43\textwidth][c]{\textbf{(b) Log scale}}
        \par\vspace{2mm}
    \begin{subfigure}[t]{0.43\textwidth}
        \centering
        \includegraphics[width=\linewidth]{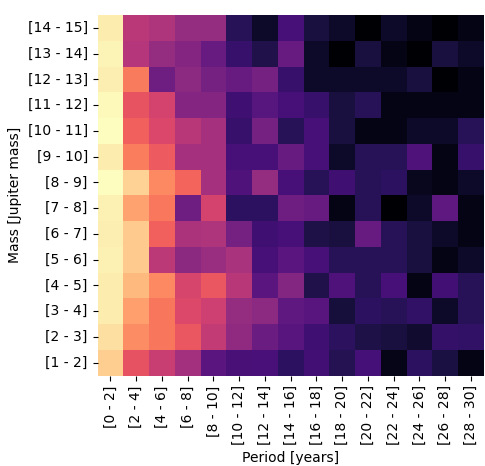}
        \label{fig:left2}
    \end{subfigure}%
    \hfill
    \begin{subfigure}[t]{0.515\textwidth}
        \centering
        \raisebox{-13.7mm}{\includegraphics[width=\linewidth]{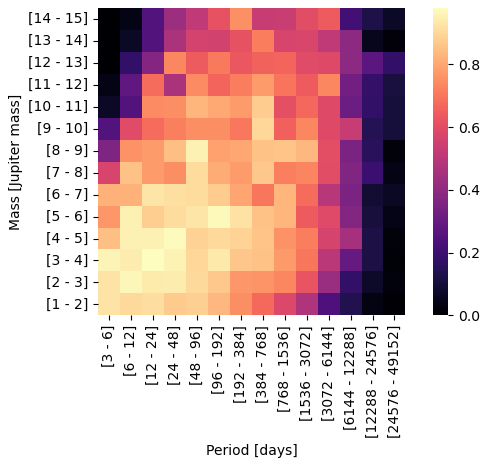}}
        \label{fig:right2}
    \end{subfigure}

    \vspace{4mm}

    \begin{subfigure}[t]{0.44\textwidth}
        \centering
        \raisebox{11.7mm}{\includegraphics[width=\linewidth]
        {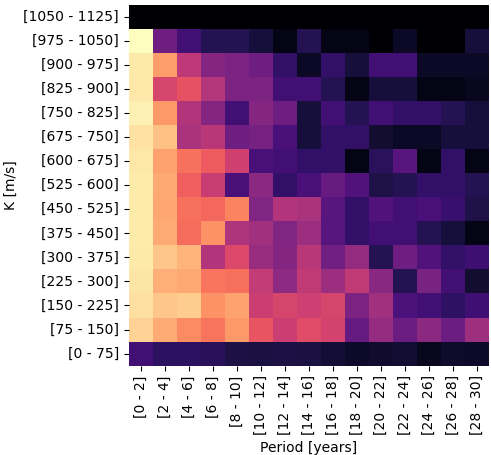}}
        \label{fig:left3}
    \end{subfigure}%
    \hfill
    \begin{subfigure}[t]{0.53\textwidth}
        \centering
        \includegraphics[width=\linewidth]{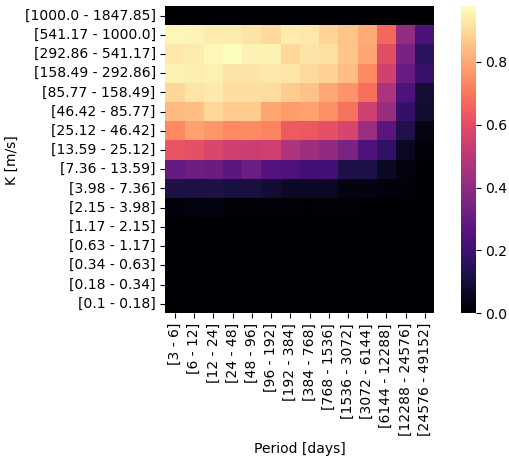}
        \label{fig:right3}
    \end{subfigure}

    \caption{Heat maps showing the detection efficiency of a \textsc{Minerva}-Australis class facility (RV precision $10 \ms$) for low cadence observations carried out over a period of three years (i.e. a short temporal baseline). The colour of each square reveals the mean detection probability for a planet in that particular mass and orbital period range, averaged over the 28 star systems studied in this work, based on simulated injection-recovery tests using 3000 injected planets. The lighter the colour, the greater the mean recovery rate. It is clear that the addition of just three years of low cadence data does not hugely improve our ability to detect Cold Jupiters over the current state of the data (presented in Figure~\ref{fig:lin-log_1}), though some small gains are made. The upper panels show the data in planet mass vs orbital period space, whilst the lower ones show the same information plotted in terms of the observed radial velocity, $K$, and the orbital period of the planet in question. In the upper right panel, it appears that no planets are detected at very high mass for very short periods. This is an artefact of the injection-recovery process used, where \texttt{RVSearch} only injects signals up to a radial velocity of 1000\ms. At the shortest orbital periods injected, this ceiling results in a maximum possible planet mass of $< 10$M$_J$ -- and so that apparent void in detections is simply a location where no planets are injected. This ceiling is clearly seen in the lower right hand panel, showing that the highest velocity planets are very efficiently detected at short orbital periods, just as one would expect.}
    \label{fig:lin-log_2}
\end{figure*}

\begin{figure*}
    \centering

    \begin{subfigure}[b]{0.48\textwidth}
        \centering
        \includegraphics[width=\linewidth]{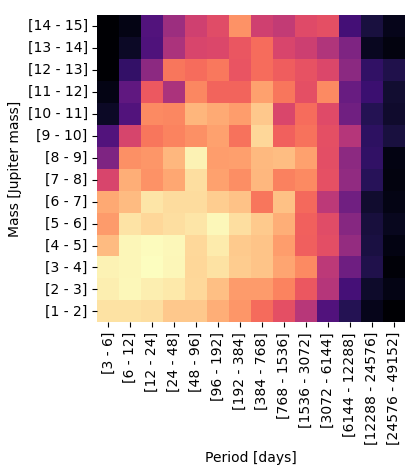}
        \caption{Heatmap for VLT/ESPRESSO (low cadence – 3 y) (log scale)}
    \end{subfigure}
    \hfill
    \begin{subfigure}[b]{0.48\textwidth}
        \centering
        \includegraphics[width=\linewidth]{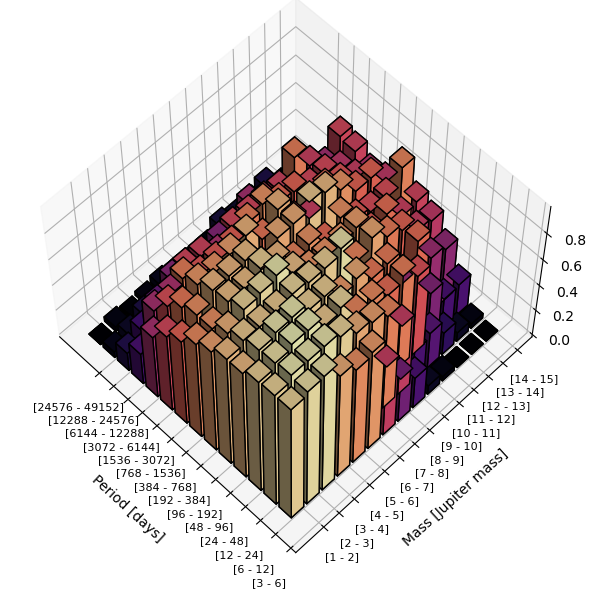}
        \caption{3D representation}
    \end{subfigure}
    
    \begin{subfigure}[b]{0.48\textwidth}
        \centering
        \includegraphics[width=\linewidth]{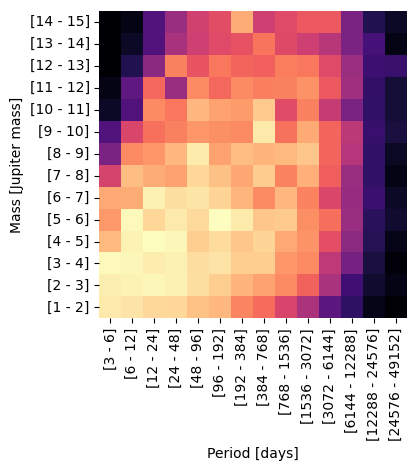}
        \caption{Heatmap for VLT/ESPRESSO (medium cadence – 6 y) (log scale)}
    \end{subfigure}
    \hfill
    \begin{subfigure}[b]{0.48\textwidth}
        \centering
        \includegraphics[width=\linewidth]{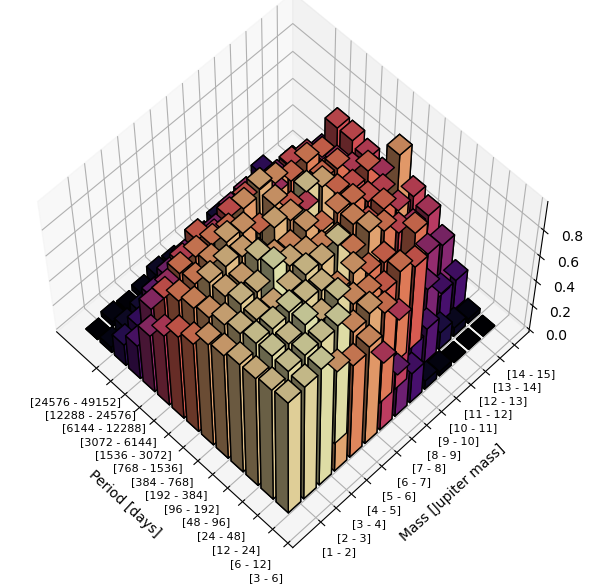}
        \caption{3D representation}
    \end{subfigure}

    \caption{Heat maps showing the efficiency with which a high-precision exoplanet survey (such as VLT/ESPRESSO) can detect exoplanets as a function of orbital period and planet mass, for three combinations of observation baseline and observation cadence. The left-hand column shows the data in two-dimensional plots, whilst the right-hand column presents the same results in a three-dimensional form, to help the reader visualise the regions where the ability of the facility to detect planets falls off. The valley at high-mass and short-period (the top left of the 2D maps) is artificial, being the result of the fact that the injection-recovery analysis using \texttt{RVSearch} only considered RV signals with amplitudes between 0.1 and 1000 ms$^-1$ -- meaning that no extremely massive planets were injected at short orbital periods. Increasing both the observational cadence and baseline of observations clearly improves the efficiency with which Jupiter-analogue planets can be detected, with the most pronounced benefits coming at long orbital periods as a result of the longer baseline used. Data for the high-cadence 12-year observations are presented in Figure~9.}
    \label{fig:6subplots}
\end{figure*}
\begin{figure*}
    \centering

    \begin{subfigure}[b]{0.48\textwidth}
        \centering
        \includegraphics[width=\linewidth]{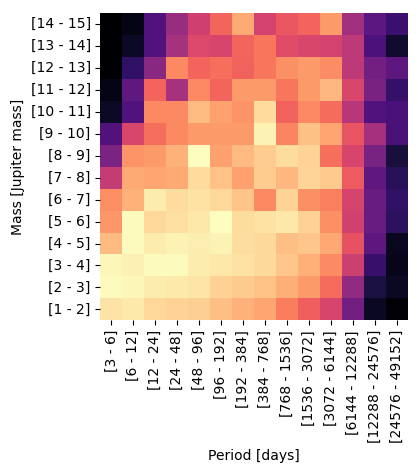}
        \caption{Heat map for VLT/ESPRESSO (high cadence – 12 y) (log scale)}
    \end{subfigure}
    \hfill
    \begin{subfigure}[b]{0.48\textwidth}
        \centering
        \includegraphics[width=\linewidth]{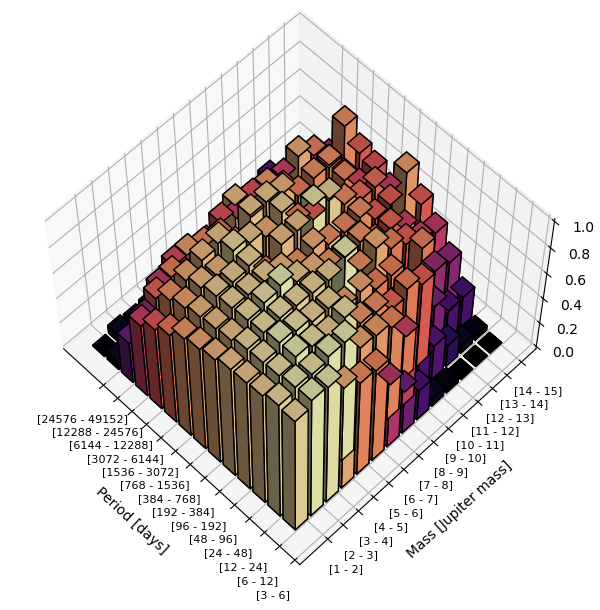}
        \caption{3D representation}
    \end{subfigure}
    \label{fig:6subplots2}
    \caption{Continuation of Figure~\ref{fig:6subplots}, showing the heat map of detection efficiency for VLT/ESPRESSO observations using a high cadence for an observational baseline of 12 years, with the left panel showing a 2D representation of the data, and the right panel showing the same information in three dimensions.}
\end{figure*}

\begin{table*}[!h]
\begin{center}
\caption{Table showing the semi-major axis (in AU) at which a detection efficiency of 50\% is achieved for 1 M$_J$ planets orbiting the 28 targets stars of this study, for a low-precision, \textsc{Minerva}-Australis analogue facility, as a function of the observational baseline and observational cadence used. In general, the longer the observation baseline, the more distant the Jupiter analogues that could be detected, regardless of the observation cadence chosen. However, it is clear that, for most targets, applying a higher cadence yields markedly better results than simply observing with low cadence.}
\label{tab:1Jmass_a_Minerva}
\begin{tabular}{| c | c | c | c | c | c | c | c | c | c | }
\hline
&\multicolumn{3}{|c|}{Low cadence}&\multicolumn{3}{|c|}{Medium cadence}&\multicolumn{3}{|c|}{High cadence} \\
\hline
&3 years&6 years&12 years&3 years&6 years&12 years&3 years&6 years&12 years\\
\hline
51Peg & 5.87 & 5.74 & 5.99 & 5.85 & 5.87 & 5.87 & 5.85 & 5.91 & 6.52 \\
\hline
BD-10 3166 & 3.0 & 2.96 & 3.96 & 3.09 & 3.91 & 5.71 & 3.51 & 4.92 & 7.37 \\
\hline
HD102956 & 0.97 & 1.08 & 0.84 & 1.14 & 0.95 & 1.2 & 0.98 & 1.1 & 1.22 \\
\hline
HD103720 & 2.56 & 2.36 & 2.56 & 2.37 & 2.49 & 2.66 & 2.43 & 2.82 & 4.8 \\
\hline
HD103774 & 0.95 & 1.06 & 1.16 & 0.96 & 1.17 & 1.15 & 1.06 & 1.16 & 1.78 \\
\hline
HD11231 & 0.1 & 0.36 & 0.46 & 0.35 & 0.54 & 0.56 & 0.64 & 1.16 & 3.14 \\
\hline
HD118203 & 0.14 & 0.15 & 0.18 & 0.15 & 0.18 & 0.19 & 0.19 & 0.18 & 0.21 \\
\hline
HD143105 & 0.49 & 0.51 & 0.55 & 0.52 & 0.55 & 0.59 & 0.65 & 0.64 & 0.91 \\
\hline
HD149026 & 2.54 & 3.27 & 4.72 & 2.51 & 4.54 & 7.76 & 4.28 & 5.96 & 8.21 \\
\hline
HD149143 & 2.73 & 2.82 & 2.82 & 2.7 & 2.94 & 3.32 & 2.92 & 2.72 & 4.79 \\
\hline
HD162020 & 1.13 & 1.15 & 0.1 & 1.08 & 0.11 & 0.1 & 0.13 & 0.1 & 0.1 \\
\hline
HD179949 & 1.15 & 1.2 & 1.21 & 1.18 & 1.14 & 1.17 & 1.23 & 1.18 & 1.1 \\
\hline
HD185269 & 3.62 & 3.32 & 3.62 & 3.4 & 1.67 & 1.84 & 1.79 & 1.84 & 1.86 \\
\hline
HD187123 & 1.85 & 10.57 & 9.97 & 2.0 & 2.09 & 2.02 & 10.29 & 9.73 & 7.62 \\
\hline
HD189733 & 0.67 & 0.78 & 0.79 & 0.9 & 0.8 & 1.05 & 0.82 & 1.2 & 1.14 \\
\hline
HD209458 & 6.18 & 6.35 & 6.12 & 6.03 & 6.25 & 7.29 & 6.21 & 6.37 & 7.72 \\
\hline
HD212301 & 0.7 & 1.08 & 1.82 & 0.91 & 1.69 & 2.62 & 1.84 & 1.97 & 5.71 \\
\hline
HD217107 & 6.41 & 8.51 & 8.91 & 8.27 & 9.31 & 9.24 & 9.16 & 9.29 & 8.79 \\
\hline
HD2638 & 0.62 & 0.69 & 0.68 & 0.66 & 0.59 & 1.65 & 0.7 & 2.58 & 7.13 \\
\hline
HD285507 & 1.36 & 1.05 & 1.04 & 1.4 & 1.77 & 4.22 & 1.78 & 1.99 & 6.55 \\
\hline
HD68988 & 2.87 & 3.43 & 3.04 & 3.23 & 3.14 & 0.26 & 3.35 & 0.2 & 0.24 \\
\hline
HD75289 & 3.23 & 4.55 & 5.48 & 4.74 & 5.1 & 6.49 & 4.88 & 6.49 & 8.4 \\
\hline
HD83443 & 7.79 & 7.22 & 6.32 & 8.15 & 7.36 & 7.98 & 7.56 & 4.4 & 5.02 \\
\hline
HD86081 & 1.2 & 1.2 & 0.89 & 1.37 & 1.19 & 1.45 & 1.25 & 2.49 & 6.28 \\
\hline
HD88133 & 4.26 & 4.73 & 5.32 & 4.3 & 5.18 & 6.76 & 4.77 & 6.29 & 8.19 \\
\hline
HIP14810 & 2.6 & 2.55 & 2.58 & 2.54 & 2.42 & 2.97 & 2.58 & 2.89 & 2.63 \\
\hline
TauBoo & 0.1 & 0.1 & 0.1 & 0.1 & 0.1 & 0.1 & 0.1 & 0.1 & 0.1 \\
\hline
upsAnd & 2.28 & 2.42 & 2.18 & 2.17 & 2.36 & 2.5 & 2.42 & 2.48 & 2.33 \\
\hline

\end{tabular}
\end{center}
\end{table*}
\clearpage

\begin{table}[!h]
\begin{center}
\caption{Table showing the semi-major axis at which a 1 M$_J$ planet would have a 50\% probability of detection around our 28 target stars, for observations carried out with a medium-precision exoplanet survey facility, such as Keck/HIRES, as a function of observational cadence and the baseline over which additional observations are carried out. As was the case for a low-precision facility (e.g. Table~\ref{tab:1Jmass_a_Minerva}), a longer baseline of observations typically leads to Jupiter analogues being detectable on longer period orbits, with higher cadence observations typically resulting in an improvement in that detection distance over lower cadence observations.}
\label{tab:1Jmass_a_Keck}
\begin{tabular}{| c | c | c | c | c | c | c | c | c | c | }
\hline
&\multicolumn{3}{|c|}{Low cadence}&\multicolumn{3}{|c|}{Medium cadence}&\multicolumn{3}{|c|}{High cadence} \\
\hline
&3 years&6 years&12 years&3 years&6 years&12 years&3 years&6 years&12 years\\
\hline
51Peg & 5.85 & 5.85 & 5.89 & 5.73 & 5.87 & 6.08 & 5.87 & 5.91 & 6.42 \\
\hline
BD-10 3166 & 2.84 & 3.05 & 4.2 & 3.13 & 4.14 & 6.4 & 3.42 & 5.77 & 8.65 \\
\hline
HD102956 & 1.13 & 0.88 & 1.18 & 1.04 & 1.12 & 1.24 & 1.25 & 1.07 & 1.35 \\
\hline
HD103720 & 2.56 & 2.17 & 2.57 & 2.56 & 2.46 & 2.56 & 2.36 & 2.78 & 4.36 \\
\hline
HD103774 & 1.05 & 1.05 & 1.06 & 0.99 & 1.17 & 1.23 & 1.82 & 1.62 & 1.49 \\
\hline
HD11231 & 0.1 & 0.1 & 0.48 & 0.47 & 0.55 & 1.01 & 0.85 & 1.11 & 2.84 \\
\hline
HD118203 & 0.14 & 0.16 & 0.18 & 0.17 & 0.18 & 0.19 & 0.19 & 0.18 & 0.19 \\
\hline
HD143105 & 0.52 & 0.48 & 0.55 & 0.52 & 0.55 & 0.56 & 0.59 & 0.63 & 0.84 \\
\hline
HD149026 & 2.12 & 3.16 & 4.8 & 2.35 & 4.52 & 7.31 & 3.61 & 6.99 & 9.43 \\
\hline
HD149143 & 2.87 & 2.82 & 2.73 & 2.82 & 3.01 & 4.43 & 3.07 & 3.28 & 3.34 \\
\hline
HD162020 & 1.13 & 0.99 & 0.1 & 1.19 & 0.11 & 0.1 & 0.13 & 0.1 & 0.1 \\
\hline
HD179949 & 1.15 & 1.15 & 1.15 & 1.15 & 1.2 & 1.17 & 1.15 & 1.18 & 1.08 \\
\hline
HD185269 & 3.62 & 3.32 & 3.61 & 3.4 & 1.62 & 1.83 & 1.67 & 1.77 & 1.85 \\
\hline
HD187123 & 9.57 & 9.03 & 10.15 & 10.97 & 9.79 & 9.03 & 9.97 & 9.03 & 7.29 \\
\hline
HD189733 & 0.6 & 0.68 & 0.83 & 0.79 & 0.71 & 0.97 & 0.86 & 1.24 & 1.1 \\
\hline
HD209458 & 5.7 & 6.08 & 6.31 & 6.18 & 6.25 & 6.76 & 6.01 & 6.37 & 7.41 \\
\hline
HD212301 & 0.91 & 1.23 & 1.67 & 1.13 & 1.59 & 2.18 & 1.72 & 3.1 & 6.57 \\
\hline
HD217107 & 6.49 & 8.51 & 8.34 & 8.12 & 9.31 & 8.83 & 8.56 & 9.83 & 9.4 \\
\hline
HD2638 & 0.63 & 0.66 & 0.71 & 0.64 & 0.63 & 1.6 & 0.74 & 2.51 & 8.69 \\
\hline
HD285507 & 1.14 & 1.14 & 0.93 & 0.87 & 1.29 & 3.81 & 1.87 & 2.01 & 5.99 \\
\hline
HD68988 & 3.16 & 2.81 & 0.25 & 3.53 & 0.17 & 0.28 & 0.25 & 0.21 & 0.47 \\
\hline
HD75289 & 5.05 & 4.68 & 5.82 & 4.36 & 5.07 & 7.38 & 4.9 & 6.25 & 7.42 \\
\hline
HD83443 & 6.96 & 7.77 & 6.81 & 8.09 & 7.3 & 5.59 & 8.0 & 4.63 & 6.2 \\
\hline
HD86081 & 1.19 & 1.19 & 0.8 & 1.2 & 1.1 & 1.45 & 1.55 & 2.66 & 9.75 \\
\hline
HD88133 & 4.3 & 4.59 & 8.14 & 4.54 & 6.75 & 10.36 & 4.98 & 7.8 & 9.79 \\
\hline
HIP14810 & 2.52 & 2.56 & 2.53 & 2.6 & 2.71 & 2.97 & 2.63 & 2.63 & 8.53 \\
\hline
TauBoo & 0.1 & 0.1 & 0.1 & 0.1 & 0.1 & 0.1 & 0.1 & 0.1 & 0.1 \\
\hline
UpsAnd & 2.16 & 2.42 & 2.17 & 2.01 & 2.04 & 2.14 & 1.95 & 2.33 & 2.6  \\
\hline

\end{tabular}
\end{center}
\end{table}
\clearpage

\begin{table}[!h]
\begin{center}
\caption{The semi-major axes at which a planet of mass 1 M$_J$ would be detected with 50\% probability orbiting our 28 target stars, for observations using a high-precision survey facility, such as VLT/ESPRESSO, as a function of observational cadence and the baseline over which observations are performed. As was the case for low- and medium-precision facilities, a longer baseline of observations again leads to Jupiter mass planets being detectable on longer period orbits, with higher cadence observations typically resulting in an improvement in that detection distance over lower cadence observations.}

\label{tab:1Jmass_a_ESPRESSO}
\begin{tabular}{| c | c | c | c | c | c | c | c | c | c | }
\hline
&\multicolumn{3}{|c|}{Low cadence}&\multicolumn{3}{|c|}{Medium cadence}&\multicolumn{3}{|c|}{High cadence} \\
\hline
&3 years&6 years&12 years&3 years&6 years&12 years&3 years&6 years&12 years\\
\hline
51Peg & 5.73 & 5.73 & 5.87 & 5.85 & 5.77 & 6.1 & 5.87 & 5.89 & 6.72 \\
\hline
BD-10 3166 & 2.92 & 3.06 & 4.44 & 2.75 & 3.78 & 7.37 & 3.7 & 5.2 & 9.51 \\
\hline
HD102956 & 0.97 & 0.84 & 1.0 & 1.16 & 0.75 & 1.12 & 1.06 & 1.34 & 1.22 \\
\hline
HD103720 & 2.56 & 2.17 & 2.39 & 2.47 & 2.31 & 2.65 & 2.44 & 2.71 & 4.8 \\
\hline
HD103774 & 0.99 & 1.05 & 1.0 & 0.98 & 1.04 & 1.22 & 1.06 & 1.63 & 1.65 \\
\hline
HD11231 & 0.1 & 0.35 & 0.4 & 0.32 & 0.5 & 0.69 & 0.53 & 0.79 & 2.7 \\
\hline
HD118203 & 0.14 & 0.15 & 0.17 & 0.17 & 0.19 & 0.19 & 0.19 & 0.17 & 0.21 \\
\hline
HD143105 & 0.44 & 0.54 & 0.52 & 0.47 & 0.56 & 0.57 & 0.62 & 0.65 & 0.89 \\
\hline
HD149026 & 2.31 & 4.9 & 5.51 & 2.6 & 4.49 & 8.07 & 2.91 & 7.06 & 8.77 \\
\hline
HD149143 & 2.72 & 2.94 & 2.57 & 2.46 & 2.84 & 3.59 & 2.73 & 3.85 & 4.53 \\
\hline
HD162020 & 1.09 & 1.0 & 0.1 & 1.19 & 0.11 & 0.1 & 0.13 & 0.1 & 0.1 \\
\hline
HD179949 & 1.15 & 1.15 & 1.16 & 1.1 & 1.21 & 1.14 & 1.21 & 1.21 & 1.08 \\
\hline
HD185269 & 3.63 & 3.32 & 3.59 & 3.4 & 1.66 & 1.84 & 1.73 & 1.82 & 1.79 \\
\hline
HD187123 & 9.11 & 10.29 & 9.78 & 9.56 & 10.35 & 9.11 & 9.74 & 9.39 & 9.1 \\
\hline
HD189733 & 0.57 & 0.64 & 0.69 & 0.74 & 0.79 & 0.87 & 0.82 & 1.23 & 1.09 \\
\hline
HD209458 & 5.88 & 5.88 & 5.88 & 5.7 & 6.23 & 6.65 & 6.01 & 6.52 & 7.56 \\
\hline
HD212301 & 0.84 & 1.37 & 1.81 & 1.34 & 1.88 & 2.62 & 2.04 & 2.23 & 5.72 \\
\hline
HD217107 & 6.72 & 8.06 & 9.0 & 7.62 & 8.5 & 8.75 & 7.76 & 9.04 & 9.09 \\
\hline
HD2638 & 0.65 & 0.68 & 0.74 & 0.62 & 0.61 & 1.56 & 0.69 & 2.59 & 8.39 \\
\hline
HD285507 & 1.73 & 1.03 & 0.97 & 1.43 & 1.5 & 4.27 & 1.66 & 2.07 & 5.98 \\
\hline
HD68988 & 2.99 & 3.01 & 3.11 & 3.32 & 0.18 & 0.17 & 0.19 & 0.22 & 0.33 \\
\hline
HD75289 & 3.57 & 4.63 & 6.0 & 3.33 & 3.35 & 6.78 & 4.73 & 6.78 & 7.97 \\
\hline
HD83443 & 7.57 & 7.95 & 7.44 & 8.39 & 7.95 & 5.75 & 7.36 & 5.4 & 6.7 \\
\hline
HD86081 & 1.18 & 1.19 & 0.84 & 1.2 & 1.06 & 1.4 & 1.73 & 2.46 & 8.82 \\
\hline
HD88133 & 4.26 & 4.56 & 6.76 & 4.5 & 6.89 & 9.39 & 5.19 & 7.46 & 11.76 \\
\hline
HIP14810 & 2.52 & 2.73 & 2.83 & 2.6 & 2.59 & 2.88 & 2.99 & 2.63 & 9.02 \\
\hline
TauBoo & 0.1 & 0.1 & 0.1 & 0.1 & 0.1 & 0.1 & 0.1 & 0.1 & 2.53 \\
\hline
UpsAnd & 2.52 & 2.73 & 6.76 & 2.6 & 2.59 & 2.88 & 2.99 & 2.63 & 0.1 \\
\hline
\end{tabular}
\end{center}
\end{table}
\clearpage

\subsection{VLT/ESPRESSO/Minerva performance comparison}\label{sec:E_M_comparison}

In the previous sections, we have considered the benefits of a variety of observational strategies in the search for Cold Jupiters for three different types of exoplanet search facilities -- those with low radial velocity precision (analogues to the \textsc{Minerva}-Australis array), with medium precision (like data obtained using Keck), and those with high precision (``extreme precision radial velocity'' or ``EPRV'') such as VLT/ESPRESSO). We have clearly shown that longer observational baselines are the key to detecting distant massive planets using radial velocity observations, regardless of the facility involved. Our results also reveal that increasing the observational cadence leads to the more efficient detection of Cold Jupiters in the vast majority of cases. In general, we also find that the use of high-precision facilities yields only marginal gains over the results obtained by smaller, more accessible telescopes (see e.g. Figure~\ref{fig:AU_for_1Jmass}). In this section, we therefore address the question: can the long-term performance of large, highly accurate telescopes in the search for Jupiter-like planets be matched by smaller, more accessible, and more affordable facilities?

To address this, we consider our two extreme cases - VLT/ESPRESSO (high precision) and \textsc{Minerva}-Australis (low precision). Rather than simply considering the detection efficiencies of the two facilities on their own, we look to make two direct comparisons. In the first, we calculate the ratio of successful detections made by VLT/ESPRESSO to those made using \textsc{Minerva}-Australis -- in other words, we divide the detection efficiency of one facility by the other. The results of this comparison are shown in Figure~\ref{fig:E-M_3x3}. In that Figure, a value of 1.0 denotes that the two facilities perform equally well. If the value is less than 1.0 (represented by reddish to black shades in the colour scale of Figure \ref{fig:E-M_3x3}), then \textsc{Minerva}-Australis outperforms VLT/ESPRESSO, whilst values greater than 1.0 (the lighter shades in Figure~\ref{fig:E-M_3x3}) denote VLT/ESPRESSO outperforming \textsc{Minerva}-Australis. Looking at Figure \ref{fig:E-M_3x3}, it is evident that both telescopes perform equally well across the majority of the phase-space considered. The right side of each plot initially appears to highlight areas where performance differences become significant, with numerous dark-shaded regions -- particularly in the last column, which corresponds to high-cadence observations. The results in this region should, however, be treated with some caution. This is the region where the detection efficiencies for all three facilities are very low (as can be seen in Figures~\ref{fig:lin-log_2} and \ref{fig:6subplots}), and so the variation seen here is most likely simply noise.

The second comparison we make is to take the results calculated above (the ratio of the detection efficiencies of the two facilities), and scale the result by the ratio of the cost of the two facilities, as detailed in Equation~\ref{eq:dollar-equivalence}. 
VLT/ESPRESSO\footnote{Estimated cost: USD 860,000,000} is a high-performance but very expensive facility compared to \textsc{Minerva}-Australis\footnote{Estimated cost: USD 4,000,000} \citep{eso_faq}, and so it is interesting to consider the cost-effectiveness of the two types of facility as tools to search for Cold Jupiters. 


\begin{equation}
\left(\frac{\text{\$VLT/ESPRESSO}}{\text{\$Minerva}}\right) \times \left(\frac{\# \text{Minerva}}{\# \text{VLT/ESPRESSO}}\right)
\label{eq:dollar-equivalence}
\end{equation}

This equation equals 1 when \textsc{Minerva}-Australis is as cost-effective as VLT/ESPRESSO, whilst values greater than 1 indicate that \textsc{Minerva}-Australis provides better value for money (not in absolute performance, but in cost per detection). We present the results of this analysis in Figure~\ref{fig:E-M_3x3_price}, where it can be clearly seen that \textsc{Minerva}-Australis represents a far more cost-effective tool for searching for Cold Jupiters than the more expensive facility. This is not a surprise - both facilities are broad equally efficient as tools for the detection of Cold Jupiters, and so it makes sense that the cheaper facility would be more cost-effective. Once again, the noisy results in the right hand columns in Figure~\ref{fig:E-M_3x3_price} are purely the result of the low detection efficiencies for the two facilities at such long orbital periods.

Finally, we note some limitations to the simple comparison of cost-effectiveness we have described here.  For example, this approach clearly breaks down for very faint targets where smaller telescopes are unable to obtain useful RV data at all.  Another subtlety we have not considered is a ``hybrid'' observing strategy whereby certain orbital phases disproportionately benefit from higher precision measurements.  Such intricacies are beyond the scope of this work; it is clear that in practice, small and large telescopes serve complementary roles for this and other science goals.

\begin{figure*}
  \centering
  \includegraphics[width=\textwidth]{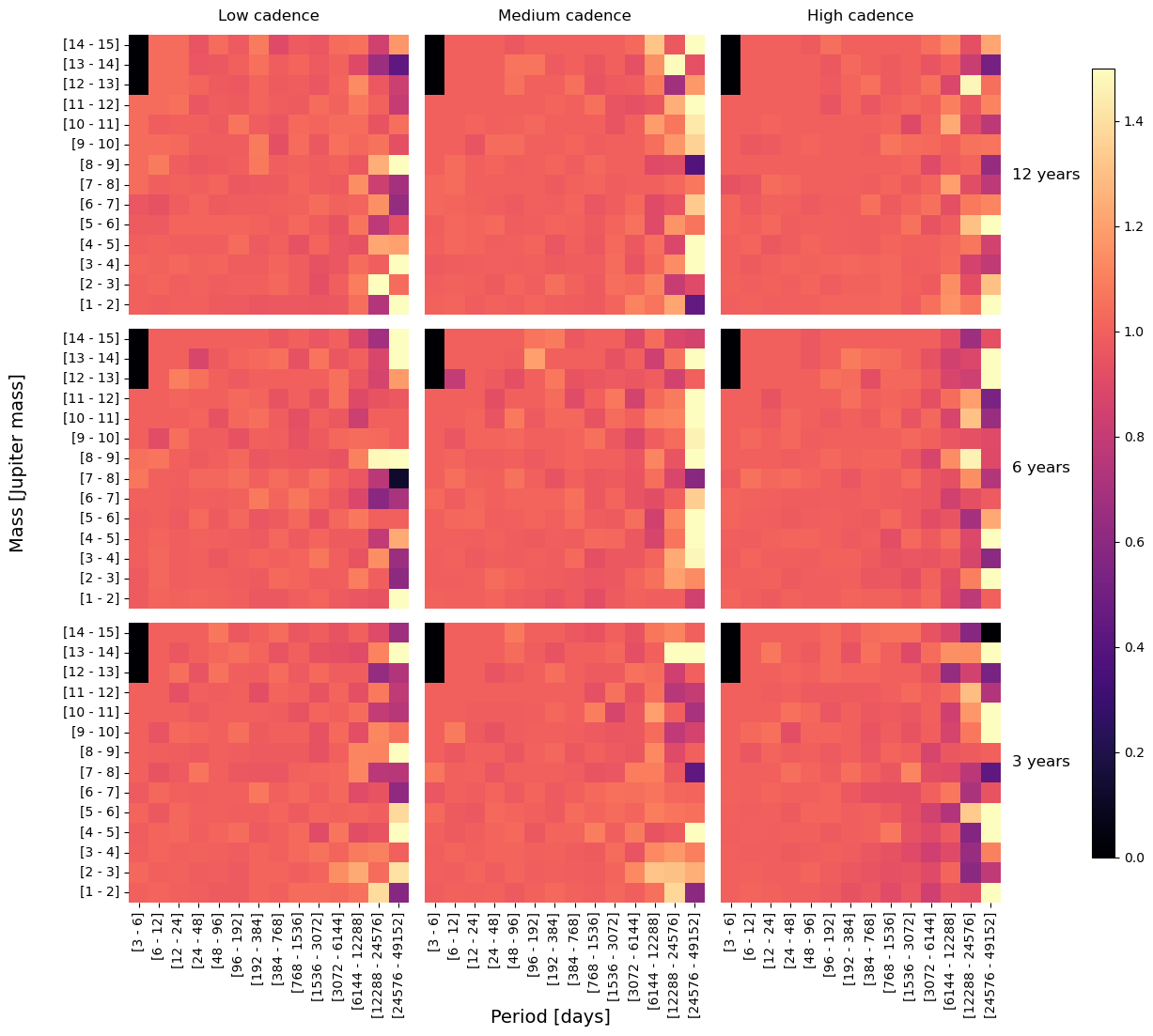} 
   \caption{A direct comparison between the detection efficiencies of a low-precision facility (\text{Minerva}-Australis) and a high-precision facility (VLT/ESPRESSO), for the different observation scenarios considered in this work. Each square of each panel shows the ratio of the detection efficiencies of the two facilities -- in other words, the efficiency of VLT/ESPRESSO divided by the efficiency of \textsc{Minerva}-Australis at that point. Values greater than 1.0 show areas where VLT/ESPRESSO can more effectively detect massive planets than \textsc{Minerva}-Australis, given the exact same observation strategy, whilst values less than 1.0 are areas where the smaller, cheaper facility proves to be more efficient. Other than at very long orbital periods, it is clear that both facilities perform approximately equally well - with the only significant differences emerging towards longer periods (~$\sim$6164 to $\sim$24576 days), where VLT/ESPRESSO appears to be slightly more effective at medium cadence, and \textsc{Minerva}-Australis performs slightly better at high cadence. The right-hand most columns are, however, regions of low detection efficiency for both facilities, meaning that the results displayed there are relatively noisy, and should be viewed with caution. }
  \label{fig:E-M_3x3}
\end{figure*}


\begin{figure*}
  \centering
  \includegraphics[width=\textwidth]{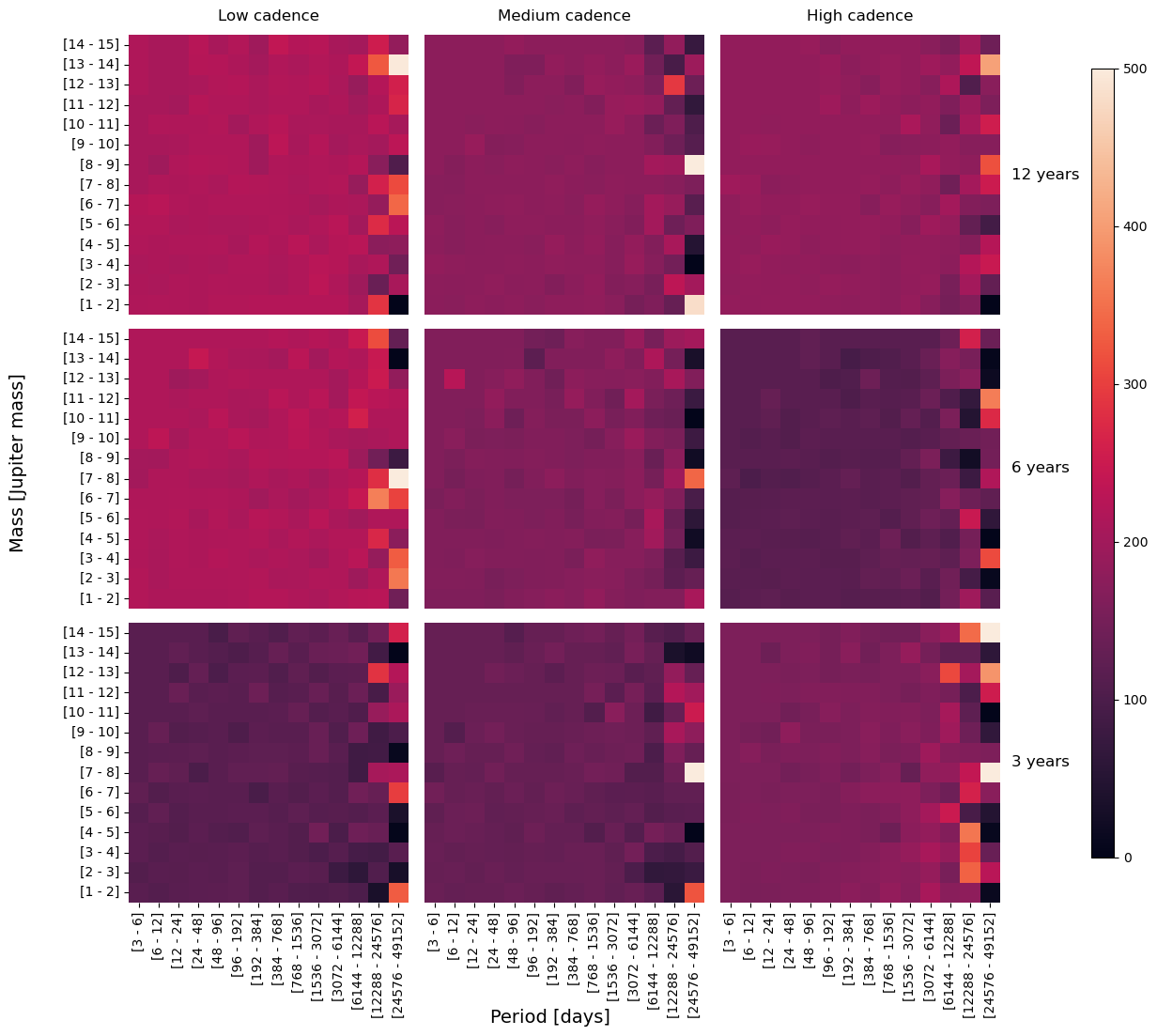} 
  \caption{Figure showing the relative cost of detection for Cold Jupiters using a low resolution, low cost facility (such as \textsc{Minerva}-Australis) compared to a high resolution, high cost facility such as VLT/ESPRESSO. This figure attempts to illustrate the most cost-effective method to search for massive planets orbiting other stars. Each panel plots the relative 'cost-effectiveness' of Cold Jupiter detections between these two facilities, for that particular observational scenario. A value of 500 (bright white) occurs when making a detection using VLT/ESPRESSO would come at a cost 500 times higher than the same detection using \textsc{Minerva}-Australis; with the values in each box determined using equation~\ref{eq:dollar-equivalence}. It is immediately apparent that the detection of Cold Jupiters is strikingly more cost-effective using facilities like \textsc{Minerva-Australis}, with the greatest difference between the cost-effectiveness being seen for the lowest observational cadences.}
  \label{fig:E-M_3x3_price}
\end{figure*}

\section{Conclusions}\label{sec:conclusions}

In this work, we have defined Hot Jupiter-Cold Jupiter "siblings" as planetary systems containing both a Hot Jupiter and a distant outer giant. Motivated by the discovery of the Hot Jupiter-Cold Jupiter "siblings" in the HD\,83443 system \citep{errico2022HD83443}; we wished to estimate how common such architectures are, given the severe observational biases against detecting these sorts of systems.  From a sample of 28 Hot Jupiter systems that had enough RV data to meaningfully contribute to this investigation, we indeed found the detection sensitivity woefully lacking, even for planets of many Jupiter masses.  We then performed an extensive suite of simulated observational strategies to determine the way forward in addressing this problem.  

Our analysis across nine scenarios and three telescopes reveals several key findings. First, the existing data, despite being gathered from well-studied and bright Hot Jupiter hosts, are generally ineffective at detecting cold Jupiters, highlighting the challenge of probing these longer-period companions.  We suspect that this arises from a human bias: for Hot Jupiter systems, once the short-period planet is well-characterised, the target's observing priority is reduced or eliminated in favour of other, ``more interesting'' targets.  The result is that Hot Jupiter systems tend to have short baselines of RV data, hindering our quest to understand the occurrence of any planetary siblings.  

We also find that, although higher instrumental precision extends sensitivity to larger semi-major axes for a given observing strategy, the total baseline over which an observation campaign is carried out is the dominant factor in our ability to detect cold Jupiters.  We find that the results are relatively insensitive to the RV measurement precision of the facility being used; the performance of a high-cadence, decade-long program using a low-cost dedicated facility such as \textsc{Minerva}-Australis matches, or even exceeds, that of shorter, lower-cadence observations by a higher cost extreme-precision RV (EPRV) facility such as VLT/ESPRESSO.  Indeed, our results show that low-cost dedicated facilities such as \textsc{Minerva}-Australis will prove to be far more cost-effective in delivering discoveries of Cold Jupiters than facilities such as Keck and VLT/ESPRESSO.

Notably, a high-cadence, decade-long Minerva program can match or exceed the performance of shorter, lower-cadence VLT/ESPRESSO observations, illustrating that modest measurement precision can be compensated for by strategic scheduling.  A cost-normalised comparison further demonstrates that smaller, more accessible facilities can deliver superior return per dollar than flagship instruments when optimised for long-term monitoring.  Our results emphasise the value of extended, regular RV campaigns for uncovering long-period exoplanets and inform the design of future surveys.  This finding underscores the continuing relevance of traditional RV facilities in an era dominated by EPRV instruments.

\clearpage

\section*{Acknowledgements}

The authors would like to express their sincere gratitude to the anonymous referee of this paper, whose comments and questions led to significant improvements in the manuscript.

{\textsc{Minerva}}-Australis is supported by Australian Research Council LIEF Grant LE160100001, Discovery Grants DP180100972 and DP220100365, Mount Cuba Astronomical Foundation and institutional partners University of Southern Queensland, UNSW Sydney, MIT, Nanjing University, George Mason University, University of Louisville, University of California Riverside, University of Florida, and The University of Texas at Austin.

This publication makes use of The Data \& Analysis Center for Exoplanets (DACE), which is a facility based at the University of Geneva (CH) dedicated to extrasolar planets data visualisation, exchange and analysis. DACE is a platform of the Swiss National Centre of Competence in Research (NCCR) PlanetS, federating the Swiss expertise in Exoplanet research. The DACE platform is available at \url{https://dace.unige.ch}.

This research has made use of the NASA Exoplanet Archive, which is operated by the California Institute of Technology, under contract with the National Aeronautics and Space Administration under the Exoplanet Exploration Program.

This work uses astropy \citep{astropy:2013, astropy:2018, astropy:2022} , scipy \citep{sciPy2020-NMeth}, numpy \citep{harris2020array}, matplotlib \citep{Hunter:2007}, pandas \citep{reback2020pandas}, seaborn \citep{waskom2021seaborn}, Jupyter (https://jupyter.org/) and RVSearch \citep{rosenthal2021rvsearch}

We respectfully acknowledge the traditional custodians of all lands throughout Australia, and recognise their continued cultural and spiritual connection to the land, waterways, cosmos, and community. We pay our deepest respects to all Elders, ancestors and descendants of the Giabal, Jarowair, and Kambuwal nations, upon whose lands the \textsc{Minerva}-Australis facility at Mt Kent observatory is situated.

\vspace{5mm}

\clearpage

\bibliography{bib}

@ARTICLE{118203DRS,
       author = {{Castro-Gonz{\'a}lez}, A. and {Lillo-Box}, J. and {Correia}, A.~C.~M. and {Santos}, N.~C. and {Barrado}, D. and {Morales-Calder{\'o}n}, M. and {Shkolnik}, E.~L.},
        title = "{Signs of magnetic star-planet interactions in HD 118203. TESS detects stellar variability that matches the orbital period of a close-in eccentric Jupiter-sized companion}",
      journal = {\aap},
         year = 2024,
        month = apr,
       volume = {684},
          eid = {A160},
        pages = {A160},
          doi = {10.1051/0004-6361/202348722},
archivePrefix = {arXiv},
       eprint = {2401.17272},
 primaryClass = {astro-ph.EP},
       adsurl = {https://ui.adsabs.harvard.edu/abs/2024A&A...684A.160C}
}

@article{addison2019minerva,
  title={\textsc{Minerva}-{A}ustralis. {I}. {D}esign, commissioning, and first photometric results},
  author={Addison, Brett and Wright, Duncan J and Wittenmyer, Robert A and Horner, Jonathan and Mengel, Matthew W and Johns, Daniel and Marti, Connor and Nicholson, Belinda and Soutter, Jack and Bowler, Brendan and others},
  journal={Publications of the Astronomical Society of the Pacific},
  volume={131},
  number={1005},
  pages={115003},
  year={2019},
  publisher={IOP Publishing}
}

@ARTICLE{addison2021TOI257b,
       author = {{Addison}, Brett C. and {Wright}, Duncan J. and {Nicholson}, Belinda A. and {Cale}, Bryson and {Mocnik}, Teo and {Huber}, Daniel and {Plavchan}, Peter and {Wittenmyer}, Robert A. and {Vanderburg}, Andrew and {Chaplin}, William J. and {Chontos}, Ashley and {Clark}, Jake T. and {Eastman}, Jason D. and {Ziegler}, Carl and {Brahm}, Rafael and {Carter}, Bradley D. and {Clerte}, Mathieu and {Espinoza}, N{\'e}stor and {Horner}, Jonathan and {Bentley}, John and {Jord{\'a}n}, Andr{\'e}s and {Kane}, Stephen R. and {Kielkopf}, John F. and {Laychock}, Emilie and {Mengel}, Matthew W. and {Okumura}, Jack and {Stassun}, Keivan G. and {Bedding}, Timothy R. and {Bowler}, Brendan P. and {Burnelis}, Andrius and {Blanco-Cuaresma}, Sergi and {Collins}, Michaela and {Crossfield}, Ian and {Davis}, Allen B. and {Evensberget}, Dag and {Heitzmann}, Alexis and {Howell}, Steve B. and {Law}, Nicholas and {Mann}, Andrew W. and {Marsden}, Stephen C. and {Matson}, Rachel A. and {O'Connor}, James H. and {Shporer}, Avi and {Stevens}, Catherine and {Tinney}, C.~G. and {Tylor}, Christopher and {Wang}, Songhu and {Zhang}, Hui and {Henning}, Thomas and {Kossakowski}, Diana and {Ricker}, George and {Sarkis}, Paula and {Schlecker}, Martin and {Torres}, Pascal and {Vanderspek}, Roland and {Latham}, David W. and {Seager}, Sara and {Winn}, Joshua N. and {Jenkins}, Jon M. and {Mireles}, Ismael and {Rowden}, Pam and {Pepper}, Joshua and {Daylan}, Tansu and {Schlieder}, Joshua E. and {Collins}, Karen A. and {Collins}, Kevin I. and {Tan}, Thiam-Guan and {Ball}, Warrick H. and {Basu}, Sarbani and {Buzasi}, Derek L. and {Campante}, Tiago L. and {Corsaro}, Enrico and {Gonz{\'a}lez-Cuesta}, L. and {Davies}, Guy R. and {de Almeida}, Leandro and {do Nascimento}, Jr., Jose-Dias and {Garc{\'\i}a}, Rafael A. and {Guo}, Zhao and {Handberg}, Rasmus and {Hekker}, Saskia and {Hey}, Daniel R. and {Kallinger}, Thomas and {Kawaler}, Steven D. and {Kayhan}, Cenk and {Kuszlewicz}, James S. and {Lund}, Mikkel N. and {Lyttle}, Alexander and {Mathur}, Savita and {Miglio}, Andrea and {Mosser}, Benoit and {Nielsen}, Martin B. and {Serenelli}, Aldo M. and {Aguirre}, Victor Silva and {Theme{\ss}l}, Nathalie},
        title = "{TOI-257b (HD 19916b): a warm sub-saturn orbiting an evolved F-type star}",
      journal = {\mnras},
         year = 2021,
        month = apr,
       volume = {502},
       number = {3},
        pages = {3704-3722},
          doi = {10.1093/mnras/staa3960},
archivePrefix = {arXiv},
       eprint = {2001.07345},
 primaryClass = {astro-ph.EP},
       adsurl = {https://ui.adsabs.harvard.edu/abs/2021MNRAS.502.3704A}
}

@article{astropy:2013,
Adsurl = {http://adsabs.harvard.edu/abs/2013A%26A...558A..33A},
Archiveprefix = {arXiv},
Author = {{Astropy Collaboration} and {Robitaille}, T.~P. and {Tollerud}, E.~J. and {Greenfield}, P. and {Droettboom}, M. and {Bray}, E. and {Aldcroft}, T. and {Davis}, M. and {Ginsburg}, A. and {Price-Whelan}, A.~M. and {Kerzendorf}, W.~E. and {Conley}, A. and {Crighton}, N. and {Barbary}, K. and {Muna}, D. and {Ferguson}, H. and {Grollier}, F. and {Parikh}, M.~M. and {Nair}, P.~H. and {Unther}, H.~M. and {Deil}, C. and {Woillez}, J. and {Conseil}, S. and {Kramer}, R. and {Turner}, J.~E.~H. and {Singer}, L. and {Fox}, R. and {Weaver}, B.~A. and {Zabalza}, V. and {Edwards}, Z.~I. and {Azalee Bostroem}, K. and {Burke}, D.~J. and {Casey}, A.~R. and {Crawford}, S.~M. and {Dencheva}, N. and {Ely}, J. and {Jenness}, T. and {Labrie}, K. and {Lim}, P.~L. and {Pierfederici}, F. and {Pontzen}, A. and {Ptak}, A. and {Refsdal}, B. and {Servillat}, M. and {Streicher}, O.},
Doi = {10.1051/0004-6361/201322068},
Eid = {A33},
Eprint = {1307.6212},
Journal = {\aap},
Month = oct,
Pages = {A33},
Primaryclass = {astro-ph.IM},
Title = {{Astropy: A community Python package for astronomy}},
Volume = 558,
Year = 2013}

@ARTICLE{astropy:2018,
       author = {{Astropy Collaboration} and {Price-Whelan}, A.~M. and
         {Sip{\H{o}}cz}, B.~M. and {G{\"u}nther}, H.~M. and {Lim}, P.~L. and
         {Crawford}, S.~M. and {Conseil}, S. and {Shupe}, D.~L. and
         {Craig}, M.~W. and {Dencheva}, N. and {Ginsburg}, A. and {Vand
        erPlas}, J.~T. and {Bradley}, L.~D. and {P{\'e}rez-Su{\'a}rez}, D. and
         {de Val-Borro}, M. and {Aldcroft}, T.~L. and {Cruz}, K.~L. and
         {Robitaille}, T.~P. and {Tollerud}, E.~J. and {Ardelean}, C. and
         {Babej}, T. and {Bach}, Y.~P. and {Bachetti}, M. and {Bakanov}, A.~V. and
         {Bamford}, S.~P. and {Barentsen}, G. and {Barmby}, P. and
         {Baumbach}, A. and {Berry}, K.~L. and {Biscani}, F. and {Boquien}, M. and
         {Bostroem}, K.~A. and {Bouma}, L.~G. and {Brammer}, G.~B. and
         {Bray}, E.~M. and {Breytenbach}, H. and {Buddelmeijer}, H. and
         {Burke}, D.~J. and {Calderone}, G. and {Cano Rodr{\'\i}guez}, J.~L. and
         {Cara}, M. and {Cardoso}, J.~V.~M. and {Cheedella}, S. and {Copin}, Y. and
         {Corrales}, L. and {Crichton}, D. and {D'Avella}, D. and {Deil}, C. and
         {Depagne}, {\'E}. and {Dietrich}, J.~P. and {Donath}, A. and
         {Droettboom}, M. and {Earl}, N. and {Erben}, T. and {Fabbro}, S. and
         {Ferreira}, L.~A. and {Finethy}, T. and {Fox}, R.~T. and
         {Garrison}, L.~H. and {Gibbons}, S.~L.~J. and {Goldstein}, D.~A. and
         {Gommers}, R. and {Greco}, J.~P. and {Greenfield}, P. and
         {Groener}, A.~M. and {Grollier}, F. and {Hagen}, A. and {Hirst}, P. and
         {Homeier}, D. and {Horton}, A.~J. and {Hosseinzadeh}, G. and {Hu}, L. and
         {Hunkeler}, J.~S. and {Ivezi{\'c}}, {\v{Z}}. and {Jain}, A. and
         {Jenness}, T. and {Kanarek}, G. and {Kendrew}, S. and {Kern}, N.~S. and
         {Kerzendorf}, W.~E. and {Khvalko}, A. and {King}, J. and {Kirkby}, D. and
         {Kulkarni}, A.~M. and {Kumar}, A. and {Lee}, A. and {Lenz}, D. and
         {Littlefair}, S.~P. and {Ma}, Z. and {Macleod}, D.~M. and
         {Mastropietro}, M. and {McCully}, C. and {Montagnac}, S. and
         {Morris}, B.~M. and {Mueller}, M. and {Mumford}, S.~J. and {Muna}, D. and
         {Murphy}, N.~A. and {Nelson}, S. and {Nguyen}, G.~H. and
         {Ninan}, J.~P. and {N{\"o}the}, M. and {Ogaz}, S. and {Oh}, S. and
         {Parejko}, J.~K. and {Parley}, N. and {Pascual}, S. and {Patil}, R. and
         {Patil}, A.~A. and {Plunkett}, A.~L. and {Prochaska}, J.~X. and
         {Rastogi}, T. and {Reddy Janga}, V. and {Sabater}, J. and
         {Sakurikar}, P. and {Seifert}, M. and {Sherbert}, L.~E. and
         {Sherwood-Taylor}, H. and {Shih}, A.~Y. and {Sick}, J. and
         {Silbiger}, M.~T. and {Singanamalla}, S. and {Singer}, L.~P. and
         {Sladen}, P.~H. and {Sooley}, K.~A. and {Sornarajah}, S. and
         {Streicher}, O. and {Teuben}, P. and {Thomas}, S.~W. and
         {Tremblay}, G.~R. and {Turner}, J.~E.~H. and {Terr{\'o}n}, V. and
         {van Kerkwijk}, M.~H. and {de la Vega}, A. and {Watkins}, L.~L. and
         {Weaver}, B.~A. and {Whitmore}, J.~B. and {Woillez}, J. and
         {Zabalza}, V. and {Astropy Contributors}},
        title = "{The Astropy Project: Building an Open-science Project and Status of the v2.0 Core Package}",
      journal = {\aj},
         year = 2018,
        month = sep,
       volume = {156},
       number = {3},
          eid = {123},
        pages = {123},
          doi = {10.3847/1538-3881/aabc4f},
archivePrefix = {arXiv},
       eprint = {1801.02634},
 primaryClass = {astro-ph.IM},
       adsurl = {https://ui.adsabs.harvard.edu/abs/2018AJ....156..123A}
}

@ARTICLE{astropy:2022,
       author = {{Astropy Collaboration} and {Price-Whelan}, Adrian M. and {Lim}, Pey Lian and {Earl}, Nicholas and {Starkman}, Nathaniel and {Bradley}, Larry and {Shupe}, David L. and {Patil}, Aarya A. and {Corrales}, Lia and {Brasseur}, C.~E. and {N{"o}the}, Maximilian and {Donath}, Axel and {Tollerud}, Erik and {Morris}, Brett M. and {Ginsburg}, Adam and {Vaher}, Eero and {Weaver}, Benjamin A. and {Tocknell}, James and {Jamieson}, William and {van Kerkwijk}, Marten H. and {Robitaille}, Thomas P. and {Merry}, Bruce and {Bachetti}, Matteo and {G{"u}nther}, H. Moritz and {Aldcroft}, Thomas L. and {Alvarado-Montes}, Jaime A. and {Archibald}, Anne M. and {B{'o}di}, Attila and {Bapat}, Shreyas and {Barentsen}, Geert and {Baz{'a}n}, Juanjo and {Biswas}, Manish and {Boquien}, M{'e}d{'e}ric and {Burke}, D.~J. and {Cara}, Daria and {Cara}, Mihai and {Conroy}, Kyle E. and {Conseil}, Simon and {Craig}, Matthew W. and {Cross}, Robert M. and {Cruz}, Kelle L. and {D'Eugenio}, Francesco and {Dencheva}, Nadia and {Devillepoix}, Hadrien A.~R. and {Dietrich}, J{"o}rg P. and {Eigenbrot}, Arthur Davis and {Erben}, Thomas and {Ferreira}, Leonardo and {Foreman-Mackey}, Daniel and {Fox}, Ryan and {Freij}, Nabil and {Garg}, Suyog and {Geda}, Robel and {Glattly}, Lauren and {Gondhalekar}, Yash and {Gordon}, Karl D. and {Grant}, David and {Greenfield}, Perry and {Groener}, Austen M. and {Guest}, Steve and {Gurovich}, Sebastian and {Handberg}, Rasmus and {Hart}, Akeem and {Hatfield-Dodds}, Zac and {Homeier}, Derek and {Hosseinzadeh}, Griffin and {Jenness}, Tim and {Jones}, Craig K. and {Joseph}, Prajwel and {Kalmbach}, J. Bryce and {Karamehmetoglu}, Emir and {Ka{l}uszy{'n}ski}, Miko{l}aj and {Kelley}, Michael S.~P. and {Kern}, Nicholas and {Kerzendorf}, Wolfgang E. and {Koch}, Eric W. and {Kulumani}, Shankar and {Lee}, Antony and {Ly}, Chun and {Ma}, Zhiyuan and {MacBride}, Conor and {Maljaars}, Jakob M. and {Muna}, Demitri and {Murphy}, N.~A. and {Norman}, Henrik and {O'Steen}, Richard and {Oman}, Kyle A. and {Pacifici}, Camilla and {Pascual}, Sergio and {Pascual-Granado}, J. and {Patil}, Rohit R. and {Perren}, Gabriel I. and {Pickering}, Timothy E. and {Rastogi}, Tanuj and {Roulston}, Benjamin R. and {Ryan}, Daniel F. and {Rykoff}, Eli S. and {Sabater}, Jose and {Sakurikar}, Parikshit and {Salgado}, Jes{'u}s and {Sanghi}, Aniket and {Saunders}, Nicholas and {Savchenko}, Volodymyr and {Schwardt}, Ludwig and {Seifert-Eckert}, Michael and {Shih}, Albert Y. and {Jain}, Anany Shrey and {Shukla}, Gyanendra and {Sick}, Jonathan and {Simpson}, Chris and {Singanamalla}, Sudheesh and {Singer}, Leo P. and {Singhal}, Jaladh and {Sinha}, Manodeep and {Sip{H{o}}cz}, Brigitta M. and {Spitler}, Lee R. and {Stansby}, David and {Streicher}, Ole and {{{S}}umak}, Jani and {Swinbank}, John D. and {Taranu}, Dan S. and {Tewary}, Nikita and {Tremblay}, Grant R. and {Val-Borro}, Miguel de and {Van Kooten}, Samuel J. and {Vasovi{'c}}, Zlatan and {Verma}, Shresth and {de Miranda Cardoso}, Jos{'e} Vin{'i}cius and {Williams}, Peter K.~G. and {Wilson}, Tom J. and {Winkel}, Benjamin and {Wood-Vasey}, W.~M. and {Xue}, Rui and {Yoachim}, Peter and {Zhang}, Chen and {Zonca}, Andrea and {Astropy Project Contributors}},
        title = "{The Astropy Project: Sustaining and Growing a Community-oriented Open-source Project and the Latest Major Release (v5.0) of the Core Package}",
      journal = {\apj},
         year = 2022,
        month = aug,
       volume = {935},
       number = {2},
          eid = {167},
        pages = {167},
          doi = {10.3847/1538-4357/ac7c74},
archivePrefix = {arXiv},
       eprint = {2206.14220},
 primaryClass = {astro-ph.IM},
       adsurl = {https://ui.adsabs.harvard.edu/abs/2022ApJ...935..167A}
}

@ARTICLE{twojupiters,
       author = {{Wittenmyer}, Robert A. and {Horner}, Jonathan and {Tinney}, C.~G. and {Butler}, R.~P. and {Jones}, H.~R.~A. and {Tuomi}, Mikko and {Salter}, G.~S. and {Carter}, B.~D. and {Koch}, F. Elliott and {O'Toole}, S.~J. and {Bailey}, J. and {Wright}, D.},
        title = "{The Anglo-Australian Planet Search. XXIII. Two New Jupiter Analogs}",
      journal = {\apj},
         year = 2014,
        month = mar,
       volume = {783},
       number = {2},
          eid = {103},
        pages = {103},
          doi = {10.1088/0004-637X/783/2/103},
archivePrefix = {arXiv},
       eprint = {1401.5525},
 primaryClass = {astro-ph.EP},
       adsurl = {https://ui.adsabs.harvard.edu/abs/2014ApJ...783..103W}
}

@ARTICLE{witt13,
       author = {{Wittenmyer}, Robert A. and {Tinney}, C.~G. and {Horner}, J. and {Butler}, R.~P. and {Jones}, H.~R.~A. and {O'Toole}, S.~J. and {Bailey}, J. and {Carter}, B.~D. and {Salter}, G.~S. and {Wright}, D.},
        title = "{Observing Strategies for the Detection of Jupiter Analogs}",
      journal = {\pasp},
         year = 2013,
        month = apr,
       volume = {125},
       number = {926},
        pages = {351},
          doi = {10.1086/670680},
archivePrefix = {arXiv},
       eprint = {1303.3336},
 primaryClass = {astro-ph.EP},
       adsurl = {https://ui.adsabs.harvard.edu/abs/2013PASP..125..351W}
}

@ARTICLE{grieves2018,
       author = {{Grieves}, N. and {Ge}, J. and {Thomas}, N. and {Willis}, K. and {Ma}, B. and {Lorenzo-Oliveira}, D. and {Queiroz}, A.~B.~A. and {Ghezzi}, L. and {Chiappini}, C. and {Anders}, F. and {Dutra-Ferreira}, L. and {Porto de Mello}, G.~F. and {Santiago}, B.~X. and {da Costa}, L.~N. and {Ogando}, R.~L.~C. and {del Peloso}, E.~F. and {Tan}, J.~C. and {Schneider}, D.~P. and {Pepper}, J. and {Stassun}, K.~G. and {Zhao}, B. and {Bizyaev}, D. and {Pan}, K.},
        title = "{Chemo-kinematics of the Milky Way from the SDSS-III MARVELS survey}",
      journal = {\mnras},
         year = 2018,
        month = dec,
       volume = {481},
       number = {3},
        pages = {3244-3265},
          doi = {10.1093/mnras/sty2431},
archivePrefix = {arXiv},
       eprint = {1803.11538},
 primaryClass = {astro-ph.SR},
       adsurl = {https://ui.adsabs.harvard.edu/abs/2018MNRAS.481.3244G}
}

@dataset{cannon1993,
       author = {{Cannon}, A.~J. and {Pickering}, E.~C.},
        title = "{VizieR Online Data Catalog: Henry Draper Catalogue and Extension (Cannon+ 1918-1924; ADC 1989)}",
 howpublished = {VizieR On-line Data Catalog: III/135A.  Originally published in: Harv. Ann. 91-100 (1918-1924)},
         year = 1993,
        month = oct,
          eid = {III/135A},
       adsurl = {https://ui.adsabs.harvard.edu/abs/1993yCat.3135....0C}
}

@ARTICLE{sweetcat2024,
       author = {{Sousa}, S.~G. and {Adibekyan}, V. and {Delgado-Mena}, E. and {Santos}, N.~C. and {Rojas-Ayala}, B. and {Soares}, B.~M.~T.~B. and {Legoinha}, H. and {Ulmer-Moll}, S. and {Camacho}, J.~D. and {Barros}, S.~C.~C. and {Demangeon}, O.~D.~S. and {Hoyer}, S. and {Israelian}, G. and {Mortier}, A. and {Tsantaki}, M. and {Monteiro}, M.~A.},
        title = "{SWEET-Cat 2.0: The Cat just got SWEETer. Higher quality spectra and precise parallaxes from Gaia eDR3}",
      journal = {\aap},
         year = 2021,
        month = dec,
       volume = {656},
          eid = {A53},
        pages = {A53},
          doi = {10.1051/0004-6361/202141584},
archivePrefix = {arXiv},
       eprint = {2109.04781},
 primaryClass = {astro-ph.EP},
       adsurl = {https://ui.adsabs.harvard.edu/abs/2021A&A...656A..53S}
}

@ARTICLE{gray2001,
       author = {{Gray}, R.~O. and {Napier}, M.~G. and {Winkler}, L.~I.},
        title = "{The Physical Basis of Luminosity Classification in the Late A-, F-, and Early G-Type Stars. I. Precise Spectral Types for 372 Stars}",
      journal = {\aj},
         year = 2001,
        month = apr,
       volume = {121},
       number = {4},
        pages = {2148-2158},
          doi = {10.1086/319956},
       adsurl = {https://ui.adsabs.harvard.edu/abs/2001AJ....121.2148G}
}

@ARTICLE{salz2015,
       author = {{Salz}, M. and {Schneider}, P.~C. and {Czesla}, S. and {Schmitt}, J.~H.~M.~M.},
        title = "{High-energy irradiation and mass loss rates of hot Jupiters in the solar neighborhood}",
      journal = {\aap},
         year = 2015,
        month = apr,
       volume = {576},
          eid = {A42},
        pages = {A42},
          doi = {10.1051/0004-6361/201425243},
archivePrefix = {arXiv},
       eprint = {1502.00576},
 primaryClass = {astro-ph.EP},
       adsurl = {https://ui.adsabs.harvard.edu/abs/2015A&A...576A..42S}
}

@ARTICLE{kraus2007,
       author = {{Kraus}, Adam L. and {Hillenbrand}, Lynne A.},
        title = "{The Stellar Populations of Praesepe and Coma Berenices}",
      journal = {\aj},
         year = 2007,
        month = dec,
       volume = {134},
       number = {6},
        pages = {2340-2352},
          doi = {10.1086/522831},
archivePrefix = {arXiv},
       eprint = {0708.2719},
 primaryClass = {astro-ph},
       adsurl = {https://ui.adsabs.harvard.edu/abs/2007AJ....134.2340K}
}

@ARTICLE{lagrange23,
       author = {{Lagrange}, A. -M. and {Philipot}, F. and {Rubini}, P. and {Meunier}, N. and {Kiefer}, F. and {Kervella}, P. and {Delorme}, P. and {Beust}, H.},
        title = "{Radial distribution of giant exoplanets at Solar System scales}",
      journal = {\aap},
         year = 2023,
        month = sep,
       volume = {677},
          eid = {A71},
        pages = {A71},
          doi = {10.1051/0004-6361/202346165},
archivePrefix = {arXiv},
       eprint = {2305.00047},
 primaryClass = {astro-ph.EP},
       adsurl = {https://ui.adsabs.harvard.edu/abs/2023A&A...677A..71L}
}

@ARTICLE{cumming08,
       author = {{Cumming}, Andrew and {Butler}, R. Paul and {Marcy}, Geoffrey W. and {Vogt}, Steven S. and {Wright}, Jason T. and {Fischer}, Debra A.},
        title = "{The Keck Planet Search: Detectability and the Minimum Mass and Orbital Period Distribution of Extrasolar Planets}",
      journal = {\pasp},
         year = 2008,
        month = may,
       volume = {120},
       number = {867},
        pages = {531},
          doi = {10.1086/588487},
archivePrefix = {arXiv},
       eprint = {0803.3357},
 primaryClass = {astro-ph},
       adsurl = {https://ui.adsabs.harvard.edu/abs/2008PASP..120..531C}
}

@ARTICLE{gan24,
       author = {{Gan}, Tianjun and {Guo}, Kangrou and {Liu}, Beibei and {Wang}, Sharon X. and {Mao}, Shude and {Buchner}, Johannes and {Fulton}, Benjamin J.},
        title = "{Relative Occurrence Rate between Hot and Cold Jupiters as an Indicator to Probe Planet Migration}",
      journal = {\apj},
         year = 2024,
        month = may,
       volume = {967},
       number = {1},
          eid = {74},
        pages = {74},
          doi = {10.3847/1538-4357/ad3deb},
archivePrefix = {arXiv},
       eprint = {2404.07033},
 primaryClass = {astro-ph.EP},
       adsurl = {https://ui.adsabs.harvard.edu/abs/2024ApJ...967...74G}
}

@ARTICLE{poleski21,
       author = {{Poleski}, R. and {Skowron}, J. and {Mr{\'o}z}, P. and {Udalski}, A. and {Szyma{\'n}ski}, M.~K. and {Pietrukowicz}, P. and {Ulaczyk}, K. and {Rybicki}, K. and {Iwanek}, P. and {Wrona}, M. and {Gromadzki}, M.},
        title = "{Wide-Orbit Exoplanets are Common. Analysis of Nearly 20 Years of OGLE Microlensing Survey Data}",
      journal = {\actaa},
         year = 2021,
        month = mar,
       volume = {71},
       number = {1},
        pages = {1-23},
          doi = {10.32023/0001-5237/71.1.1},
archivePrefix = {arXiv},
       eprint = {2104.02079},
 primaryClass = {astro-ph.EP},
       adsurl = {https://ui.adsabs.harvard.edu/abs/2021AcA....71....1P}
}

@ARTICLE{bryan16,
       author = {{Bryan}, Marta L. and {Knutson}, Heather A. and {Howard}, Andrew W. and {Ngo}, Henry and {Batygin}, Konstantin and {Crepp}, Justin R. and {Fulton}, B.~J. and {Hinkley}, Sasha and {Isaacson}, Howard and {Johnson}, John A. and {Marcy}, Geoffry W. and {Wright}, Jason T.},
        title = "{Statistics of Long Period Gas Giant Planets in Known Planetary Systems}",
      journal = {\apj},
         year = 2016,
        month = apr,
       volume = {821},
       number = {2},
          eid = {89},
        pages = {89},
          doi = {10.3847/0004-637X/821/2/89},
archivePrefix = {arXiv},
       eprint = {1601.07595},
 primaryClass = {astro-ph.EP},
       adsurl = {https://ui.adsabs.harvard.edu/abs/2016ApJ...821...89B}
}

@ARTICLE{nielsen19,
       author = {{Nielsen}, Eric L. and {De Rosa}, Robert J. and {Macintosh}, Bruce and {Wang}, Jason J. and {Ruffio}, Jean-Baptiste and {Chiang}, Eugene and {Marley}, Mark S. and {Saumon}, Didier and {Savransky}, Dmitry and {Ammons}, S. Mark and {Bailey}, Vanessa P. and {Barman}, Travis and {Blain}, C{\'e}lia and {Bulger}, Joanna and {Burrows}, Adam and {Chilcote}, Jeffrey and {Cotten}, Tara and {Czekala}, Ian and {Doyon}, Rene and {Duch{\^e}ne}, Gaspard and {Esposito}, Thomas M. and {Fabrycky}, Daniel and {Fitzgerald}, Michael P. and {Follette}, Katherine B. and {Fortney}, Jonathan J. and {Gerard}, Benjamin L. and {Goodsell}, Stephen J. and {Graham}, James R. and {Greenbaum}, Alexandra Z. and {Hibon}, Pascale and {Hinkley}, Sasha and {Hirsch}, Lea A. and {Hom}, Justin and {Hung}, Li-Wei and {Dawson}, Rebekah Ilene and {Ingraham}, Patrick and {Kalas}, Paul and {Konopacky}, Quinn and {Larkin}, James E. and {Lee}, Eve J. and {Lin}, Jonathan W. and {Maire}, J{\'e}r{\^o}me and {Marchis}, Franck and {Marois}, Christian and {Metchev}, Stanimir and {Millar-Blanchaer}, Maxwell A. and {Morzinski}, Katie M. and {Oppenheimer}, Rebecca and {Palmer}, David and {Patience}, Jennifer and {Perrin}, Marshall and {Poyneer}, Lisa and {Pueyo}, Laurent and {Rafikov}, Roman R. and {Rajan}, Abhijith and {Rameau}, Julien and {Rantakyr{\"o}}, Fredrik T. and {Ren}, Bin and {Schneider}, Adam C. and {Sivaramakrishnan}, Anand and {Song}, Inseok and {Soummer}, Remi and {Tallis}, Melisa and {Thomas}, Sandrine and {Ward-Duong}, Kimberly and {Wolff}, Schuyler},
        title = "{The Gemini Planet Imager Exoplanet Survey: Giant Planet and Brown Dwarf Demographics from 10 to 100 au}",
      journal = {\aj},
         year = 2019,
        month = jul,
       volume = {158},
       number = {1},
          eid = {13},
        pages = {13},
          doi = {10.3847/1538-3881/ab16e9},
archivePrefix = {arXiv},
       eprint = {1904.05358},
 primaryClass = {astro-ph.EP},
       adsurl = {https://ui.adsabs.harvard.edu/abs/2019AJ....158...13N}
}

@ARTICLE{butler01,
       author = {{Butler}, R. Paul and {Tinney}, C.~G. and {Marcy}, Geoffrey W. and {Jones}, Hugh R.~A. and {Penny}, Alan J. and {Apps}, Kevin},
        title = "{Two New Planets from the Anglo-Australian Planet Search}",
      journal = {\apj},
         year = 2001,
        month = jul,
       volume = {555},
       number = {1},
        pages = {410-417},
          doi = {10.1086/321467},
       adsurl = {https://ui.adsabs.harvard.edu/abs/2001ApJ...555..410B}
}

@ARTICLE{Diverse1,
       author = {{Adams}, Elisabeth R. and {Jackson}, Brian and {Endl}, Michael},
        title = "{Ultra-short-period Planets in K2 SuPerPiG Results for Campaigns 0-5}",
      journal = {\aj},
         year = 2016,
        month = aug,
       volume = {152},
       number = {2},
          eid = {47},
        pages = {47},
          doi = {10.3847/0004-6256/152/2/47},
archivePrefix = {arXiv},
       eprint = {1603.06488},
 primaryClass = {astro-ph.EP},
       adsurl = {https://ui.adsabs.harvard.edu/abs/2016AJ....152...47A}
}

@ARTICLE{Diverse2,
       author = {{Zhu}, Wei and {Petrovich}, Cristobal and {Wu}, Yanqin and {Dong}, Subo and {Xie}, Jiwei},
        title = "{About 30\% of Sun-like Stars Have Kepler-like Planetary Systems: A Study of Their Intrinsic Architecture}",
      journal = {\apj},
         year = 2018,
        month = jun,
       volume = {860},
       number = {2},
          eid = {101},
        pages = {101},
          doi = {10.3847/1538-4357/aac6d5},
archivePrefix = {arXiv},
       eprint = {1802.09526},
 primaryClass = {astro-ph.EP},
       adsurl = {https://ui.adsabs.harvard.edu/abs/2018ApJ...860..101Z}
}

@ARTICLE{Diverse3,
       author = {{Kunimoto}, Michelle and {Matthews}, Jaymie M.},
        title = "{Searching the Entirety of Kepler Data. II. Occurrence Rate Estimates for FGK Stars}",
      journal = {\aj},
         year = 2020,
        month = jun,
       volume = {159},
       number = {6},
          eid = {248},
        pages = {248},
          doi = {10.3847/1538-3881/ab88b0},
archivePrefix = {arXiv},
       eprint = {2004.05296},
 primaryClass = {astro-ph.EP},
       adsurl = {https://ui.adsabs.harvard.edu/abs/2020AJ....159..248K}
}

@ARTICLE{howardfulton2016,
       author = {{Howard}, Andrew W. and {Fulton}, Benjamin J.},
        title = "{Limits on Planetary Companions from Doppler Surveys of Nearby Stars}",
      journal = {\pasp},
         year = 2016,
        month = nov,
       volume = {128},
       number = {969},
        pages = {114401},
          doi = {10.1088/1538-3873/128/969/114401},
archivePrefix = {arXiv},
       eprint = {1606.03134},
 primaryClass = {astro-ph.EP},
       adsurl = {https://ui.adsabs.harvard.edu/abs/2016PASP..128k4401H}
}

@ARTICLE{tinney01,
       author = {{Tinney}, C.~G. and {Butler}, R. Paul and {Marcy}, Geoffrey W. and {Jones}, Hugh R.~A. and {Penny}, Alan J. and {Vogt}, Steven S. and {Apps}, Kevin and {Henry}, Gregory W.},
        title = "{First Results from the Anglo-Australian Planet Search: A Brown Dwarf Candidate and a 51 Peg-like Planet}",
      journal = {\apj},
         year = 2001,
        month = apr,
       volume = {551},
       number = {1},
        pages = {507-511},
          doi = {10.1086/320097},
archivePrefix = {arXiv},
       eprint = {astro-ph/0012204},
 primaryClass = {astro-ph},
       adsurl = {https://ui.adsabs.harvard.edu/abs/2001ApJ...551..507T}
}

@INPROCEEDINGS{vogt1994,
       author = {{Vogt}, S.~S. and {Allen}, S.~L. and {Bigelow}, B.~C. and {Bresee}, L. and {Brown}, B. and {Cantrall}, T. and {Conrad}, A. and {Couture}, M. and {Delaney}, C. and {Epps}, H.~W. and {Hilyard}, D. and {Hilyard}, D.~F. and {Horn}, E. and {Jern}, N. and {Kanto}, D. and {Keane}, M.~J. and {Kibrick}, R.~I. and {Lewis}, J.~W. and {Osborne}, J. and {Pardeilhan}, G.~H. and {Pfister}, T. and {Ricketts}, T. and {Robinson}, L.~B. and {Stover}, R.~J. and {Tucker}, D. and {Ward}, J. and {Wei}, M.~Z.},
        title = "{HIRES: the high-resolution echelle spectrometer on the Keck 10-m Telescope}",
    booktitle = {Instrumentation in Astronomy VIII},
         year = 1994,
       editor = {{Crawford}, David L. and {Craine}, Eric R.},
       series = {Society of Photo-Optical Instrumentation Engineers (SPIE) Conference Series},
       volume = {2198},
        month = jun,
        pages = {362},
          doi = {10.1117/12.176725},
       adsurl = {https://ui.adsabs.harvard.edu/abs/1994SPIE.2198..362V}
}

@ARTICLE{vogt2002ten,
       author = {{Vogt}, Steven S. and {Butler}, R. Paul and {Marcy}, Geoffrey W. and {Fischer}, Debra A. and {Pourbaix}, Dimitri and {Apps}, Kevin and {Laughlin}, Gregory},
        title = "{Ten Low-Mass Companions from the Keck Precision Velocity Survey}",
      journal = {\apj},
         year = 2002,
        month = mar,
       volume = {568},
       number = {1},
        pages = {352-362},
          doi = {10.1086/338768},
archivePrefix = {arXiv},
       eprint = {astro-ph/0110378},
 primaryClass = {astro-ph},
       adsurl = {https://ui.adsabs.harvard.edu/abs/2002ApJ...568..352V}
}

@ARTICLE{vogt2005stars,
       author = {{Vogt}, Steven S. and {Butler}, R. Paul and {Marcy}, Geoffrey W. and {Fischer}, Debra A. and {Henry}, Gregory W. and {Laughlin}, Greg and {Wright}, Jason T. and {Johnson}, John A.},
        title = "{Five New Multicomponent Planetary Systems}",
      journal = {\apj},
         year = 2005,
        month = oct,
       volume = {632},
       number = {1},
        pages = {638-658},
          doi = {10.1086/432901},
       adsurl = {https://ui.adsabs.harvard.edu/abs/2005ApJ...632..638V}
}

@ARTICLE{SpaceWeatherHabitability,
       author = {{Airapetian}, V.~S. and {Barnes}, R. and {Cohen}, O. and {Collinson}, G.~A. and {Danchi}, W.~C. and {Dong}, C.~F. and {Del Genio}, A.~D. and {France}, K. and {Garcia-Sage}, K. and {Glocer}, A. and {Gopalswamy}, N. and {Grenfell}, J.~L. and {Gronoff}, G. and {G{\"u}del}, M. and {Herbst}, K. and {Henning}, W.~G. and {Jackman}, C.~H. and {Jin}, M. and {Johnstone}, C.~P. and {Kaltenegger}, L. and {Kay}, C.~D. and {Kobayashi}, K. and {Kuang}, W. and {Li}, G. and {Lynch}, B.~J. and {L{\"u}ftinger}, T. and {Luhmann}, J.~G. and {Maehara}, H. and {Mlynczak}, M.~G. and {Notsu}, Y. and {Osten}, R.~A. and {Ramirez}, R.~M. and {Rugheimer}, S. and {Scheucher}, M. and {Schlieder}, J.~E. and {Shibata}, K. and {Sousa-Silva}, C. and {Stamenkovi{\'c}}, V. and {Strangeway}, R.~J. and {Usmanov}, A.~V. and {Vergados}, P. and {Verkhoglyadova}, O.~P. and {Vidotto}, A.~A. and {Voytek}, M. and {Way}, M.~J. and {Zank}, G.~P. and {Yamashiki}, Y.},
        title = "{Impact of space weather on climate and habitability of terrestrial-type exoplanets}",
      journal = {International Journal of Astrobiology},
         year = 2020,
        month = apr,
       volume = {19},
       number = {2},
        pages = {136-194},
          doi = {10.1017/S1473550419000132},
archivePrefix = {arXiv},
       eprint = {1905.05093},
 primaryClass = {astro-ph.EP},
       adsurl = {https://ui.adsabs.harvard.edu/abs/2020IJAsB..19..136A}
}

@ARTICLE{vogt2000six,
       author = {{Vogt}, Steven S. and {Marcy}, Geoffrey W. and {Butler}, R. Paul and {Apps}, Kevin},
        title = "{Six New Planets from the Keck Precision Velocity Survey}",
      journal = {\apj},
         year = 2000,
        month = jun,
       volume = {536},
       number = {2},
        pages = {902-914},
          doi = {10.1086/308981},
archivePrefix = {arXiv},
       eprint = {astro-ph/9911506},
 primaryClass = {astro-ph},
       adsurl = {https://ui.adsabs.harvard.edu/abs/2000ApJ...536..902V}
}

@ARTICLE{bonomo2023cold,
       author = {{Bonomo}, A.~S. and {Dumusque}, X. and {Massa}, A. and 
{Mortier}, A. and {Bongiolatti}, R. and {Malavolta}, L. and {Sozzetti}, A. 
and {Buchhave}, L.~A. and {Damasso}, M. and {Haywood}, R.~D. and 
{Morbidelli}, A. and {Latham}, D.~W. and {Molinari}, E. and {Pepe}, F. and 
{Poretti}, E. and {Udry}, S. and {Affer}, L. and {Boschin}, W. and 
{Charbonneau}, D. and {Cosentino}, R. and {Cretignier}, M. and {Ghedina}, 
A. and {Lega}, E. and {L{\'o}pez-Morales}, M. and {Margini}, M. and 
{Mart{\'\i}nez Fiorenzano}, A.~F. and {Mayor}, M. and {Micela}, G. and 
{Pedani}, M. and {Pinamonti}, M. and {Rice}, K. and {Sasselov}, D. and 
{Tronsgaard}, R. and {Vanderburg}, A.},
        title = "{Cold Jupiters and improved masses in 38 Kepler and K2 
small-planet systems from 3661 high-precision HARPS-N radial velocities. 
No excess of cold Jupiters in small-planet systems}",
      journal = {arXiv e-prints},
         year = 2023,
        month = apr,
          eid = {arXiv:2304.05773},
        pages = {arXiv:2304.05773},
          doi = {10.48550/arXiv.2304.05773},
archivePrefix = {arXiv},
       eprint = {2304.05773},
 primaryClass = {astro-ph.EP},
       adsurl = {https://ui.adsabs.harvard.edu/abs/2023arXiv230405773B}
}

@ARTICLE{bouchy2005elodie,
       author = {{Bouchy}, F. and {Udry}, S. and {Mayor}, M. and {Moutou}, C. and {Pont}, F. and {Iribarne}, N. and {da Silva}, R. and {Ilovaisky}, S. and {Queloz}, D. and {Santos}, N.~C. and {S{\'e}gransan}, D. and {Zucker}, S.},
        title = "{ELODIE metallicity-biased search for transiting Hot Jupiters. II. A very hot Jupiter transiting the bright K star HD 189733}",
      journal = {\aap},
         year = 2005,
        month = dec,
       volume = {444},
       number = {1},
        pages = {L15-L19},
          doi = {10.1051/0004-6361:200500201},
archivePrefix = {arXiv},
       eprint = {astro-ph/0510119},
 primaryClass = {astro-ph},
       adsurl = {https://ui.adsabs.harvard.edu/abs/2005A&A...444L..15B}
}

@ARTICLE{bryson2021occurrence,
       author = {{Bryson}, Steve and {Kunimoto}, Michelle and {Kopparapu}, 
Ravi K. and {Coughlin}, Jeffrey L. and {Borucki}, William J. and {Koch}, 
David and {Aguirre}, Victor Silva and {Allen}, Christopher and 
{Barentsen}, Geert and {Batalha}, Natalie M. and {Berger}, Travis and 
{Boss}, Alan and {Buchhave}, Lars A. and {Burke}, Christopher J. and 
{Caldwell}, Douglas A. and {Campbell}, Jennifer R. and {Catanzarite}, 
Joseph and {Chandrasekaran}, Hema and {Chaplin}, William J. and 
{Christiansen}, Jessie L. and {Christensen-Dalsgaard}, J{\o}rgen and 
{Ciardi}, David R. and {Clarke}, Bruce D. and {Cochran}, William D. and 
{Dotson}, Jessie L. and {Doyle}, Laurance R. and {Duarte}, Eduardo 
Seperuelo and {Dunham}, Edward W. and {Dupree}, Andrea K. and {Endl}, 
Michael and {Fanson}, James L. and {Ford}, Eric B. and {Fujieh}, Maura and 
{Gautier}, Thomas N., III and {Geary}, John C. and {Gilliland}, Ronald L. 
and {Girouard}, Forrest R. and {Gould}, Alan and {Haas}, Michael R. and 
{Henze}, Christopher E. and {Holman}, Matthew J. and {Howard}, Andrew W. 
and {Howell}, Steve B. and {Huber}, Daniel and {Hunter}, Roger C. and 
{Jenkins}, Jon M. and {Kjeldsen}, Hans and {Kolodziejczak}, Jeffery and 
{Larson}, Kipp and {Latham}, David W. and {Li}, Jie and {Mathur}, Savita 
and {Meibom}, S{\o}ren and {Middour}, Chris and {Morris}, Robert L. and 
{Morton}, Timothy D. and {Mullally}, Fergal and {Mullally}, Susan E. and 
{Pletcher}, David and {Prsa}, Andrej and {Quinn}, Samuel N. and 
{Quintana}, Elisa V. and {Ragozzine}, Darin and {Ramirez}, Solange V. and 
{Sanderfer}, Dwight T. and {Sasselov}, Dimitar and {Seader}, Shawn E. and 
{Shabram}, Megan and {Shporer}, Avi and {Smith}, Jeffrey C. and {Steffen}, 
Jason H. and {Still}, Martin and {Torres}, Guillermo and {Troeltzsch}, 
John and {Twicken}, Joseph D. and {Uddin}, Akm Kamal and {Van Cleve}, 
Jeffrey E. and {Voss}, Janice and {Weiss}, Lauren M. and {Welsh}, William 
F. and {Wohler}, Bill and {Zamudio}, Khadeejah A.},
        title = "{The Occurrence of Rocky Habitable-zone Planets around 
Solar-like Stars from Kepler Data}",
      journal = {\aj},
         year = 2021,
        month = jan,
       volume = {161},
       number = {1},
          eid = {36},
        pages = {36},
          doi = {10.3847/1538-3881/abc418},
archivePrefix = {arXiv},
       eprint = {2010.14812},
 primaryClass = {astro-ph.EP},
       adsurl = {https://ui.adsabs.harvard.edu/abs/2021AJ....161...36B}
}

@ARTICLE{butler1997three,
       author = {{Butler}, R. Paul and {Marcy}, Geoffrey W. and {Williams}, Eric and {Hauser}, Heather and {Shirts}, Phil},
        title = "{Three New ``51 Pegasi-Type'' Planets}",
      journal = {\apjl},
         year = 1997,
        month = jan,
       volume = {474},
       number = {2},
        pages = {L115-L118},
          doi = {10.1086/310444},
       adsurl = {https://ui.adsabs.harvard.edu/abs/1997ApJ...474L.115B}
}

@ARTICLE{butler1998planet,
       author = {{Butler}, R. Paul and {Marcy}, Geoffrey W. and {Vogt}, Steven S. and {Apps}, Kevin},
        title = "{A Planet with a 3.1 Day Period around a Solar Twin}",
      journal = {\pasp},
         year = 1998,
        month = dec,
       volume = {110},
       number = {754},
        pages = {1389-1393},
          doi = {10.1086/316287},
       adsurl = {https://ui.adsabs.harvard.edu/abs/1998PASP..110.1389B}
}

@ARTICLE{butler2000planetary,
       author = {{Butler}, R. Paul and {Vogt}, Steven S. and {Marcy}, Geoffrey W. and {Fischer}, Debra A. and {Henry}, Gregory W. and {Apps}, Kevin},
        title = "{Planetary Companions to the Metal-rich Stars BD -10{\textdegree}3166 and HD 52265}",
      journal = {\apj},
         year = 2000,
        month = dec,
       volume = {545},
       number = {1},
        pages = {504-511},
          doi = {10.1086/317796},
       adsurl = {https://ui.adsabs.harvard.edu/abs/2000ApJ...545..504B}
}

@article{butler2001two,
  title={Two new planets from the {A}nglo-{A}ustralian {P}lanet {S}earch},
  author={Butler, R Paul and Tinney, CG and Marcy, Geoffrey W and Jones, Hugh RA and Penny, Alan J and Apps, Kevin},
  journal={The Astrophysical Journal},
  volume={555},
  number={1},
  pages={410},
  year={2001},
  publisher={IOP Publishing}
}

@ARTICLE{butler2002on,
       author = {{Butler}, R. Paul and {Marcy}, Geoffrey W. and {Vogt}, Steven S. and {Tinney}, C.~G. and {Jones}, Hugh R.~A. and {McCarthy}, Chris and {Penny}, Alan J. and {Apps}, Kevin and {Carter}, Brad D.},
        title = "{On the Double-Planet System around HD 83443}",
      journal = {\apj},
         year = 2002,
        month = oct,
       volume = {578},
       number = {1},
        pages = {565-572},
          doi = {10.1086/342471},
archivePrefix = {arXiv},
       eprint = {astro-ph/0206178},
 primaryClass = {astro-ph},
       adsurl = {https://ui.adsabs.harvard.edu/abs/2002ApJ...578..565B}
}

@ARTICLE{butler2006catalog,
       author = {{Butler}, R.~P. and {Wright}, J.~T. and {Marcy}, G.~W. and {Fischer}, D.~A. and {Vogt}, S.~S. and {Tinney}, C.~G. and {Jones}, H.~R.~A. and {Carter}, B.~D. and {Johnson}, J.~A. and {McCarthy}, C. and {Penny}, A.~J.},
        title = "{Catalog of Nearby Exoplanets}",
      journal = {\apj},
         year = 2006,
        month = jul,
       volume = {646},
       number = {1},
        pages = {505-522},
          doi = {10.1086/504701},
archivePrefix = {arXiv},
       eprint = {astro-ph/0607493},
 primaryClass = {astro-ph},
       adsurl = {https://ui.adsabs.harvard.edu/abs/2006ApJ...646..505B}
}

@article{butler2017lces,
  title={The \textsc{LCES HIRES}/{K}eck precision radial velocity exoplanet survey},
  author={Butler, R Paul and Vogt, Steven S and Laughlin, Gregory and Burt, Jennifer A and Rivera, Eugenio J and Tuomi, Mikko and Teske, Johanna and Arriagada, Pamela and Diaz, Matias and Holden, Brad and others},
  journal={The Astronomical Journal},
  volume={153},
  number={5},
  pages={208},
  year={2017},
  publisher={IOP Publishing}
}

@ARTICLE{Jessie25,
       author = {{Christiansen}, Jessie L. and {McElroy}, Douglas L. and {Harbut}, Marcy and {Ciardi}, David R. and {Crane}, Megan and {Good}, John and {Hardegree-Ullman}, Kevin K. and {Kesseli}, Aurora Y. and {Lund}, Michael B. and {Lynn}, Meca and {Muthiar}, Ananda and {Nilsson}, Ricky and {Oluyide}, Toba and {Papin}, Michael and {Rivera}, Amalia and {Swain}, Melanie and {Susemiehl}, Nicholas D. and {Tam}, Raymond and {van Eyken}, Julian and {Beichman}, Charles},
        title = "{The NASA Exoplanet Archive and Exoplanet Follow-up Observing Program: Data, Tools, and Usage}",
      journal = {PSJ},
         year = 2025,
        month = aug,
       volume = {6},
       number = {8},
          eid = {186},
        pages = {186},
          doi = {10.3847/PSJ/ade3c2},
archivePrefix = {arXiv},
       eprint = {2506.03299},
 primaryClass = {astro-ph.EP},
       adsurl = {https://ui.adsabs.harvard.edu/abs/2025PSJ.....6..186C}
}

@ARTICLE{locurto2015,
       author = {{Lo Curto}, G. and {Pepe}, F. and {Avila}, G. and {Boffin}, H. and {Bovay}, S. and {Chazelas}, B. and {Coffinet}, A. and {Fleury}, M. and {Hughes}, I. and {Lovis}, C. and {Maire}, C. and {Manescau}, A. and {Pasquini}, L. and {Rihs}, S. and {Sinclaire}, P. and {Udry}, S.},
        title = "{HARPS Gets New Fibres After 12 Years of Operations}",
      journal = {The Messenger},
         year = 2015,
        month = dec,
       volume = {162},
        pages = {9-15},
       adsurl = {https://ui.adsabs.harvard.edu/abs/2015Msngr.162....9L}
}

@ARTICLE{Charbonneau2000,
       author = {{Charbonneau}, David and {Brown}, Timothy M. and {Latham}, David W. and {Mayor}, Michel},
        title = "{Detection of Planetary Transits Across a Sun-like Star}",
      journal = {\apjl},
         year = 2000,
        month = jan,
       volume = {529},
       number = {1},
        pages = {L45-L48},
          doi = {10.1086/312457},
archivePrefix = {arXiv},
       eprint = {astro-ph/9911436},
 primaryClass = {astro-ph},
       adsurl = {https://ui.adsabs.harvard.edu/abs/2000ApJ...529L..45C}
}

@ARTICLE{RVref1,
       author = {{Polanski}, Alex S. and {Lubin}, Jack and {Beard}, Corey and {Akana Murphy}, Joseph M. and {Rubenzahl}, Ryan and {Hill}, Michelle L. and {Crossfield}, Ian J.~M. and {Chontos}, Ashley and {Robertson}, Paul and {Isaacson}, Howard and {Kane}, Stephen R. and {Ciardi}, David R. and {Batalha}, Natalie M. and {Dressing}, Courtney and {Fulton}, Benjamin and {Howard}, Andrew W. and {Huber}, Daniel and {Petigura}, Erik A. and {Weiss}, Lauren M. and {Angelo}, Isabel and {Behmard}, Aida and {Blunt}, Sarah and {Brinkman}, Casey L. and {Dai}, Fei and {Dalba}, Paul A. and {Fetherolf}, Tara and {Giacalone}, Steven and {Hirsch}, Lea A. and {Holcomb}, Rae and {Kosiarek}, Molly R. and {Mayo}, Andrew W. and {MacDougall}, Mason G. and {Mo{\v{c}}nik}, Teo and {Pidhorodetska}, Daria and {Rice}, Malena and {Rosenthal}, Lee J. and {Scarsdale}, Nicholas and {Turtelboom}, Emma V. and {Tyler}, Dakotah and {Van Zandt}, Judah and {Yee}, Samuel W. and {Coria}, David R. and {Dulz}, Shannon D. and {Hartman}, Joel D. and {Householder}, Aaron and {Lange}, Sarah and {Langford}, Andrew and {Louden}, Emma M. and {Siegel}, Jared C. and {Gilbert}, Emily A. and {Gonzales}, Erica J. and {Schlieder}, Joshua E. and {Boyle}, Andrew W. and {Christiansen}, Jessie L. and {Clark}, Catherine A. and {Fernandes}, Rachel B. and {Lund}, Michael B. and {Savel}, Arjun B. and {Gill}, Holden and {Beichman}, Charles and {Matson}, Rachel and {Matthews}, Elisabeth C. and {Furlan}, E. and {Howell}, Steve B. and {Scott}, Nicholas J. and {Everett}, Mark E. and {Livingston}, John H. and {Ershova}, Irina O. and {Cheryasov}, Dmitry V. and {Safonov}, Boris and {Lillo-Box}, Jorge and {Barrado}, David and {Morales-Calder{\'o}n}, Mar{\'\i}a},
        title = "{The TESS-Keck Survey. XX. 15 New TESS Planets and a Uniform RV Analysis of All Survey Targets}",
      journal = {\apjs},
         year = 2024,
        month = jun,
       volume = {272},
       number = {2},
          eid = {32},
        pages = {32},
          doi = {10.3847/1538-4365/ad4484},
archivePrefix = {arXiv},
       eprint = {2405.14786},
 primaryClass = {astro-ph.EP},
       adsurl = {https://ui.adsabs.harvard.edu/abs/2024ApJS..272...32P}
}

@ARTICLE{RVref2,
       author = {{Zhang}, Jingwen and {Weiss}, Lauren M. and {Huber}, Daniel and {Xuan}, Jerry W. and {Bottom}, Michael and {Fulton}, Benjamin J. and {Isaacson}, Howard and {MacDougall}, Mason G. and {Saunders}, Nicholas},
        title = "{Discovery of a Jupiter Analog Misaligned to the Inner Planetary System in HD 73344}",
      journal = {\aj},
         year = 2025,
        month = apr,
       volume = {169},
       number = {4},
          eid = {200},
        pages = {200},
          doi = {10.3847/1538-3881/ada60a},
archivePrefix = {arXiv},
       eprint = {2408.09614},
 primaryClass = {astro-ph.EP},
       adsurl = {https://ui.adsabs.harvard.edu/abs/2025AJ....169..200Z}
}

@ARTICLE{RVref3,
       author = {{Van Zandt}, Judah and {Petigura}, Erik A. and {Lubin}, Jack and {Weiss}, Lauren M. and {Turtelboom}, Emma V. and {Fetherolf}, Tara and {Murphy}, Joseph M. Akana and {Crossfield}, Ian J.~M. and {Gilbert}, Gregory J. and {Mo{\v{c}}nik}, Teo and {Batalha}, Natalie M. and {Dressing}, Courtney and {Fulton}, Benjamin and {Howard}, Andrew W. and {Huber}, Daniel and {Isaacson}, Howard and {Kane}, Stephen R. and {Robertson}, Paul and {Roy}, Arpita and {Angelo}, Isabel and {Behmard}, Aida and {Beard}, Corey and {Chontos}, Ashley and {Dai}, Fei and {Giacalone}, Steven and {Hill}, Michelle L. and {Holcomb}, Rae and {Howell}, Steve B. and {Mayo}, Andrew W. and {Pidhorodetska}, Daria and {Polanski}, Alex S. and {Rogers}, James and {Rosenthal}, Lee J. and {Rubenzahl}, Ryan A. and {Scarsdale}, Nicholas and {Tyler}, Dakotah and {Yee}, Samuel W. and {Zink}, Jon},
        title = "{The TESS{\textendash}Keck Survey. XXIV. Outer Giants May Be More Prevalent in the Presence of Inner Small Planets}",
      journal = {\aj},
         year = 2025,
        month = may,
       volume = {169},
       number = {5},
          eid = {235},
        pages = {235},
          doi = {10.3847/1538-3881/adbbed},
archivePrefix = {arXiv},
       eprint = {2501.06342},
 primaryClass = {astro-ph.EP},
       adsurl = {https://ui.adsabs.harvard.edu/abs/2025AJ....169..235V}
}

@ARTICLE{RVref4,
       author = {{Laliotis}, Katherine and {Burt}, Jennifer A. and {Mamajek}, Eric E. and {Li}, Zhexing and {Perdelwitz}, Volker and {Zhao}, Jinglin and {Butler}, R. Paul and {Holden}, Bradford and {Rosenthal}, Lee and {Fulton}, B.~J. and {Feng}, Fabo and {Kane}, Stephen R. and {Bailey}, Jeremy and {Carter}, Brad and {Crane}, Jeffrey D. and {Furlan}, Elise and {Gnilka}, Crystal L. and {Howell}, Steve B. and {Laughlin}, Gregory and {Shectman}, Stephen A. and {Teske}, Johanna K. and {Tinney}, C.~G. and {Vogt}, Steven S. and {Wang}, Sharon Xuesong and {Wittenmyer}, Robert A.},
        title = "{Doppler Constraints on Planetary Companions to Nearby Sun-like Stars: An Archival Radial Velocity Survey of Southern Targets for Proposed NASA Direct Imaging Missions}",
      journal = {\aj},
         year = 2023,
        month = apr,
       volume = {165},
       number = {4},
          eid = {176},
        pages = {176},
          doi = {10.3847/1538-3881/acc067},
archivePrefix = {arXiv},
       eprint = {2302.10310},
 primaryClass = {astro-ph.EP},
       adsurl = {https://ui.adsabs.harvard.edu/abs/2023AJ....165..176L}
}

@ARTICLE{RVref5,
       author = {{Harada}, Caleb K. and {Dressing}, Courtney D. and {Kane}, Stephen R. and {Blunt}, Sarah and {Dietrich}, Jamie and {Hinkel}, Natalie R. and {Li}, Zhexing and {Mamajek}, Eric and {Rice}, Malena and {Tuchow}, Noah W. and {Turtelboom}, Emma V. and {Wittenmyer}, Robert A.},
        title = "{SPORES-HWO. II. Limits on Planetary Companions of Future High-contrast Imaging Targets from $>$20 Years of HIRES and HARPS Radial Velocities}",
      journal = {arXiv e-prints},
         year = 2024,
        month = sep,
          eid = {arXiv:2409.10679},
        pages = {arXiv:2409.10679},
          doi = {10.48550/arXiv.2409.10679},
archivePrefix = {arXiv},
       eprint = {2409.10679},
 primaryClass = {astro-ph.EP},
       adsurl = {https://ui.adsabs.harvard.edu/abs/2024arXiv240910679H}
}

@ARTICLE{li25,
       author = {{Li}, Zhexing and {Kane}, Stephen R. and {Blunt}, Sarah and {Harada}, Caleb K.},
        title = "{Radial Velocity Strategies for the Orbital Refinement of Exoplanet Direct Imaging Targets}",
      journal = {arXiv e-prints},
         year = 2025,
        month = sep,
          eid = {arXiv:2509.17169},
        pages = {arXiv:2509.17169},
          doi = {10.48550/arXiv.2509.17169},
archivePrefix = {arXiv},
       eprint = {2509.17169},
 primaryClass = {astro-ph.EP},
       adsurl = {https://ui.adsabs.harvard.edu/abs/2025arXiv250917169L}
}

@ARTICLE{RVref6,
       author = {{Wittenmyer}, Robert A. and {Errico}, Adriana and {Holt}, Timothy R. and {Horner}, Jonathan and {Harada}, Caleb K. and {Kane}, Stephen R. and {Li}, Zhexing and {Fetherolf}, Tara},
        title = "{Optimizing radial velocity detection limits for Southern Habitable Worlds Observatory targets}",
      journal = {\mnras},
         year = 2025,
        month = may,
       volume = {539},
       number = {1},
        pages = {457-462},
          doi = {10.1093/mnras/staf503},
archivePrefix = {arXiv},
       eprint = {2503.20121},
 primaryClass = {astro-ph.EP},
       adsurl = {https://ui.adsabs.harvard.edu/abs/2025MNRAS.539..457W}
}

@ARTICLE{dasilva2006elodie,
       author = {{da Silva}, R. and {Udry}, S. and {Bouchy}, F. and {Mayor}, M. and {Moutou}, C. and {Pont}, F. and {Queloz}, D. and {Santos}, N.~C. and {S{\'e}gransan}, D. and {Zucker}, S.},
        title = "{Elodie metallicity-biased search for transiting Hot Jupiters. I. Two Hot Jupiters orbiting the slightly evolved stars <ASTROBJ>HD 118203</ASTROBJ> and <ASTROBJ>HD 149143</ASTROBJ>}",
      journal = {\aap},
         year = 2006,
        month = feb,
       volume = {446},
       number = {2},
        pages = {717-722},
          doi = {10.1051/0004-6361:20054116},
archivePrefix = {arXiv},
       eprint = {astro-ph/0510048},
 primaryClass = {astro-ph},
       adsurl = {https://ui.adsabs.harvard.edu/abs/2006A&A...446..717D}
}

@ARTICLE{dressing2015occurrence,
       author = {{Dressing}, Courtney D. and {Charbonneau}, David},
        title = "{The Occurrence of Potentially Habitable Planets Orbiting 
M Dwarfs Estimated from the Full Kepler Dataset and an Empirical 
Measurement of the Detection Sensitivity}",
      journal = {\apj},
         year = 2015,
        month = jul,
       volume = {807},
       number = {1},
          eid = {45},
        pages = {45},
          doi = {10.1088/0004-637X/807/1/45},
archivePrefix = {arXiv},
       eprint = {1501.01623},
 primaryClass = {astro-ph.EP},
       adsurl = {https://ui.adsabs.harvard.edu/abs/2015ApJ...807...45D}
}

@ARTICLE{errico2022HD83443,
       author = {{Errico}, Adriana and {Wittenmyer}, Robert A. and {Horner}, Jonathan and {Li}, Zhexing and {Brandt}, G. Mirek and {Kane}, Stephen R. and {Fetherolf}, Tara and {Holt}, Timothy R. and {Carter}, Brad and {Clark}, Jake T. and {Butler}, R.~P. and {Tinney}, C.~G. and {Ballard}, Sarah and {Bowler}, Brendan P. and {Kielkopf}, John and {Liu}, Huigen and {Plavchan}, Peter P. and {Shporer}, Avi and {Zhang}, Hui and {Wright}, Duncan J. and {Addison}, Brett C. and {Mengel}, Matthew W. and {Okumura}, Jack},
        title = "{HD 83443c: A Highly Eccentric Giant Planet on a 22 yr Orbit}",
      journal = {\aj},
         year = 2022,
        month = jun,
       volume = {163},
       number = {6},
          eid = {273},
        pages = {273},
          doi = {10.3847/1538-3881/ac6589},
archivePrefix = {arXiv},
       eprint = {2204.05711},
        primaryClass = {astro-ph.EP},
       adsurl = {https://ui.adsabs.harvard.edu/abs/2022AJ....163..273E}
}

@misc{eso_faq,
  author       = {{European Southern Observatory}},
  title        = {FAQ – VLT and Paranal},
  howpublished = {\url{https://www.eso.org/public/about-eso/faq/faq-vlt-paranal/#25}},
  note         = {Accessed: 27 April 2025},
  year         = {n.d.}
}

@ARTICLE{Feng15,
       author = {{Feng}, Y. Katherina and {Wright}, Jason T. and {Nelson}, Benjamin and {Wang}, Sharon X. and {Ford}, Eric B. and {Marcy}, Geoffrey W. and {Isaacson}, Howard and {Howard}, Andrew W.},
        title = "{The California Planet Survey IV: A Planet Orbiting the Giant Star HD 145934 and Updates to Seven Systems with Long-period Planets}",
      journal = {\apj},
         year = 2015,
        month = feb,
       volume = {800},
       number = {1},
          eid = {22},
        pages = {22},
          doi = {10.1088/0004-637X/800/1/22},
archivePrefix = {arXiv},
       eprint = {1501.00633},
 primaryClass = {astro-ph.EP},
       adsurl = {https://ui.adsabs.harvard.edu/abs/2015ApJ...800...22F}
}

@ARTICLE{fischer1999planetary,
       author = {{Fischer}, Debra A. and {Marcy}, Geoffrey W. and {Butler}, R. Paul and {Vogt}, Steven S. and {Apps}, Kevin},
        title = "{Planetary Companions around Two Solar-Type Stars: HD 195019 and HD 217107}",
      journal = {\pasp},
         year = 1999,
        month = jan,
       volume = {111},
       number = {755},
        pages = {50-56},
          doi = {10.1086/316304},
archivePrefix = {arXiv},
       eprint = {astro-ph/9810420},
 primaryClass = {astro-ph},
       adsurl = {https://ui.adsabs.harvard.edu/abs/1999PASP..111...50F}
}

@ARTICLE{fischer1999stars,
       author = {{Fischer}, Debra A. and {Marcy}, Geoffrey W. and {Butler}, R. Paul and {Vogt}, Steven S. and {Apps}, Kevin},
        title = "{Planetary Companions around Two Solar-Type Stars: HD 195019 and HD 217107}",
      journal = {\pasp},
         year = 1999,
        month = jan,
       volume = {111},
       number = {755},
        pages = {50-56},
          doi = {10.1086/316304},
archivePrefix = {arXiv},
       eprint = {astro-ph/9810420},
 primaryClass = {astro-ph},
       adsurl = {https://ui.adsabs.harvard.edu/abs/1999PASP..111...50F}
}

@ARTICLE{fischer2003planetary,
       author = {{Fischer}, Debra A. and {Marcy}, Geoffrey W. and {Butler}, R. Paul and {Vogt}, Steven S. and {Henry}, Gregory W. and {Pourbaix}, Dimitri and {Walp}, Bernard and {Misch}, Anthony A. and {Wright}, Jason T.},
        title = "{A Planetary Companion to HD 40979 and Additional Planets Orbiting HD 12661 and HD 38529}",
      journal = {\apj},
         year = 2003,
        month = apr,
       volume = {586},
       number = {2},
        pages = {1394-1408},
          doi = {10.1086/367889},
       adsurl = {https://ui.adsabs.harvard.edu/abs/2003ApJ...586.1394F}
}

@ARTICLE{fischer2005n2k,
       author = {{Fischer}, Debra A. and {Laughlin}, Greg and {Butler}, Paul and {Marcy}, Geoff and {Johnson}, John and {Henry}, Greg and {Valenti}, Jeff and {Vogt}, Steve and {Ammons}, Mark and {Robinson}, Sarah and {Spear}, Greg and {Strader}, Jay and {Driscoll}, Peter and {Fuller}, Abby and {Johnson}, Teresa and {Manrao}, Elizabeth and {McCarthy}, Chris and {Mu{\~n}oz}, Melesio and {Tah}, K.~L. and {Wright}, Jason and {Ida}, Shigeru and {Sato}, Bun'ei and {Toyota}, Eri and {Minniti}, Dante},
        title = "{The N2K Consortium. I. A Hot Saturn Planet Orbiting HD 88133}",
      journal = {\apj},
         year = 2005,
        month = feb,
       volume = {620},
       number = {1},
        pages = {481-486},
          doi = {10.1086/426810},
       adsurl = {https://ui.adsabs.harvard.edu/abs/2005ApJ...620..481F}
}

@ARTICLE{fischer2006,
       author = {{Fischer}, Debra A. and {Laughlin}, Gregory and {Marcy}, Geoffrey W. and {Butler}, R. Paul and {Vogt}, Steven S. and {Johnson}, John A. and {Henry}, Gregory W. and {McCarthy}, Chris and {Ammons}, Mark and {Robinson}, Sarah and {Strader}, Jay and {Valenti}, Jeff A. and {McCullough}, P.~R. and {Charbonneau}, David and {Haislip}, Joshua and {Knutson}, Heather A. and {Reichart}, Daniel E. and {McGee}, Padric and {Monard}, Berto and {Wright}, Jason T. and {Ida}, Shigeru and {Sato}, Bun'ei and {Minniti}, Dante},
        title = "{The N2K Consortium. III. Short-Period Planets Orbiting HD 149143 and HD 109749}",
      journal = {\apj},
         year = 2006,
        month = feb,
       volume = {637},
       number = {2},
        pages = {1094-1101},
          doi = {10.1086/498557},
       adsurl = {https://ui.adsabs.harvard.edu/abs/2006ApJ...637.1094F}
}

@article{fulton2018radvel,
  title={RadVel: the radial velocity modeling toolkit},
  author={Fulton, Benjamin J and Petigura, Erik A and Blunt, Sarah and Sinukoff, Evan},
  journal={Publications of the Astronomical Society of the Pacific},
  volume={130},
  number={986},
  pages={044504},
  year={2018},
  publisher={IOP Publishing}
}

@ARTICLE{FujiBS,
       author = {{Fujii}, Yuka and {Angerhausen}, Daniel and {Deitrick}, Russell and {Domagal-Goldman}, Shawn and {Grenfell}, John Lee and {Hori}, Yasunori and {Kane}, Stephen R. and {Pall{\'e}}, Enric and {Rauer}, Heike and {Siegler}, Nicholas and {Stapelfeldt}, Karl and {Stevenson}, Kevin B.},
        title = "{Exoplanet Biosignatures: Observational Prospects}",
      journal = {Astrobiology},
         year = 2018,
        month = jun,
       volume = {18},
       number = {6},
        pages = {739-778},
          doi = {10.1089/ast.2017.1733},
archivePrefix = {arXiv},
       eprint = {1705.07098},
 primaryClass = {astro-ph.EP},
       adsurl = {https://ui.adsabs.harvard.edu/abs/2018AsBio..18..739F}
}

@ARTICLE{gilbert2023second,
       author = {{Gilbert}, Emily A. and {Vanderburg}, Andrew and 
{Rodriguez}, Joseph E. and {Hord}, Benjamin J. and {Clement}, Matthew S. 
and {Barclay}, Thomas and {Quintana}, Elisa V. and {Schlieder}, Joshua E. 
and {Kane}, Stephen R. and {Jenkins}, Jon M. and {Twicken}, Joseph D. and 
{Kunimoto}, Michelle and {Vanderspek}, Roland and {Arney}, Giada N. and 
{Charbonneau}, David and {G{\"u}nther}, Maximilian N. and {Huang}, Chelsea 
X. and {Isopi}, Giovanni and {Kostov}, Veselin B. and {Kristiansen}, 
Martti H. and {Latham}, David W. and {Mallia}, Franco and {Mamajek}, Eric 
E. and {Mireles}, Ismael and {Quinn}, Samuel N. and {Ricker}, George R. 
and {Schulte}, Jack and {Seager}, S. and {Suissa}, Gabrielle and {Winn}, 
Joshua N. and {Youngblood}, Allison and {Zapparata}, Aldo},
        title = "{A Second Earth-sized Planet in the Habitable Zone of the 
M Dwarf, TOI-700}",
      journal = {\apjl},
         year = 2023,
        month = feb,
       volume = {944},
       number = {2},
          eid = {L35},
        pages = {L35},
          doi = {10.3847/2041-8213/acb599},
archivePrefix = {arXiv},
       eprint = {2301.03617},
 primaryClass = {astro-ph.EP},
       adsurl = {https://ui.adsabs.harvard.edu/abs/2023ApJ...944L..35G}
}

@ARTICLE{gillon2017seven,
       author = {{Gillon}, Micha{\"e}l and {Triaud}, Amaury H.~M.~J. and 
{Demory}, Brice-Olivier and {Jehin}, Emmanu{\"e}l and {Agol}, Eric and 
{Deck}, Katherine M. and {Lederer}, Susan M. and {de Wit}, Julien and 
{Burdanov}, Artem and {Ingalls}, James G. and {Bolmont}, Emeline and 
{Leconte}, Jeremy and {Raymond}, Sean N. and {Selsis}, Franck and 
{Turbet}, Martin and {Barkaoui}, Khalid and {Burgasser}, Adam and 
{Burleigh}, Matthew R. and {Carey}, Sean J. and {Chaushev}, Aleksander and 
{Copperwheat}, Chris M. and {Delrez}, Laetitia and {Fernandes}, Catarina 
S. and {Holdsworth}, Daniel L. and {Kotze}, Enrico J. and {Van Grootel}, 
Val{\'e}rie and {Almleaky}, Yaseen and {Benkhaldoun}, Zouhair and 
{Magain}, Pierre and {Queloz}, Didier},
        title = "{Seven temperate terrestrial planets around the nearby 
ultracool dwarf star TRAPPIST-1}",
      journal = {\nat},
         year = 2017,
        month = feb,
       volume = {542},
       number = {7642},
        pages = {456-460},
          doi = {10.1038/nature21360},
archivePrefix = {arXiv},
       eprint = {1703.01424},
 primaryClass = {astro-ph.EP},
       adsurl = {https://ui.adsabs.harvard.edu/abs/2017Natur.542..456G}
}

@ARTICLE{lhb1,
       author = {{Gomes}, R. and {Levison}, H.~F. and {Tsiganis}, K. and {Morbidelli}, A.},
        title = "{Origin of the cataclysmic Late Heavy Bombardment period of the terrestrial planets}",
      journal = {\nat},
         year = 2005,
        month = may,
       volume = {435},
       number = {7041},
        pages = {466-469},
          doi = {10.1038/nature03676},
       adsurl = {https://ui.adsabs.harvard.edu/abs/2005Natur.435..466G}
}

@ARTICLE{HarHab,
       author = {{Harada}, Caleb K. and {Dressing}, Courtney D. and {Kane}, Stephen R. and {Ardestani}, Bahareh Adami},
        title = "{Setting the Stage for the Search for Life with the Habitable Worlds Observatory: Properties of 164 Promising Planet-survey Targets}",
      journal = {\apjs},
         year = 2024,
        month = jun,
       volume = {272},
       number = {2},
          eid = {30},
        pages = {30},
          doi = {10.3847/1538-4365/ad3e81},
archivePrefix = {arXiv},
       eprint = {2401.03047},
 primaryClass = {astro-ph.EP},
       adsurl = {https://ui.adsabs.harvard.edu/abs/2024ApJS..272...30H}
}

@Article{harris2020array,
 title         = {Array programming with {NumPy}},
 author        = {Charles R. Harris and K. Jarrod Millman and St{\'{e}}fan J.
                 van der Walt and Ralf Gommers and Pauli Virtanen and David
                 Cournapeau and Eric Wieser and Julian Taylor and Sebastian
                 Berg and Nathaniel J. Smith and Robert Kern and Matti Picus
                 and Stephan Hoyer and Marten H. van Kerkwijk and Matthew
                 Brett and Allan Haldane and Jaime Fern{\'{a}}ndez del
                 R{\'{i}}o and Mark Wiebe and Pearu Peterson and Pierre
                 G{\'{e}}rard-Marchant and Kevin Sheppard and Tyler Reddy and
                 Warren Weckesser and Hameer Abbasi and Christoph Gohlke and
                 Travis E. Oliphant},
 year          = {2020},
 month         = sep,
 journal       = {Nature},
 volume        = {585},
 number        = {7825},
 pages         = {357--362},
 doi           = {10.1038/s41586-020-2649-2},
 publisher     = {Springer Science and Business Media {LLC}},
 url           = {https://doi.org/10.1038/s41586-020-2649-2}
}

@ARTICLE{haswell2020dispersed,
       author = {{Haswell}, Carole A. and {Staab}, Daniel and {Barnes}, John R. and {Anglada-Escud{\'e}}, Guillem and {Fossati}, Luca and {Jenkins}, James S. and {Norton}, Andrew J. and {Doherty}, James P.~J. and {Cooper}, Joseph},
        title = "{Dispersed Matter Planet Project discoveries of ablating planets orbiting nearby bright stars}",
      journal = {Nature Astronomy},
         year = 2020,
        month = jan,
       volume = {4},
        pages = {408-418},
          doi = {10.1038/s41550-019-0973-y},
archivePrefix = {arXiv},
       eprint = {1912.10874},
 primaryClass = {astro-ph.EP},
       adsurl = {https://ui.adsabs.harvard.edu/abs/2020NatAs...4..408H}
}

@ARTICLE{hebrard2016sophie,
       author = {{H{\'e}brard}, G. and {Arnold}, L. and {Forveille}, T. and {Correia}, A.~C.~M. and {Laskar}, J. and {Bonfils}, X. and {Boisse}, I. and {D{\'\i}az}, R.~F. and {Hagelberg}, J. and {Sahlmann}, J. and {Santos}, N.~C. and {Astudillo-Defru}, N. and {Borgniet}, S. and {Bouchy}, F. and {Bourrier}, V. and {Courcol}, B. and {Delfosse}, X. and {Deleuil}, M. and {Demangeon}, O. and {Ehrenreich}, D. and {Gregorio}, J. and {Jovanovic}, N. and {Labrevoir}, O. and {Lagrange}, A. -M. and {Lovis}, C. and {Lozi}, J. and {Moutou}, C. and {Montagnier}, G. and {Pepe}, F. and {Rey}, J. and {Santerne}, A. and {S{\'e}gransan}, D. and {Udry}, S. and {Vanhuysse}, M. and {Vigan}, A. and {Wilson}, P.~A.},
        title = "{The SOPHIE search for northern extrasolar planets. X. Detection and characterization of giant planets by the dozen}",
      journal = {\aap},
         year = 2016,
        month = apr,
       volume = {588},
          eid = {A145},
        pages = {A145},
          doi = {10.1051/0004-6361/201527585},
archivePrefix = {arXiv},
       eprint = {1602.04622},
 primaryClass = {astro-ph.EP},
       adsurl = {https://ui.adsabs.harvard.edu/abs/2016A&A...588A.145H}
}

@ARTICLE{henry2000transiting,
       author = {{Henry}, Gregory W. and {Marcy}, Geoffrey W. and {Butler}, R. Paul and {Vogt}, Steven S.},
        title = "{A Transiting ``51 Peg-like'' Planet}",
      journal = {\apjl},
         year = 2000,
        month = jan,
       volume = {529},
       number = {1},
        pages = {L41-L44},
          doi = {10.1086/312458},
       adsurl = {https://ui.adsabs.harvard.edu/abs/2000ApJ...529L..41H}
}

@ARTICLE{LV3,
       author = {{Horner}, J. and {Mousis}, O. and {Petit}, J. -M. and {Jones}, B.~W.},
        title = "{Differences between the impact regimes of the terrestrial planets: Implications for primordial D:H ratios}",
      journal = {\planss},
         year = 2009,
        month = oct,
       volume = {57},
       number = {12},
        pages = {1338-1345},
          doi = {10.1016/j.pss.2009.06.006},
       adsurl = {https://ui.adsabs.harvard.edu/abs/2009P&SS...57.1338H}
}

@ARTICLE{HabRev,
       author = {{Horner}, J. and {Jones}, B.~W.},
        title = "{Determining habitability: which exoEarths should we search for life?}",
      journal = {International Journal of Astrobiology},
         year = 2010,
        month = oct,
       volume = {9},
       number = {4},
        pages = {273-291},
          doi = {10.1017/S1473550410000261},
archivePrefix = {arXiv},
       eprint = {1007.3413},
 primaryClass = {astro-ph.EP},
       adsurl = {https://ui.adsabs.harvard.edu/abs/2010IJAsB...9..273H}
}

@ARTICLE{FoF1,
       author = {{Horner}, J. and {Jones}, B.~W.},
        title = "{Jupiter {\textendash} friend or foe? I: The asteroids}",
      journal = {International Journal of Astrobiology},
         year = 2008,
        month = oct,
       volume = {7},
       number = {3-4},
        pages = {251-261},
          doi = {10.1017/S1473550408004187},
archivePrefix = {arXiv},
       eprint = {0806.2795},
 primaryClass = {astro-ph},
       adsurl = {https://ui.adsabs.harvard.edu/abs/2008IJAsB...7..251H}
}

@ARTICLE{FoF2,
       author = {{Horner}, J. and {Jones}, B.~W.},
        title = "{Jupiter - friend or foe? II: the Centaurs}",
      journal = {International Journal of Astrobiology},
         year = 2009,
        month = apr,
       volume = {8},
       number = {2},
        pages = {75-80},
          doi = {10.1017/S1473550408004357},
archivePrefix = {arXiv},
       eprint = {0903.3305},
 primaryClass = {astro-ph.EP},
       adsurl = {https://ui.adsabs.harvard.edu/abs/2009IJAsB...8...75H}
}

@ARTICLE{FoF3,
       author = {{Horner}, J. and {Jones}, B.~W. and {Chambers}, J.},
        title = "{Jupiter - friend or foe? III: the Oort cloud comets}",
      journal = {International Journal of Astrobiology},
         year = 2010,
        month = jan,
       volume = {9},
       number = {1},
        pages = {1-10},
          doi = {10.1017/S1473550409990346},
archivePrefix = {arXiv},
       eprint = {0911.4381},
 primaryClass = {astro-ph.EP},
       adsurl = {https://ui.adsabs.harvard.edu/abs/2010IJAsB...9....1H}
}

@ARTICLE{FoF4,
       author = {{Horner}, J. and {Jones}, B.~W.},
        title = "{Jupiter - friend or foe? IV: the influence of orbital eccentricity and inclination}",
      journal = {International Journal of Astrobiology},
         year = 2012,
        month = jul,
       volume = {11},
       number = {3},
        pages = {147-156},
          doi = {10.1017/S1473550412000043},
archivePrefix = {arXiv},
       eprint = {1111.3144},
 primaryClass = {astro-ph.EP},
       adsurl = {https://ui.adsabs.harvard.edu/abs/2012IJAsB..11..147H}
}

@ARTICLE{Graz16,
       author = {{Grazier}, Kevin R.},
        title = "{Jupiter: Cosmic Jekyll and Hyde}",
      journal = {Astrobiology},
         year = 2016,
        month = jan,
       volume = {16},
       number = {1},
        pages = {23-38},
          doi = {10.1089/ast.2015.1321},
       adsurl = {https://ui.adsabs.harvard.edu/abs/2016AsBio..16...23G}
}

@ARTICLE{Milank20,
       author = {{Horner}, Jonathan and {Vervoort}, Pam and {Kane}, Stephen R. and {Ceja}, Alma Y. and {Waltham}, David and {Gilmore}, James and {Kirtland Turner}, Sandra},
        title = "{Quantifying the Influence of Jupiter on the Earth{\textquoteright}s Orbital Cycles}",
      journal = {\aj},
         year = 2020,
        month = jan,
       volume = {159},
       number = {1},
          eid = {10},
        pages = {10},
          doi = {10.3847/1538-3881/ab5365},
archivePrefix = {arXiv},
       eprint = {1910.14250},
 primaryClass = {astro-ph.EP},
       adsurl = {https://ui.adsabs.harvard.edu/abs/2020AJ....159...10H}
}

@ARTICLE{SSRev,
       author = {{Horner}, J. and {Kane}, S.~R. and {Marshall}, J.~P. and {Dalba}, P.~A. and {Holt}, T.~R. and {Wood}, J. and {Maynard-Casely}, H.~E. and {Wittenmyer}, R. and {Lykawka}, P.~S. and {Hill}, M. and {Salmeron}, R. and {Bailey}, J. and {L{\"o}hne}, T. and {Agnew}, M. and {Carter}, B.~D. and {Tylor}, C.~C.~E.},
        title = "{Solar System Physics for Exoplanet Research}",
      journal = {\pasp},
         year = 2020,
        month = oct,
       volume = {132},
       number = {1016},
          eid = {102001},
        pages = {102001},
          doi = {10.1088/1538-3873/ab8eb9},
archivePrefix = {arXiv},
       eprint = {2004.13209},
 primaryClass = {astro-ph.EP},
       adsurl = {https://ui.adsabs.harvard.edu/abs/2020PASP..132j2001H}
}

@Article{Hunter:2007,
  Author    = {Hunter, J. D.},
  Title     = {Matplotlib: A 2D graphics environment},
  Journal   = {Computing in Science \& Engineering},
  Volume    = {9},
  Number    = {3},
  Pages     = {90--95},
  publisher = {IEEE COMPUTER SOC},
  doi       = {10.1109/MCSE.2007.55},
  year      = 2007
}

@ARTICLE{johnson2010hot,
       author = {{Johnson}, John Asher and {Bowler}, Brendan P. and {Howard}, Andrew W. and {Henry}, Gregory W. and {Marcy}, Geoffrey W. and {Isaacson}, Howard and {Brewer}, John Michael and {Fischer}, Debra A. and {Morton}, Timothy D. and {Crepp}, Justin R.},
        title = "{A Hot Jupiter Orbiting the 1.7 M $_{sun}$ Subgiant HD 102956}",
      journal = {\apjl},
         year = 2010,
        month = oct,
       volume = {721},
       number = {2},
        pages = {L153-L157},
          doi = {10.1088/2041-8205/721/2/L153},
archivePrefix = {arXiv},
       eprint = {1007.4555},
 primaryClass = {astro-ph.EP},
       adsurl = {https://ui.adsabs.harvard.edu/abs/2010ApJ...721L.153J}
}

@ARTICLE{johnson2006eccentric,
       author = {{Johnson}, John Asher and {Marcy}, Geoffrey W. and {Fischer}, Debra A. and {Henry}, Gregory W. and {Wright}, Jason T. and {Isaacson}, Howard and {McCarthy}, Chris},
        title = "{An Eccentric Hot Jupiter Orbiting the Subgiant HD 185269}",
      journal = {\apj},
         year = 2006,
        month = dec,
       volume = {652},
       number = {2},
        pages = {1724-1728},
          doi = {10.1086/508255},
archivePrefix = {arXiv},
       eprint = {astro-ph/0608035},
 primaryClass = {astro-ph},
       adsurl = {https://ui.adsabs.harvard.edu/abs/2006ApJ...652.1724J}
}

@ARTICLE{johnson2006n2k,
       author = {{Johnson}, John Asher and {Marcy}, Geoffrey W. and {Fischer}, Debra A. and {Laughlin}, Gregory and {Butler}, R. Paul and {Henry}, Gregory W. and {Valenti}, Jeff A. and {Ford}, Eric B. and {Vogt}, Steven S. and {Wright}, Jason T.},
        title = "{The N2K Consortium. VI. Doppler Shifts without Templates and Three New Short-Period Planets}",
      journal = {\apj},
         year = 2006,
        month = aug,
       volume = {647},
       number = {1},
        pages = {600-611},
          doi = {10.1086/505173},
archivePrefix = {arXiv},
       eprint = {astro-ph/0604348},
 primaryClass = {astro-ph},
       adsurl = {https://ui.adsabs.harvard.edu/abs/2006ApJ...647..600J}
}

@ARTICLE{kipping2013arametrizing,
       author = {{Kipping}, D.~M.},
        title = "{Parametrizing the exoplanet eccentricity distribution with the beta  distribution.}",
      journal = {\mnras},
         year = 2013,
        month = jul,
       volume = {434},
        pages = {L51-L55},
          doi = {10.1093/mnrasl/slt075},
archivePrefix = {arXiv},
       eprint = {1306.4982},
 primaryClass = {astro-ph.EP},
       adsurl = {https://ui.adsabs.harvard.edu/abs/2013MNRAS.434L..51K}
}

@ARTICLE{knutson2014friends,
       author = {{Knutson}, Heather A. and {Fulton}, Benjamin J. and {Montet}, Benjamin T. and {Kao}, Melodie and {Ngo}, Henry and {Howard}, Andrew W. and {Crepp}, Justin R. and {Hinkley}, Sasha and {Bakos}, Gaspar {\'A}. and {Batygin}, Konstantin and {Johnson}, John Asher and {Morton}, Timothy D. and {Muirhead}, Philip S.},
        title = "{Friends of Hot Jupiters. I. A Radial Velocity Search for Massive, Long-period Companions to Close-in Gas Giant Planets}",
      journal = {\apj},
         year = 2014,
        month = apr,
       volume = {785},
       number = {2},
          eid = {126},
        pages = {126},
          doi = {10.1088/0004-637X/785/2/126},
archivePrefix = {arXiv},
       eprint = {1312.2954},
 primaryClass = {astro-ph.EP},
       adsurl = {https://ui.adsabs.harvard.edu/abs/2014ApJ...785..126K}
}

@ARTICLE{kunimoto2020searching,
       author = {{Kunimoto}, Michelle and {Matthews}, Jaymie M.},
        title = "{Searching the Entirety of Kepler Data. II. Occurrence 
Rate Estimates for FGK Stars}",
      journal = {\aj},
         year = 2020,
        month = jun,
       volume = {159},
       number = {6},
          eid = {248},
        pages = {248},
          doi = {10.3847/1538-3881/ab88b0},
archivePrefix = {arXiv},
       eprint = {2004.05296},
 primaryClass = {astro-ph.EP},
       adsurl = {https://ui.adsabs.harvard.edu/abs/2020AJ....159..248K}
}

@ARTICLE{lhb2,
       author = {{Levison}, Harold F. and {Morbidelli}, Alessandro and {Tsiganis}, Kleomenis and {Nesvorn{\'y}}, David and {Gomes}, Rodney},
        title = "{Late Orbital Instabilities in the Outer Planets Induced by Interaction with a Self-gravitating Planetesimal Disk}",
      journal = {\aj},
         year = 2011,
        month = nov,
       volume = {142},
       number = {5},
          eid = {152},
        pages = {152},
          doi = {10.1088/0004-6256/142/5/152},
       adsurl = {https://ui.adsabs.harvard.edu/abs/2011AJ....142..152L}
}

@ARTICLE{locurto2006harps,
       author = {{Lo Curto}, G. and {Mayor}, M. and {Clausen}, J.~V. and {Benz}, W. and {Bouchy}, F. and {Lovis}, C. and {Moutou}, C. and {Naef}, D. and {Pepe}, F. and {Queloz}, D. and {Santos}, N.~C. and {Sivan}, J. -P. and {Udry}, S. and {Bonfils}, X. and {Delfosse}, X. and {Mordasini}, C. and {Fouqu{\'e}}, P. and {Olsen}, E.~H. and {Pritchard}, J.~D.},
        title = "{The HARPS search for southern extra-solar planets. VII. A very hot Jupiter orbiting HD{\,}212301}",
      journal = {\aap},
         year = 2006,
        month = may,
       volume = {451},
       number = {1},
        pages = {345-350},
          doi = {10.1051/0004-6361:20054083},
       adsurl = {https://ui.adsabs.harvard.edu/abs/2006A&A...451..345L}
}

@ARTICLE{locurto2013harps,
       author = {{Lo Curto}, G. and {Mayor}, M. and {Benz}, W. and {Bouchy}, F. and {H{\'e}brard}, G. and {Lovis}, C. and {Moutou}, C. and {Naef}, D. and {Pepe}, F. and {Queloz}, D. and {Santos}, N.~C. and {Segransan}, D. and {Udry}, S.},
        title = "{The HARPS search for southern extra-solar planets . XXXII. New multi-planet systems in the HARPS volume limited sample: a super-Earth and a Neptune in the habitable zone}",
      journal = {\aap},
         year = 2013,
        month = mar,
       volume = {551},
          eid = {A59},
        pages = {A59},
          doi = {10.1051/0004-6361/201220415},
archivePrefix = {arXiv},
       eprint = {1301.2741},
 primaryClass = {astro-ph.EP},
       adsurl = {https://ui.adsabs.harvard.edu/abs/2013A&A...551A..59L}
}

@ARTICLE{maciejewski2024tracking,
       author = {{Maciejewski}, G. and {Niedzielski}, A. and {Go{\'z}dziewski}, K. and {Wolszczan}, A. and {Villaver}, E. and {Fern{\'a}ndez}, M. and {Adam{\'o}w}, M. and {Sierzputowska}, J.},
        title = "{Tracking Advanced Planetary Systems (TAPAS) with HARPS-N. VIII. A wide-orbit planetary companion in the hot-Jupiter system HD 118203}",
      journal = {\aap},
         year = 2024,
        month = aug,
       volume = {688},
          eid = {A172},
        pages = {A172},
          doi = {10.1051/0004-6361/202451084},
archivePrefix = {arXiv},
       eprint = {2407.11706},
 primaryClass = {astro-ph.EP},
       adsurl = {https://ui.adsabs.harvard.edu/abs/2024A&A...688A.172M}
}

@ARTICLE{malavolta2016gaps,
       author = {{Malavolta}, L. and {Nascimbeni}, V. and {Piotto}, G. and {Quinn}, S.~N. and {Borsato}, L. and {Granata}, V. and {Bonomo}, A.~S. and {Marzari}, F. and {Bedin}, L.~R. and {Rainer}, M. and {Desidera}, S. and {Lanza}, A.~F. and {Poretti}, E. and {Sozzetti}, A. and {White}, R.~J. and {Latham}, D.~W. and {Cunial}, A. and {Libralato}, M. and {Nardiello}, D. and {Boccato}, C. and {Claudi}, R.~U. and {Cosentino}, R. and {Covino}, E. and {Gratton}, R. and {Maggio}, A. and {Micela}, G. and {Molinari}, E. and {Pagano}, I. and {Smareglia}, R. and {Affer}, L. and {Andreuzzi}, G. and {Aparicio}, A. and {Benatti}, S. and {Bignamini}, A. and {Borsa}, F. and {Damasso}, M. and {Di Fabrizio}, L. and {Harutyunyan}, A. and {Esposito}, M. and {Fiorenzano}, A.~F.~M. and {Gandolfi}, D. and {Giacobbe}, P. and {Gonz{\'a}lez Hern{\'a}ndez}, J.~I. and {Maldonado}, J. and {Masiero}, S. and {Molinaro}, M. and {Pedani}, M. and {Scandariato}, G.},
        title = "{The GAPS programme with HARPS-N at TNG. XI. Pr 0211 in M 44: the first multi-planet system in an open cluster}",
      journal = {\aap},
         year = 2016,
        month = apr,
       volume = {588},
          eid = {A118},
        pages = {A118},
          doi = {10.1051/0004-6361/201527933},
archivePrefix = {arXiv},
       eprint = {1602.00009},
 primaryClass = {astro-ph.EP},
       adsurl = {https://ui.adsabs.harvard.edu/abs/2016A&A...588A.118M}
}

@ARTICLE{marcy1997planet,
       author = {{Marcy}, Geoffrey W. and {Butler}, R. Paul and {Williams}, Eric and {Bildsten}, Lars and {Graham}, James R. and {Ghez}, Andrea M. and {Jernigan}, J. Garrett},
        title = "{The Planet around 51 Pegasi}",
      journal = {\apj},
         year = 1997,
        month = may,
       volume = {481},
       number = {2},
        pages = {926-935},
          doi = {10.1086/304088},
       adsurl = {https://ui.adsabs.harvard.edu/abs/1997ApJ...481..926M}
}

@article{mayor1995jupiter,
  title={A {J}upiter-mass companion to a solar-type star},
  author={Mayor, Michel and Queloz, Didier},
  journal={Nature},
  volume={378},
  number={6555},
  pages={355--359},
  year={1995},
  publisher={Nature Publishing Group}
}

@ARTICLE{LV2,
       author = {{Morbidelli}, A. and {Chambers}, J. and {Lunine}, J.~I. and {Petit}, J.~M. and {Robert}, F. and {Valsecchi}, G.~B. and {Cyr}, K.~E.},
        title = "{Source regions and time scales for the delivery of water to Earth}",
      journal = {Meteoritics \& Planetary Science},
         year = 2000,
        month = nov,
       volume = {35},
       number = {6},
        pages = {1309-1320},
          doi = {10.1111/j.1945-5100.2000.tb01518.x},
       adsurl = {https://ui.adsabs.harvard.edu/abs/2000M&PS...35.1309M}
}

@article{moutou2015harps,
  title={The \textsc{HARPS} search for southern extra-solar planets-{XXXVII}. {F}ive new long-period giant planets and a system update},
  author={Moutou, C and Curto, G Lo and Mayor, M and Bouchy, F and Benz, Willy and Lovis, C and Naef, D and Pepe, F and Queloz, D and Santos, NC and others},
  journal={Astronomy \& Astrophysics},
  volume={576},
  pages={A48},
  year={2015},
  publisher={EDP Sciences}
}

@ARTICLE{moutou2005harps,
       author = {{Moutou}, C. and {Mayor}, M. and {Bouchy}, F. and {Lovis}, C. and {Pepe}, F. and {Queloz}, D. and {Santos}, N.~C. and {Udry}, S. and {Benz}, W. and {Lo Curto}, G. and {Naef}, D. and {S{\'e}gransan}, D. and {Sivan}, J. -P.},
        title = "{The HARPS search for southern extra-solar planets. IV. Three close-in planets around HD{\,}2638, HD{\,}27894 and HD{\,}63454}",
      journal = {\aap},
         year = 2005,
        month = aug,
       volume = {439},
       number = {1},
        pages = {367-373},
          doi = {10.1051/0004-6361:20052826},
       adsurl = {https://ui.adsabs.harvard.edu/abs/2005A&A...439..367M}
}

@ARTICLE{naef2001coralie,
       author = {{Naef}, D. and {Mayor}, M. and {Pepe}, F. and {Queloz}, D. and {Santos}, N.~C. and {Udry}, S. and {Burnet}, M.},
        title = "{The CORALIE survey for southern extrasolar planets. V. 3 new extrasolar planets}",
      journal = {\aap},
         year = 2001,
        month = aug,
       volume = {375},
        pages = {205-218},
          doi = {10.1051/0004-6361:20010841},
archivePrefix = {arXiv},
       eprint = {astro-ph/0106255},
 primaryClass = {astro-ph},
       adsurl = {https://ui.adsabs.harvard.edu/abs/2001A&A...375..205N}
}

@ARTICLE{naef2004elodie,
       author = {{Naef}, D. and {Mayor}, M. and {Beuzit}, J.~L. and {Perrier}, C. and {Queloz}, D. and {Sivan}, J.~P. and {Udry}, S.},
        title = "{The ELODIE survey for northern extra-solar planets. III. Three planetary candidates detected with ELODIE}",
      journal = {\aap},
         year = 2004,
        month = jan,
       volume = {414},
        pages = {351-359},
          doi = {10.1051/0004-6361:20034091},
archivePrefix = {arXiv},
       eprint = {astro-ph/0310261},
 primaryClass = {astro-ph},
       adsurl = {https://ui.adsabs.harvard.edu/abs/2004A&A...414..351N}
}

@ARTICLE{nesv18,
       author = {{Nesvorn{\'y}}, David},
        title = "{Dynamical Evolution of the Early Solar System}",
      journal = {\araa},
         year = 2018,
        month = sep,
       volume = {56},
        pages = {137-174},
          doi = {10.1146/annurev-astro-081817-052028},
archivePrefix = {arXiv},
       eprint = {1807.06647},
 primaryClass = {astro-ph.EP},
       adsurl = {https://ui.adsabs.harvard.edu/abs/2018ARA&A..56..137N}
}

@ARTICLE{LV4,
       author = {{O'Brien}, David P. and {Walsh}, Kevin J. and {Morbidelli}, Alessandro and {Raymond}, Sean N. and {Mandell}, Avi M.},
        title = "{Water delivery and giant impacts in the {\textquoteleft}Grand Tack{\textquoteright} scenario}",
      journal = {\icarus},
         year = 2014,
        month = sep,
       volume = {239},
        pages = {74-84},
          doi = {10.1016/j.icarus.2014.05.009},
archivePrefix = {arXiv},
       eprint = {1407.3290},
 primaryClass = {astro-ph.EP},
       adsurl = {https://ui.adsabs.harvard.edu/abs/2014Icar..239...74O}
}

@ARTICLE{LV1,
       author = {{Owen}, Tobias and {Bar-Nun}, Akiva},
        title = "{Comets, Impacts, and Atmospheres}",
      journal = {\icarus},
         year = 1995,
        month = jan,
       volume = {116},
       number = {2},
        pages = {215-226},
          doi = {10.1006/icar.1995.1122},
       adsurl = {https://ui.adsabs.harvard.edu/abs/1995Icar..116..215O}
}

@ARTICLE{pepe2014espresso,
       author = {{Pepe}, F. and {Molaro}, P. and {Cristiani}, S. and {Rebolo}, R. and {Santos}, N.~C. and {Dekker}, H. and {M{\'e}gevand}, D. and {Zerbi}, F.~M. and {Cabral}, A. and {Di Marcantonio}, P. and {Abreu}, M. and {Affolter}, M. and {Aliverti}, M. and {Allende Prieto}, C. and {Amate}, M. and {Avila}, G. and {Baldini}, V. and {Bristow}, P. and {Broeg}, C. and {Cirami}, R. and {Coelho}, J. and {Conconi}, P. and {Coretti}, I. and {Cupani}, G. and {D'Odorico}, V. and {De Caprio}, V. and {Delabre}, B. and {Dorn}, R. and {Figueira}, P. and {Fragoso}, A. and {Galeotta}, S. and {Genolet}, L. and {Gomes}, R. and {Gonz{\'a}lez Hern{\'a}ndez}, J.~I. and {Hughes}, I. and {Iwert}, O. and {Kerber}, F. and {Landoni}, M. and {Lizon}, J. -L. and {Lovis}, C. and {Maire}, C. and {Mannetta}, M. and {Martins}, C. and {Monteiro}, M. and {Oliveira}, A. and {Poretti}, E. and {Rasilla}, J.~L. and {Riva}, M. and {Santana Tschudi}, S. and {Santos}, P. and {Sosnowska}, D. and {Sousa}, S. and {Span{\'o}}, P. and {Tenegi}, F. and {Toso}, G. and {Vanzella}, E. and {Viel}, M. and {Zapatero Osorio}, M.~R.},
        title = "{ESPRESSO: The next European exoplanet hunter}",
      journal = {Astronomische Nachrichten},
         year = 2014,
        month = jan,
       volume = {335},
       number = {1},
        pages = {8},
          doi = {12.1002/asna.201312004},
       adsurl = {https://ui.adsabs.harvard.edu/abs/2014AN....335....8P}
}

@book{perryman2018exoplanet,
  title={The exoplanet handbook},
  author={Perryman, Michael},
  year={2018},
  publisher={Cambridge University Press}
}


\end{document}